\documentclass[12pt]{article}
\usepackage{epsf}
\usepackage{aas_macros}
\usepackage{amssymb}
\usepackage{comment}
\usepackage{amsmath}
\usepackage{braket}
\usepackage{mathrsfs}
\usepackage{mathtools}
\usepackage{array}
\usepackage{fancyhdr}
\usepackage[dvipdfmx]{graphicx}
\usepackage{color}
\usepackage{cite}
\usepackage{bm}
\usepackage{url}
\usepackage{fancyhdr}
\usepackage[colorlinks,citecolor=blue]{hyperref}
\usepackage[hang,small,bf]{caption}
\usepackage[subrefformat=parens]{subcaption}
\newcommand{\rmv}[1]{%
}

\makeatletter

\@addtoreset{equation}{section}
\makeatother

\def\be{\begin{equation}}
\def\ee{\end{equation}}
\def\ben{\begin{eqnarray}}
\def\een{\end{eqnarray}}

\begin{document}


\begin{center}

\vskip .75in

{\Large \bf Weak lensing of gravitational waves in wave optics: Beyond the Born approximation}

\vskip .75in

{\large
Morifumi Mizuno$\,^1$,  Teruaki Suyama$\,^1$
}

\vskip 0.25in

{\em
$^{1}$Department of Physics, Tokyo Institute of Technology, 2-12-1 Ookayama, Meguro-ku,
Tokyo 152-8551, Japan
}

\end{center}
\vskip .5in

\begin{abstract}
The Universe's matter inhomogeneity gravitationally affects the propagation of 
gravitational waves (GWs), causing the lensing effect.
Particularly, the weak lensing of GWs has been studied within the range of the Born approximation to constrain the small-scale power spectrum.
In this work, 
the validity of the Born approximation is investigated by accounting for the higher-order terms in the gravitational potential $\Phi$. 
To do so, we formulate the post-Born approximation and derive the magnification $K$ and the phase modulation $S$ up to third order in $\Phi$.
We find that the average of $S$ and $K$ is non-zero and that the average of $S$ depends on the size of the point mass.
Due to this size dependency, the signal is enhanced, and the number of GW events required for detecting the average of $S$ decreases.
We find that this number can become comparable to or even smaller than the number required for detecting the variance of $S$ in certain scenarios.
In addition, it is verified that, for lensing by dark low-mass halos, the post-Born corrections are a few orders of magnitude smaller than the Born approximation at $f\geq0.01$~Hz.
However, in the presence of the point mass, there is a condition under which the Born approximation fails.
We derive the correction terms to the Born approximation and identify the condition under which the Born approximation no longer holds.
For the magnification, the Born approximation is valid as long as the wavelength of GWs is larger than the Schwarzschild radius of lenses, 
while for the phase modulation, this condition is modified due to the physical size of the point mass.

\end{abstract}

\section{Introduction}
When light travels across the Universe, its trajectory is bent
by the gravitational potential of intervening massive objects. 
This phenomenon called gravitational lensing (GL) is quite useful in astrophysics and cosmology
(e.g., \cite{Bartelmann:2010fz, Mandelbaum:2017jpr, Oguri:2019fix}). 
For instance, it can be used to measure the cosmological parameters. 
It can probe the abundance of dark compact objects. 

According to general relativity, GL also occurs for the gravitational waves (GWs) \cite{MTW}.
One notable feature of the GL of GWs is that geometrical optics,
which is a perfect approximation in most cases for light, no longer
holds for GWs in some cases since the wavelength of GWs is typically
much larger than that of light and diffraction effect becomes important \cite{Ohanian:1974ys, Takahashi:2003ix, Nakamura:1997sw}.
In such cases, wave optics must be used to deal with the GL.
In wave optics, contrary to the geometrical optics where the 
starting point is the lens equation, the lensing signal is represented by the so-called amplification factor defined as a ratio of the lensed waveform to the unlensed one (e.g., \cite{Nakamura:1999}).
This quantity is a complex number and all the information of the lensing is encoded in it.
Its absolute value and argument represent the amplification and phase modulation of the lensed wave, respectively.  

In \cite{Takahashi:2005ug}, GL of GWs caused by dark matter fluctuations was studied.
It was shown that there is a length scale of the matter power spectrum
below which the contribution to the lensing signal is suppressed due to wavy nature. 
This scale, {\it Fresnel scale}, depends on the GW frequency.
Thus, by measuring the lensing signal at multiple frequencies and its frequency dependence, 
we can probe the matter power spectrum at the Fresnel scale.
This idea has been investigated in more detail by updating the matter power spectrum
as well as adding the compact objects in \cite{Oguri:2020ldf}.
In \cite{Oguri:2022zpn}, it was shown that the lensing signal of 
the dark matter fluctuations is hugely amplified by a massive object located 
on the line of sight which itself causes strong lensing.

When the lensing signal is weak, it is natural to keep only the terms 
first order in the gravitational potential $\Phi$ (i.e. Born approximation).
The contributions of the higher order terms are expected to be suppressed 
compared to the leading order contribution. 
In geometrical optics,
this has been explicitly demonstrated in \cite{Shapiro:2006em, Hilbert:2008kb, Krause:2009yr, Pratten:2016dsm, Petri:2016qya}. 
One naively expects that a similar conclusion can be drawn for the case of 
wave optics. 

In \cite{Takahashi:2005ug}, the amplification factor sourced by the dark matter fluctuations
was obtained under the Born approximation (thus the variance of the lensing signal
is second order in $\Phi$).
Typical magnitude of the amplitude and the phase fluctuations was found to be ${\cal O}(10^{-2}-10^{-3})$.
Thus, the lensing signal is weak and this would naturally justify the validity of the Born approximation.
However, there are two issues that need to be investigated regarding the Born approximation.
Firstly, although the post-Born corrections are expected to be small,
it is not known how much they are suppressed actually.
When the measurements of the lensing signal become available in the future,
quantitative computation of the magnitude of the post-Born corrections is indispensable to correctly
extract the matter power spectrum as well as to understand the level of the precision under consideration.
Secondly, the Born approximation used in \cite{Takahashi:2005ug} apparently breaks down at large wave frequency
since the gravitational potential in the wave equation is associated with the frequency.
Notice that this issue does not appear in the geometrical optics since the lens equation
is independent of the frequency of light. 
While it is known how the lens equation emerges in the wave optics, it is not obvious
how the breakdown of the Born approximation for the large frequency in the wave optics 
is reconciled with the Born approximation in the geometric optics. 
In this paper, we take a first step towards addressing these issues by extending the 
previous studies to next higher orders in the gravitational potential.
We first reformulate the wave equation to make its structure more tractable.
We then derive the expression of the lensing signal up to third order in the gravitational potential.
Expansion to this order is necessary to evaluate the variance of the post-Born corrections.
As we will demonstrate, the post-Born corrections are suppressed by a few orders of magnitude  compared to the leading 
order signal except in a high frequency region.
Interestingly, when the post-Born corrections are included, the average of the lensing signal does
not vanish. This average depends on the frequency in a non-trivial manner and thus cannot be
absorbed into the change of the parameters characterizing the unlensed waveform.
Our analysis suggests an interesting possibility to make use of the average of the lensing signal
as an additional observable to probe the matter power spectrum.

\section{Formulation}
\subsection{Lensing signal beyond the Born approximation}
In this section, we reformulate the wave equation and show how the 
post-Born corrections are derived.
Throughout this paper, 
we assume that the gravitational potential is small ($\Phi\ll1$) and the Universe is flat.
We also ignore the polarization of GWs since the polarization tensor
in the geometrical optics is parallel transported along the null geodesics
\cite{Schneider}
and hence the change of the polarization tensor would be suppressed by
a factor of ${\cal O}(\Phi)$ and observationally irrelevant.

The presence of mass fluctuation creates the distortion on spacetime, 
causing the deviation from the Friedmann–Lemaître–Robertson–Walker(FLRW) metric.
This effect is small in most of the astrophysical situations and it is a good approximation 
to write the metric as \cite{Dodelson}
\begin{align}
    \label{eq:FRW metric}
    ds^{2}=&g_{\mu\nu}^{B}dx^{\mu}dx^{\nu}
    =a^{2}(\eta)[-\left(1+2\Phi\right)d\eta^{2}+\left(1-2\Phi\right)d{\bm x}^{2}],
\end{align}
where $\eta$ and $\bm x$ is a conformal time and a comoving coordinate, 
and $a(\eta)$ is a scale factor. 
If the wavelength of GWs is much smaller than the typical radius of the curvature of the background metric, 
the propagation of GWs becomes the same as the wave equation 
of the massless scalar field $\phi$:
$\partial_{\mu}\left(\sqrt{-g^{B}}g^{\mu\nu}_{B}\partial_{\nu}\phi\right)=0$.
The expansion of the Universe causes attenuation of $\phi$ as $\phi \propto 1/a$.
We extract this effect by redefining the GW amplitude $\phi$ as $\phi \to \phi/a$.
Then, the wave equation becomes \cite{Nakamura:1999}
\begin{align}
    \label{eq:wave equation frequency}
    \left(\nabla^{2}+\omega^{2}\right)\tilde{\phi}=4\omega^{2}\Phi\tilde{\phi},
\end{align}
in the frequency space. 
$\tilde{\phi}(\omega, \bm x)$ is the Fourier transform of $\phi(\eta, \bm x)$
\footnote{It is defined by $\phi(\eta, \bm x)=\int \frac{d\omega}{2\pi} e^{-i\omega \eta}\tilde{\phi}(\omega, \bm x)$.
Thus, $\omega$ is the comoving (angular) frequency.} and 
the higher order terms in $\Phi$ have been ignored. 
It is common to represent the lensed waveform in terms of 
the amplification factor, 
which is the ratio of the lensed and unlensed waveform,
namely $F=\tilde{\phi}/\tilde{\phi}_{0}$ \cite{Nakamura:1999},
where the unlensed waveform is given by $\tilde{\phi}_{0}=e^{i\omega \chi}/\chi$
in terms of $\chi$ which is the (comoving) distance from the source. 
Using the amplification factor $F$, 
Eq.~(\ref{eq:wave equation frequency}) is rewritten as
\begin{align}
    \label{eq:amplification factor F}
    2i\omega \frac{\partial F}{\partial \chi}+\frac{1}{\chi^{2}}\nabla^{2}_{\theta}F=4\omega^{2} \Phi F,
\end{align}
where the polar coordinate ($\chi, \theta, \phi$) is used and 
$\nabla^{2}_{\theta}=\partial^{2} /\partial \theta^{2}+\sin\theta^{-1}\partial/\partial \theta+\sin\theta^{-2}\partial /\partial \phi^{2}$ 
is the 2 dimensional Laplace operator on 2-sphere. 
In Eq.~(\ref{eq:amplification factor F}), 
we have assumed GWs propagate along the line of sight 
and confined in the region $\theta\ll1$.
Therefore, $\nabla_{\theta}$ can be interpreted as the operator on 2-dimensional flat surface 
perpendicular to the line of sight.

In order to evaluate the effects of the post-Born approximation,
we find it convenient to deal with a new variable $J$ defined as $F=e^{i\omega J}$.
Using this new variable $J$, Eq.~(\ref{eq:amplification factor F}) becomes
\begin{align}
    \label{eq:J}
    \left(\frac{\partial }{\partial \chi}-\frac{i}{2\omega 
    \chi^{2}}\nabla_{\theta}^{2}\right)J=-2\Phi-\frac{1}{2\chi^{2}}(\nabla_{\theta}J)^{2}.
\end{align}
This differential equation can be written as an integral equation by the Green function of the linear operator 
acting on the right-hand side,
\begin{align}
    \label{eq: J integral}
    J(\chi_{s},\bm \theta)=&\int_{0}^{\chi_{s}} d\chi 
    \exp{\left[i\frac{W(\chi,\chi_{s})\nabla^{2}_{\theta}}{2\omega}\right]}\left(-
    2\Phi(\chi,\bm \theta)-\frac{1}{2\chi^{2}}(\nabla_{\theta}J)^{2}\right),
\end{align}
where $W(\chi,\chi_{s})=\frac{1}{\chi}-\frac{1}{\chi_{s}}$. 
In geometric optics, $W(\chi,\chi_{s})$ is sometimes called 
the lensing efficiency function\cite{Pratten:2016dsm}.
The change of variable to $J$ allows us to partially take into account 
the higher order terms
in the gravitational potential which are not included in the previous studies
\cite{Takahashi:2005ug, Oguri:2020ldf, Inamori:2021tlx}. 
Defining  $J^{(n)}$ as the term proportional to n-th order of the gravitational potential, 
$J^{(n)}$ can be calculated iteratively order by order as
\begin{align}
    \label{eq:J 1st order a }
    J^{(1)}(\chi_{s},\bm \theta)=&\int_{0}^{\chi_{s}} d\chi 
    \exp{\left[i\frac{W(\chi,\chi_{s})\nabla^{2}_{\theta}}{2\omega}\right]}(-2\Phi
    (\chi,\bm\theta)),
    \\
    \label{eq:J 2nd order a}
     J^{(2)}(\chi_{s},\bm\theta)=&-\int_{0}^{\chi_{s}} d\chi 
     \exp{\left[i\frac{W(\chi,\chi_{s})\nabla^{2}_{\theta}}{2\omega}\right]}\frac{
     (\nabla_{\theta}J^{(1)}(\chi,\bm\theta))^{2}}{2\chi^{2}},
     \\
     \label{eq:J 3rd order a}
     J^{(3)}(\chi_{s},\bm\theta)=&-\int_{0}^{\chi_{s}} d\chi 
     \exp{\left[i\frac{W(\chi,\chi_{s})\nabla^{2}_{\theta}}{2\omega}\right]}\frac{
     \nabla_{\theta}J^{(1)}(\chi,\bm\theta)\cdot 
     \nabla_{\theta}J^{(2)}(\chi,\bm\theta)}{\chi^{2}},
     \\
     \vdots&\notag
\end{align}
In the geometrical optics limit (i.e., large $\omega$),
$\lim_{\omega \to \infty} J^{(n)}$ of any $n$ becomes real and the
correction term at ${\cal O}(1/\omega)$ becomes imaginary.
Thus, $\lim_{\omega \to \infty} J$ is nothing but the difference between 
the arrival time of geodesic under the influence of $\Phi$ and the one without $\Phi$.
At the leading order, this reduces to the standard expression of the Shapiro time delay.
At ${\cal O}(1/\omega)$, $J$ gives the magnification in geometrical optics.
In particular, $i\omega J^{(1)}$ reduces to the standard formula of the convergence 
(e.g., \cite{Bartelmann:2010fz, Mandelbaum:2017jpr}).

In the literature, the Born approximation refers to the approximation to truncate the expansion of
$F$ up to first order in $\Phi$.
Meanwhile, since our expansion is performed for $J$,
even the truncation at $J^{(1)}$ partially captures the higher order terms not
included in the previous studies (see also the footnote \ref{f1}). 
In spite of such a difference at the conceptual level, there is practically no difference as to
whether the Born approximation refers to the first-order truncation for $F$ or $J$
since the variation of $F$ in the former case is nothing but $J^{(1)}$.

Our aim is to investigate the leading correction to the Born approximation
of the lensing signal caused by the dark matter fluctuations. 
To this end, we compute the average and the variance of $J$ and
investigate how they are affected by the post-Born approximation 
by treating $\Phi$ as a random variable.
The average trivially vanishes in the Born approximation.
Since the average of $J^{(2)}$ does not vanish in general, we truncate the evaluation
of the average at this order.
The leading post-Born correction to the variance comes from the cross term
$J^{(1)} J^{(2)}$ and thus it is ${\cal O}(\Phi^3)$.
This is non-vanishing only when $\Phi$ is non-Gaussian. 
The next leading correction, which is ${\cal O}(\Phi^4)$,
remains finite even when $\Phi$ is Gaussian.
Thus, the correction at ${\cal O}(\Phi^4)$ may dominate over the one
at ${\cal O}(\Phi^3)$ in some cases, especially when $\Phi$ is nearly Gaussian.
Because of this reason, we compute the variance up to ${\cal O}(\Phi^4)$.
To make our calculation consistent up to this order, we need to keep the expansion
up to $J^{(3)}$ since the cross term $J^{(1)}J^{(3)}$ is ${\cal O}(\Phi^4)$.

For clarity, we define new differential operators $(W\nabla)^{(2)}$ and $(W\nabla)^{(3)}$ as
\begin{align}
    \label{eq: 2nd order W nabla }
    (W\nabla)^{(2)}=&
    W(\chi,\chi_{s})\nabla^{2}_{\theta 12}
    +W(\chi_{1},\chi)\nabla^{2}_{\theta1}
    +W(\chi_{2},\chi)\nabla^{2}_{\theta2},
    \\
    \label{eq: 3rd order W nabla }
    (W\nabla)^{(3)}=&
    W(\chi,\chi_{s})\nabla_{\theta123}^{2}
    +W(\chi_{3},\chi)\nabla_{\theta3}^{2}
    +W(\chi',\chi)\nabla^{2}_{\theta12}
    \notag
    \\
    &\qquad\qquad\qquad\qquad
    +W(\chi_{1},\chi')\nabla^{2}_{\theta1}
    +W(\chi_{2},\chi')\nabla^{2}_{\theta2}.
\end{align}
Using these notations, we obtain the following expressions of $J$ up to third order:
\begin{align}
    \label{eq:J 1st order b}
    J^{(1)}(\chi_{s},\bm\theta)=&-2\int_{0}^{\chi_{s}} d\chi 
    \exp{\left[i\frac{W(\chi,\chi_{s})\nabla^{2}_{\theta}}{2\omega}\right]}\Phi(\chi,\bm\theta),
    \\
    \label{eq:J 2nd order b}
     J^{(2)}(\chi_{s},\bm\theta)=&-2\int_{0}^{\chi_{s}} 
     \frac{d\chi}{\chi^{2}}\int_{0}^{\chi}d\chi_{1}\int_{0}^{\chi}d\chi_{2} 
     \exp{\left[i\frac{(W\nabla)^{(2)}}{2\omega}\right]}\nabla_{\theta1}\Phi_{1}\cdot\nabla_{\theta2}\Phi_{2},
     \\
     \label{eq:J 3rd order b}
     J^{(3)}(\chi_{s},\bm\theta)=&-4\int_{0}^{\chi_{s}} 
     \frac{d\chi}{\chi^{2}}\int_{0}^{\chi}d\chi_{3}\int_{0}^{\chi} 
     \frac{d\chi'}{\chi'^{2}}\int_{0}^{\chi'}d\chi_{1}\int_{0}^{\chi'}d\chi_{2}\notag
     \\
     &\qquad\qquad\times\exp{\left[i\frac{(W\nabla)^{(3)}}{2\omega}\right]}\nabla_
     {\theta12}(\nabla_{\theta1}\Phi_{1}\cdot\nabla_{\theta2}\Phi_{2})\cdot\nabla_
     {\theta3}\Phi_{3}.
\end{align}
Note that $\Phi_{i}=\Phi(\chi_{i},\bm\theta)$ and $\nabla_{\theta i}$ only acts on $\Phi_{i}$ (when there are more than 2 subscript numbers at the corner of $\nabla_{\theta}$, 
it means the operator acts on the gravitational potentials that have the corresponding subscripts.). $J^{(1)}$ corresponds to the Born approximation and subsequent terms ($J^{(2)}$ and $J^{(3)}$) are the post-Born corrections. 

The information about the phase and the magnification of GWs is encoded 
in the real and imaginary part of $J$, respectively. 
Conventionally, the phase modulation and the magnification are denoted as $S $ and $K$ \cite{Takahashi:2005ug}, and we follow the same notation in this paper. 
$S$ and $K$ are related to $J$ as
\begin{align}
    \label{eq: phase S}
    S(\omega)=&\omega \text{Re}(J),
    \\
    \label{eq: magnification}
    K(\omega)=&-\omega \text{Im}(J).
\end{align}
In this definition, 
the amplification factor is written as $F(\omega)=e^{K(\omega)}e^{iS(\omega)}$
\footnote{In \cite{Takahashi:2005ug, Oguri:2020ldf, Inamori:2021tlx}, $F$ was written as $F=1+K+iS$,
then $K$ and $S$ were obtained up to first order in $\Phi$, and finally, exponentiation $F \approx (1+K)e^{iS}$ was done. In our approach, the exponentiation procedure is naturally
incorporated from the outset by using the variable $J$.\label{f1}}.
As we have already pointed out,
the Shapiro time delay describes the time lag caused by the gravitational potential.
In the observation of GWs, the Shapiro time delay is not measurable. 
Therefore, this degree of freedom needs to be removed from the phase modulation. 
We redefine the physical phase modulation as
\begin{align}
    \label{eq: phase S minus shapiro}
    \frac{S_{\text{ph}}(\omega)}{\omega}= \frac{S(\omega)}{\omega}-\lim_{\omega\to\infty}\frac{S(\omega)}{\omega}.
\end{align}
From now on, the term phase modulation always means
this physical quantity even if it is not explicitly mentioned.
With these in mind, 
the phase modulation and the magnification are then explicitly given by
\begin{align}
    \label{eq: phase S 1st order}
    S^{(1)}=&-2\omega
    \int_{0}^{\chi_{s}}d\chi 
    \left[\cos{\left[\frac{W(\chi,\chi_{s})\nabla^{2}_{\theta}}{2\omega}\right]}-1\right]
    \Phi,
    \\
    \label{eq: phase S 2nd order}
    S^{(2)}=&-2\omega
    \int_{0}^{\chi_{s}}\frac{d\chi}{\chi^{2}}
    \int_{0}^{\chi}d\chi_{1}
    \int_{0}^{\chi}d\chi_{2} 
    \left[\cos{\left[\frac{(W\nabla)^{(2)}}{2\omega}\right]}-1\right]\nabla_{\theta1}
    \Phi_{1}\cdot\nabla_{\theta2}\Phi_{2},
    \\
    \label{eq: phase S 3rd order}
    S^{(3)}=&-4\omega
    \int_{0}^{\chi_{s}} \frac{d\chi}{\chi^{2}}
    \int_{0}^{\chi}d\chi_{3}
    \int_{0}^{\chi} 
    \frac{d\chi'}{\chi'^{2}}
    \int_{0}^{\chi'}d\chi_{1}
    \int_{0}^{\chi'}d\chi_{2}
    \notag
    \\
    &\qquad\qquad\times\left[\cos{\left[\frac{(W\nabla)^{(3)}}{2\omega}\right]}-1\right]
    \nabla_{\theta12}(\nabla_{\theta1}\Phi_{1}\cdot\nabla_{\theta2}\Phi_{2})\cdot\nabla_{\theta3}\Phi_{3},\\
     \label{eq: mag K 1st order}
    K^{(1)}=&2\omega
    \int_{0}^{\chi_{s}}d\chi 
    \sin{\left[\frac{W(\chi,\chi_{s})\nabla^{2}_{\theta}}{2\omega}\right]}
    \Phi,
    \\
    \label{eq: mag K 2nd order}
    K^{(2)}=&2\omega
    \int_{0}^{\chi_{s}}\frac{d\chi}{\chi^{2}}
    \int_{0}^{\chi}d\chi_{1}
    \int_{0}^{\chi}d\chi_{2} 
    \sin{\left[\frac{(W\nabla)^{(2)}}{2\omega}\right]}
    \nabla_{\theta1}\Phi_{1}\cdot\nabla_{\theta2}\Phi_{2},
    \\
    \label{eq: mag K 3rd order}
    K^{(3)}=&4\omega
    \int_{0}^{\chi_{s}} \frac{d\chi}{\chi^{2}}
    \int_{0}^{\chi}d\chi_{3}
    \int_{0}^{\chi} 
    \frac{d\chi'}{\chi'^{2}}
    \int_{0}^{\chi'}d\chi_{1}
    \int_{0}^{\chi'}d\chi_{2}
    \notag
    \\
    &\qquad\qquad\times\sin{\left[\frac{(W\nabla)^{(3)}}{2\omega}\right]}
    \nabla_{\theta12}(\nabla_{\theta1}\Phi_{1}\cdot\nabla_{\theta2}\Phi_{2})\cdot\nabla_{\theta3}\Phi_{3}.
\end{align}
$S^{(1)}$ and $K^{(1)}$ have been derived in the previous works \cite{Takahashi:2005ug, Oguri:2020ldf}
and ours reproduce their results.
To the best of our knowledge, higher order terms $S^{(2)}, S^{(3)}, K^{(2)}, K^{(3)}$ are the new results.
In the high-frequency limit of Eqs.~(\ref{eq: mag K 1st order}) and (\ref{eq: mag K 2nd order}),
the magnification computed from the above expressions reproduces the result derived in \cite{Shapiro:2006em, Hilbert:2008kb, Krause:2009yr, Pratten:2016dsm, Petri:2016qya} under the post-Born approximation in geometric optics, which is demonstrated in appendix \ref{app: geometric optics limit}.

\subsection{Statistics of $K$ and $S$}
The situation we have in mind is the lensing caused by the dark matter inhomogeneities randomly
distributed in the whole Universe.
This means that the $K$ and $S$ behave in a stochastic manner for individual GW events.
Thus, the comparison between the theoretical prediction and observation is possible only for the statistical quantities.
This motivates us to compute the average and the variance of the lensing signal.

To this end, we first notice that for the ensemble average of the functions of the gravitational potential, the following equations hold under the Limber approximation. 
For arbitrary functions $F(x), G(x), H(x), I(x)$ of differential operator $x$, we have
\begin{align}
    \label{eq: 2 point correlation}
    &\braket{
    F(\nabla_{\theta1})\Phi_{1}
    G(\nabla_{\theta2})\Phi_{2}}
    \notag
    \\
    &\qquad\qquad=
    \delta^{D}(\chi_{1}-\chi_{2})
    \int \frac{d^{2}\bm k_{\perp}}{(2\pi)^{2}}
    F(i\chi_{1}\bm k_{\perp})
    G(-i\chi_{1}\bm k_{\perp})
    P_{\Phi}(k_{\perp},\chi_1).
    \\\notag
    \\
    \label{eq: 3 point correlation}
    &\braket{
    F(\nabla_{\theta1})\Phi_{1}
    G(\nabla_{\theta2})\Phi_{2}
    H(\nabla_{\theta3})\Phi_{3}}_{c}
    \notag
    \\
    &\quad=
    \delta^{D}(\chi_{1}-\chi_{3})
    \delta^{D}(\chi_{2}-\chi_{3})
    \int \frac{d^{2}\bm k_{1\perp}}{(2\pi)^{2}}
    \int \frac{d^{2}\bm k_{2\perp}}{(2\pi)^{2}}
    \notag
    \\
    &\qquad\qquad\times
    F(i\chi_{3}\bm k_{1\perp})
    G(i\chi_{3}\bm k_{2\perp})
    H(-i\chi_{3}\bm k_{1\perp}-i\chi_{3}\bm k_{2\perp})
    B_{\Phi}(k_{1\perp},k_{2\perp},|\bm k_{1\perp}+\bm k_{2\perp}|,\chi_1)
    \\\notag
    \\
     \label{eq: 4 point correlation}
    &\braket{
    F(\nabla_{\theta1})\Phi_{1}
    G(\nabla_{\theta2})\Phi_{2}
    H(\nabla_{\theta3})\Phi_{3}
    I(\nabla_{\theta4})\Phi_{4}}_{c}\notag
    \\
    &\quad=
    \delta^{D}(\chi_{1}-\chi_{4})
    \delta^{D}(\chi_{2}-\chi_{4})
    \delta^{D}(\chi_{3}-\chi_{4})
    \int \frac{d^{2}\bm k_{1\perp}}{(2\pi)^{2}}
    \int \frac{d^{2}\bm k_{2\perp}}{(2\pi)^{2}}
    \int \frac{d^{2}\bm k_{3\perp}}{(2\pi)^{2}}
    \notag
    \\
    &\qquad\times
    F(i\chi_{4}\bm k_{1\perp})
    G(i\chi_{4}\bm k_{2\perp})
    H(i\chi_{4}\bm k_{3\perp})
    I(-i\chi_{4}\bm k_{1\perp}-i\chi_{4}\bm k_{2\perp}-i\chi_{4}\bm k_{3\perp})
    \notag
    \\
    &\qquad\times
    T_{\Phi}(\bm k_{1\perp},\bm k_{2\perp},\bm k_{3\perp},-\bm k_{1\perp}-\bm k_{2\perp}-\bm k_{3\perp},\chi_1).
\end{align}
Here $P_{\Phi}, B_{\Phi}, T_{\Phi}$ are the power spectrum, bispectrum, and trispecrum of $\Phi$,
and $\braket{\cdots}_{c}$ indicates the connected term.
They are characterized by 
\begin{align}
    \label{eq: potential power spectrum}
    \braket{\tilde{\Phi}(\bm k_{1},\chi)\tilde{\Phi}(\bm k_{2},\chi)}
    &=(2\pi)^{3}
    \delta^{D}(\bm k_{1}+\bm k_{2})
    P_{\Phi}(k_{1},\chi),
    \\
    \label{eq: potential bispectrum}
    \braket{\tilde{\Phi}(\bm k_{1},\chi)\tilde{\Phi}(\bm k_{2},\chi)\tilde{\Phi}(\bm k_{3},\chi)}_{c}
    &=(2\pi)^{3}
    \delta^{D}(\bm k_{1}+\bm k_{2}+\bm k_{3} )
    B_{\Phi}(k_{1},k_{2},k_{3},\chi),
    \\
    \label{eq: potential trispectrum}
    \braket{\tilde{\Phi}(\bm k_{1},\chi)\tilde{\Phi}(\bm k_{2},\chi)\tilde{\Phi}(\bm k_{3},\chi)\tilde{\Phi}(\bm k_{4},\chi)}_{c}
    &=(2\pi)^{3}
    \delta^{D}(\bm k_{1}+\bm k_{2}+\bm k_{3}+\bm k_{4} )
    T_{\Phi}(\bm k_{1},\bm k_{2},\bm k_{3},\bm k_{4},\chi),
\end{align}
where $\tilde{\Phi}(\bm k)$ is the Fourier transform of $\Phi$.
With these definitions, we are ready to derive the average and the variance of
the post-Born corrections, which we will address in the following.

\subsubsection{Average}
At the level of the Born approximation,
the average of $K$ and $S$ is zero.
This does not happen beyond the Born approximation.
Thus, the average of $K$ and $S$ fully represents the effects of the 
post-Born corrections.
The leading order correction is ${\cal O}(\Phi^2)$, and we evaluate $\braket{K}, \braket{S}$ at this order.
From Eqs~(\ref{eq: phase S 2nd order}), (\ref{eq: mag K 2nd order}), and (\ref{eq: 2 point correlation}), 
we obtain the following expressions:
\begin{align}
    \braket{S}=&2\omega
    \int_{0}^{\chi_{s}} 
    \frac{d\chi}{\chi^{2}}
    \int_{0}^{\chi}d\chi_{1} 
    \chi_{1}^{2}
    \int\frac{d^{2}\bm k_{\perp}}{(2\pi)^{2}} 
    k_{\perp}^{2}
    \left(1-\cos{\left[\frac{(\chi-\chi_{1})\chi_{1}}{\chi\omega}k_{\perp}^{2}\right]}\right)
    P_{\Phi}(k_{\perp},\chi_1), \label{ave-S}
    \\
    \braket{K}=&-2\omega
    \int_{0}^{\chi_{s}} 
    \frac{d\chi}{\chi^{2}}
    \int_{0}^{\chi}d\chi_{1}\chi_{1}^{2}
    \int\frac{d^{2}\bm k_{\perp}}{(2\pi)^{2}} 
    k_{\perp}^{2}
    \sin{\left[\frac{(\chi-\chi_{1})\chi_{1}}{\chi\omega}k_{\perp}^{2}\right]}
    P_{\Phi}(k_{\perp},\chi_1). \label{ave-K}
\end{align}
At this stage, there are three things worth mentioning.
Firstly, it is suggestive to rewrite the above relations in terms of 
the filter functions $F^{(2)}_S, F^{(2)}_K$ as
\begin{align}
     \label{eq: phase average}
    \braket{S}=&2 \int_0^{\chi_{s}} \frac{d\chi}{\chi^3} 
    \int_0^\chi d\chi' \chi'^3 (\chi-\chi') \int \frac{dk}{2\pi} F^{(2)}_S k^5 P_\Phi (k,\chi'), \\
    \label{eq: magnification average}
    \braket{K}=&-2 \int_0^{\chi_{s}} \frac{d\chi}{\chi^3} 
    \int_0^\chi d\chi' \chi'^3 (\chi-\chi') \int \frac{dk}{2\pi} F^{(2)}_K k^5 P_\Phi (k,\chi'),
\end{align}
where
\begin{equation}
    F^{(2)}_S=\frac{1-\cos k^2 r_F^2}{k^2 r_F^2},~~~~~
    F^{(2)}_K=\frac{\sin k^2 r_F^2}{k^2 r_F^2},
\end{equation}
and $r_F$ defined by $r_F^2=\chi'(\chi-\chi')/(\omega \chi)$ is the Fresnel scale \cite{Takahashi:2005ug}.
By writing in this way, it is manifest that the frequency dependence of $\braket{S}$ and $\braket{K}$ 
is solely encoded in the filter functions.
These filter functions are suppressed below the Fresnel scale $k^{-1} <r_F$.
Physically, the filter functions describe the diffraction effect that lowers the lensing signal when the size of matter fluctuations is below this scale. 
In \cite{Takahashi:2005ug}, it was argued that $\braket{S^2}$ and $\braket{K^2}$(within the
Born approximation) are 
insensitive to the matter fluctuations below the Fresnel scale.
Our result demonstrates that a similar conclusion holds for $\braket{S}, \braket{K}$.
Secondly, since, unlike in the case of geometric optics,
both $\braket{S}$ and $\braket{K}$ depend on the GW frequency due to the frequency dependence of the
Fresnel scale, 
we can extract the matter power spectrum at the Fresnel scale by measuring $\braket{S}$ and $\braket{K}$
at multiple frequencies and how they vary as the frequency is changed.
This suggests a possibility that, in addition to $\braket{S^2}$ and $\braket{K^2}$, $\braket{S}$ and $\braket{K}$ can be used as new observables to
probe the matter power spectrum at the Fresnel scale. 
Notice that, contrary to the case of the cosmological perturbations where
the average of the perturbations is absorbed into the FLRW background,
$\braket{S}$ and $\braket{K}$ cannot be absorbed into the unlensed waveform
since i) the frequency dependence of the average is different from that of
the unlensed waveform and ii) each merger event has a different unlensed waveform.
Thirdly, $\braket{S}$ is a positive definite for any $\omega$. 
Thus, if the measurement of $\braket{S}$ gives a negative value,
we can robustly conclude that it is not due to the lensing by the matter fluctuations but due to something else.

We also derive the expressions of the next leading-order contributions to the average coming from the higher-order statistical quantities(bispectrum).
\begin{align}
    \label{eq: SaveBi}
    \Braket{S^{(3)}} =&
    -4\omega\int_{0}^{\chi_{s}}\frac{d\chi}{\chi^{2}}\int_{0}^{\chi}\frac{d\chi'}{\chi'^{2}}
    \int_{0}^{\chi'}\chi_{3}^{4}d\chi_{3}
    \int\frac{d\bm k_{1\perp}}{(2\pi)^{2}}
    \int\frac{d\bm k_{2\perp}}{(2\pi)^{2}}
    \frac{(\bm k_{\perp 1}\cdot \bm k_{\perp 2} )B_{\Phi}(k_{1},k_{2},k_{3},\chi)}{k_{\perp 1}^{2}k_{\perp 2}^{2}}
    \notag
    \\
    &\times
    \left(
    1- \cos{\left[
    \frac{\chi_{3}(\chi-\chi_{3})}{2\omega\chi}(k_{\perp 1}^{2}+k_{\perp 2}^{2}+ (\bm k_{\perp 1}+\bm k_{\perp 2})^{2})
    +\frac{\chi_{3}^{2}(\chi'-\chi)}{\omega \chi \chi'}\bm k_{\perp 1}\cdot\bm k_{\perp 2}
    \right]}
    \right)
    \\
    \label{eq: KaveBi}
    \Braket{K^{(3)}} =&
    -4\omega\int_{0}^{\chi_{s}}\frac{d\chi}{\chi^{2}}\int_{0}^{\chi}\frac{d\chi'}{\chi'^{2}}
    \int_{0}^{\chi'}\chi_{3}^{4}d\chi_{3}
    \int\frac{d\bm k_{1\perp}}{(2\pi)^{2}}
    \int\frac{d\bm k_{2\perp}}{(2\pi)^{2}}
    \frac{(\bm k_{\perp 1}\cdot \bm k_{\perp 2} )B_{\Phi}(k_{1},k_{2},k_{3},\chi)}{k_{\perp 1}^{2}k_{\perp 2}^{2}},
    \notag
    \\
    &\times
    \sin{\left[
    \frac{\chi_{3}(\chi-\chi_{3})}{2\omega\chi}(k_{\perp 1}^{2}+k_{\perp 2}^{2}+ (\bm k_{\perp 1}+\bm k_{\perp 2})^{2})
    +\frac{\chi_{3}^{2}(\chi'-\chi)}{\omega \chi \chi'}\bm k_{\perp 1}\cdot\bm k_{\perp 2}
    \right]}.
\end{align}
By incorporating the bispectrum contributions to the average of $S$ and $K$, we are able to assess whether the effects of the non-Gaussianity terms on $\braket{S}$ and $\braket{K}$ are significant.

\subsubsection{Variance}
In the same way, the rms of the magnification up to fourth order 
in $\Phi$ is given by
\begin{align}
\label{var-K}
    \braket{K^{2}}=&\braket{(K^{(1)})^{2}}+2\braket{K^{(1)}K^{(2)}}+2\braket{K^{(1)}K^{(3)}}+\braket{(K^{(2)})^{2}}.
\end{align}
The variance of the magnification up to the same order is then written as 
$\Delta_ {K}^{2}=\braket{K^{2}}-\braket{K^{(2)}}^{2}$.
We define the post-Born corrections to the variance as $\Delta_{K}^{2}=\braket{K_{\rm Born}^{2}}+\delta_{K^{2}}$. 
In this definition, it is possible that $\delta_{K^{2}}<0$.
As we mentioned earlier, the third-order term in $\Phi$ is necessary because it couples with the first-order term.
At this order, the result will depend on whether $\Phi$ is Gaussian or non-Gaussian.
In the diagrammatic language, the variance contains both disconnected ($\delta_ {K^{2},\rm dc}$) 
and connected($\delta_ {K^{2},\rm c}$) parts:
\begin{equation}
    \delta_ {K^{2}}=\delta_ {K^{2},\rm dc}+\delta_{K^{2},\rm c}
\end{equation}
As for the disconnected part, we find that it consists of three distinct terms:
\begin{equation}
    \label{eq: corrcKVar}
    \delta_ {K^{2},\rm dc}=
    \braket{(K^{(2)})^{2}}_{\rm dc}
    +2\braket{K^{(1)}K^{(3)}}_{\rm dc}-\braket{K^{(2)}}_{\rm dc}^{2},
\end{equation}
where the subscript ${\rm dc}$ should be understood that the corresponding quantity
is obtained by treating $\Phi$ as a Gaussian variable.
The connected part also consists of three terms:
\begin{equation}
\label{connectedK}
    \delta_ {K^{2},\rm c}=
    2\braket{K^{(1)}K^{(2)}}_{\rm c}+2\braket{K^{(1)}K^{(3)}}_{\rm c}+\braket{(K^{(2)})^{2}}_{\rm c}.
\end{equation}

The first term in Eq.~(\ref{var-K}) is nothing but the variance in the Born approximation and has been already
derived in the literature \cite{Takahashi:2005ug}.
For completeness, we will provide its expression below. 
For the Gaussian variable, the n-point correlation function is completely specified by the two-point function, i.e., the matter power spectrum.
Using this fact, we find that each term can be written as
\begin{align}
    \label{eq: magn vari Born}
    \braket{(K^{(1)})^{2}}=&4\omega^{2}
    \int_{0}^{\chi_{s}}d\chi
    \int\frac{d^{2}\bm k_{\perp}}{(2\pi)^{2}}
    \sin^{2}{\left[\frac{(\chi_{s}-\chi)\chi}{2\chi_{s}\omega}k_{\perp}^{2}\right]}
    P_{\Phi}(k_{\perp}),
    \\
    \label{eq: mag vari 22 post Born}
    \braket{(K^{(2)})^{2}}_{\rm dc}=&\braket{K^{(2)}}_{\rm dc}^{2}+16\omega^{2}
    \int_{0}^{\chi_{s}}\frac{d\chi}{\chi^{2}}
    \int_{0}^{\chi}\frac{d\chi'}{\chi'^{2}}
    \int_{0}^{\chi'}d\chi_{1}
    \int_{0}^{\chi'}d\chi_{2}
    \int\frac{d^{2}\bm k_{1\perp}}{(2\pi)^{2}}
    \int\frac{d^{2}\bm k_{2\perp}}{(2\pi)^{2}} \notag
    \\
    &\times\sin{\left[\frac{(\chi_{s}-\chi_{1})\chi_{1}}{2\chi_{s}\omega}k_{1\perp}^{2}
    +\frac{(\chi_{s}-\chi_{2})\chi_{2}}{2\chi_{s}\omega}k_{2\perp}^{2}
    +\frac{(\chi_{s}-\chi)\chi_{1}\chi_{2}}{\omega \chi_{s} \chi}\bm k_{1\perp}\cdot \bm k_{2\perp}\right]} \notag
    \\
    &\times\sin{\left[\frac{(\chi_{s}-\chi_{1})\chi_{1}}{2\chi_{s}\omega}k_{1\perp}^{2}
    +\frac{(\chi_{s}-\chi_{2})\chi_{2}}{2\chi_{s}\omega}k_{2\perp}^{2}
    +\frac{(\chi_{s}-\chi')\chi_{1}\chi_{2}}{\omega \chi_{s} \chi'}\bm k_{1\perp}\cdot \bm k_{2\perp}\right]} \notag
    \\
    &\times\chi_{1}^{2}\chi_{2}^{2}(\bm k_{1\perp}\cdot\bm k_{2\perp})^{2}
    P_{\Phi}(k_{1\perp})P_{\Phi}(k_{2\perp}),
    \\
    \label{eq: mag vari 13 post Born}
    \braket{K^{(1)}K^{(3)}}_{\rm dc}=&-16\omega^{2}
    \int_{0}^{\chi_{s}}\frac{d\chi}{\chi^{2}}
    \int_{0}^{\chi}\frac{d\chi'}{\chi'^{2}}
    \int_{0}^{\chi'}d\chi_{1}
    \int_{0}^{\chi'}d\chi_{2}
    \int\frac{d^{2}\bm k_{1\perp}}{(2\pi)^{2}}
    \int\frac{d^{2}\bm k_{2\perp}}{(2\pi)^{2}} \notag
    \\
    \times\sin&{\left[\frac{(\chi_{s}-\chi_{2})\chi_{2}}{2\chi_{s}\omega}k_{2\perp}^{2}\right]} 
    \sin{\left[\frac{(\chi_{s}-\chi_{2})\chi_{2}}{2\chi_{s}\omega}k_{2\perp}^{2}
    +\frac{(\chi-\chi')\chi_{1}\chi_{2}}{\omega \chi \chi'}\bm k_{1\perp}\cdot \bm k_{2\perp}
    +\frac{(\chi-\chi_{1})\chi_{1}}{\chi\omega}k_{1\perp}^{2}\right]} \notag
    \\
    &\times\left(\chi_{1}^{2}\chi_{2}^{2}(\bm k_{1\perp}\cdot \bm k_{2\perp})^{2}
    +\chi_{1}^{3}\chi_{2}k_{1\perp}^{2}(\bm k_{1\perp}\cdot \bm k_{2\perp})\right)
    P_{\Phi}(k_{1\perp})P_{\Phi}(k_{2\perp}).
\end{align}
As these expressions show, the computation of $\delta_{K^2,\rm dc}$ requires 
multiple integrations in eight variables.
Among these eight variables, the integral with respect to the angle between $\bm k_{1\perp}$ and $\bm k_{2\perp}$ can be analytically performed
and the result is written in terms of the Bessel functions.
Thus, practically, the number of variables in the integration is six.
The concrete expression of $\delta_{K^2,\rm dc}$ which we will evaluate numerically 
in the next section
is given in Appendix \ref{app:postBornvariance}.

The connected part $\delta_{K^2,\rm c}$ given by Eq.~(\ref{connectedK}) 
comes from the non-Gaussianity of the matter fluctuations:
the matter bispectrum and trispectrum, and so forth. 
The first term in Eq.~(\ref{connectedK}),
which is ${\cal O}(\Phi^3)$, is written in terms of the bispectrum as
\begin{align}
    \label{eq; magnification 12 bispectrum}
    &\braket{K^{(1)}K^{(2)}}_{c}=
    -4\omega^{2}
    \int_{0}^{\chi_{s}}\frac{d\chi}{\chi^{2}}
    \int_{0}^{\chi}d\chi_{3}\chi_{3}^{2}
    \int\frac{d^{2}\bm k_{1\perp}}{(2\pi)^{2}}
    \int\frac{d^{2}\bm k_{2\perp}}{(2\pi)^{2}}
    \notag
    \\
    &\qquad\times
    \sin\left[\frac{(\chi_{s}-\chi_{3})\chi_{3}}{2\omega\chi_{s}}|\bm k_{1\perp}
    +\bm k_{2\perp}|^{2}\right]
    \sin\left[\frac{(\chi_{s}-\chi_{3})\chi_{3}}{2\omega\chi_{s}}
    (k_{1\perp}^{2}+k_{2\perp}^{2})
    +\frac{(\chi_{s}-\chi)\chi^{2}_{3}}{\omega\chi_{s}\chi}\bm k_{1\perp}\cdot \bm k_{2\perp}\right]
    \notag
    \\
    &\qquad\times
    (\bm k_{1\perp}\cdot \bm k_{2\perp})B_{\Phi}(k_{1\perp},k_{2\perp},|\bm k_{1\perp}+\bm k_{2\perp}|).
\end{align}
The other two terms,
which are ${\cal O}(\Phi^4)$, are written in terms of the trispectrum as 
\begin{align}
    \label{eq; magnification 22 trispectrum}
    &\braket{(K^{(2)})^{2}}_{c}
    =-8\omega^{2}
    \int_{0}^{\chi_{s}}\frac{d\chi}{\chi^{2}}
    \int_{0}^{\chi}\frac{d\chi'}{\chi'^{2}}
    \int_{0}^{\chi'}d\chi_{4}\chi_{4}^{4}
    \int\frac{d^{2}\bm k_{1\perp}}{(2\pi)^{2}}
    \int\frac{d^{2}\bm k_{2\perp}}{(2\pi)^{2}}
    \int\frac{d^{2}\bm k_{3\perp}}{(2\pi)^{2}}
    \notag
    \\
    &\qquad\times
    \sin\left[\frac{(\chi_{s}-\chi)\chi_{4}^{2}}{2\omega\chi_{s}\chi}|\bm k_{1\perp}+\bm k_{2\perp}|^{2}
    +\frac{(\chi-\chi_{4})\chi_{4}}{2\omega \chi}(k_{1\perp}^{2}+k_{2\perp}^{2})\right]
    \notag
    \\
    &\qquad\times
    \sin\left[\frac{(\chi_{s}-\chi')\chi_{4}^{2}}{2\omega\chi_{s}\chi'}|\bm k_{1\perp}+\bm k_{2\perp}|^{2}
    +\frac{(\chi'-\chi_{4})\chi_{4}}{2\omega \chi'}(k_{3\perp}^{2}+|\bm k_{1\perp}+\bm k_{2\perp}+\bm k_{3\perp}|^{2})\right]
    \notag
    \\
    &\qquad\times
    (\bm k_{1\perp}\cdot \bm k_{2\perp})\left[\bm k_{3\perp}\cdot(\bm k_{1\perp}+\bm k_{2\perp}+\bm k_{3\perp})\right]
    T_{\Phi}(\bm k_{1\perp},\bm k_{2\perp},\bm k_{3\perp}, -\bm k_{1\perp}-\bm k_{2\perp}-\bm k_{3\perp}).
    \\
    \label{eq; magnification 13 trispectrum}
    &\braket{K^{(1)}K^{(3)}}_{c}=
    8\omega^{2}
    \int_{0}^{\chi_{s}}\frac{d\chi}{\chi^{2}}
    \int_{0}^{\chi}\frac{d\chi'}{\chi'^{2}}
    \int_{0}^{\chi'}d\chi_{4}\chi_{4}^{4}
    \int\frac{d^{2}\bm k_{1\perp}}{(2\pi)^{2}}
    \int\frac{d^{2}\bm k_{2\perp}}{(2\pi)^{2}}
    \int\frac{d^{2}\bm k_{3\perp}}{(2\pi)^{2}}
    \notag
    \\
    &\qquad\times
    \sin\left[\frac{(\chi_{s}-\chi_{4})\chi_{4}}{2\omega\chi_{s}}|\bm k_{1\perp}+\bm k_{2\perp}+\bm k_{3\perp}|^{2}\right]
    \notag
    \\
    &\qquad\times
    \sin\left[\frac{(\chi_{s}-\chi)\chi_{4}^{2}}{2\omega\chi_{s}\chi}|\bm k_{1\perp}+\bm k_{2\perp}+\bm k_{3\perp}|^{2}
    +\frac{(\chi-\chi_{4})\chi_{4}}{2\omega\chi}k_{3\perp}^{3}\right.
    \notag
    \\
    &\left.\qquad\qquad\qquad+\frac{(\chi-\chi')\chi_{4}^{2}}{2\omega\chi\chi'}|\bm k_{1\perp}+\bm k_{2\perp}|^{2}
    +\frac{(\chi'-\chi_{4})\chi_{4}}{2\omega \chi'}(k_{1\perp}^{2}+k_{2\perp}^{2})
    \right]
    \notag
    \\
    &\qquad\times
    (\bm k_{1\perp}\cdot \bm k_{2\perp})\left[(\bm k_{1\perp}+\bm k_{2\perp})\cdot \bm k_{3\perp}\right]
    T_{\Phi}(\bm k_{1\perp},\bm k_{2\perp},\bm k_{3\perp}, -\bm k_{1\perp}-\bm k_{2\perp}-\bm k_{3\perp}).
\end{align}
In order to evaluate the connected part, we need to determine the
bispectrum and the trispectrum of the matter fluctuations.
In this paper,  we only focus on the lowest-order term in the non-Gaussianity part, which is the contribution from the bispectrum.

The above formulation is for the variance of $K$. 
The variance of $S$ can be formulated in exactly the same manner.
The quantities of $S$ corresponding to Eqs.~(\ref{eq: magn vari Born}), (\ref{eq: mag vari 22 post Born}), (\ref{eq: mag vari 13 post Born}), (\ref{eq; magnification 12 bispectrum}), (\ref{eq; magnification 22 trispectrum}), and (\ref{eq; magnification 13 trispectrum}) are obtained by 
replacing all the sine functions as $\sin (\cdots) \to 1-\cos (\cdots)$.

\section{Halo model}
Our results in the previous section are described by the power spectrum and bispectrum of the potential $P_{\Phi}(k,\chi)$ and $B_{\Phi}(k_{1},k_{2},k_{3},\chi)$. 
As mentioned, the bispectrum is the only term considered in this paper to capture the non-Gaussianity.
In the actual computations of the average and the variance
of $S$ and $K$, we need the spectra of matter instead of the gravitational potential and they are obtained through the Poisson equation:
\begin{align}
    \label{eq: Ppotential Pmatter}
    P_{\Phi}(k,\chi)=&
    \left(
    \frac{3H_{0}^{2}\Omega_{m}}{2}
    \right)^{2}
    \frac{1}{a^{2}(\chi)k^{4}}P_{\delta}(k,\chi),
    \\
    \label{eq: Bpotential Bmatter}
    B_{\phi}(k_{1},k_{2},k_{3}, \chi)=&-\left(
    \frac{3H_{0}^{2}\Omega_{m}}{2}
    \right)^{3}
    \frac{1}{a^{3}(\chi)k_{1}^{2}k_{2}^{2}k_{3}^{2}}B_{\delta}(k_{1},k_{2},k_{3},\chi)
\end{align}

The precise dependence of the matter power spectrum at a small scale, which is important for the frequency range of our interest, is very difficult to compute from the first principle due to complex physical processes such as baryonic physics. 
To circumvent this issue, 
we adopt the formulation of the halo model described in \cite{Oguri:2020ldf} which provides a useful phenomenological approach with reasonable computational costs. 

While it is advisable to refer to \cite{Oguri:2020ldf} for more detail, we would like to give a brief summary of their findings.
In \cite{Oguri:2020ldf}, the power spectrum was computed using their halo model, which incorporates subhalos and baryonic matter such as galaxies and stars. 
Their model's power spectrum shows a close match to the one computed through hydrodynamical simulations within the range covered by the simulation ($k < 30 h\mathrm{Mpc}^{-1}$), confirming the reliability of the model at least within this range.
Then, they compute the power spectrum at an even smaller scale (up to as small as $k\sim 10^{8}h\mathrm{Mpc}^{-1}$) with their halo model assuming that the halo model approach provides a reasonable estimation of the power spectrum. 
Under this assumption, they found that the power spectrum at $k\sim 10^{6}h\mathrm{Mpc}^{-1}$ is predominantly determined by the two components: dark low-mass halos ($1h^{-1}M_{\odot}\lesssim M \lesssim10^{4}h^{-1}M_{\odot}$, no stars within them, thus dark) and the point mass (ordinary stars, neutron stars, and black holes which causes the shot noise in the signal). 
Considering that the weak lensing effect on GWs is highly sensitive to the matter power spectrum at the corresponding scale\cite{Takahashi:2005ug},
they show that, within a certain frequency range of GWs 
($10 \mathrm{Hz}\lesssim f\lesssim 100\mathrm{Hz}$), the lensing dispersion is also sensitive to the abundance of dark low-mass halos and PBHs (point mass) and can be used to probe them.

Following their result that dark low-mass halos and the point mass are important,
we only focus on their contributions to the matter power spectrum and bispectrum, and ignore the effect from other sources such as the density distribution within a galaxy.
In addition, we separately compute the lensing signal from the halos and the point mass (shot noise) due to the following two reasons. 
Firstly, the power spectrum and bispectrum for these two components exhibit different characteristics(the power spectrum with subhalos included is strongly suppressed at small scales, while the shot noise contributes to all scales of the spectra equally).
By analyzing their effects separately, we can gain a clearer understanding of the individual contributions.
Secondly, there is a technical aspect to consider.
While analytical computations can be partially carried out for the shot noise contribution, full numerical computation poses challenges due to rapid oscillation in the integration processes. 
On the other hand, the subhalo contribution cannot be evaluated analytically due to the lack of a simple analytical form for the power spectrum and bispectrum, leaving numerical calculations the only option.
Consequently, separating the calculation of these two components allows us to effectively address the computational challenges associated with each contribution (the contribution from the shot noise is separately studied in \ref{shot-noise}).

Following \cite{Oguri:2020ldf},
we estimate the effect of subhalos by computing the analytic subhalo mass function based on the extended Press-Schechter theory with the tidal stripping and the dynamical friction effects included.

\begin{figure}[tb!]
\centering
\includegraphics[width=10cm]{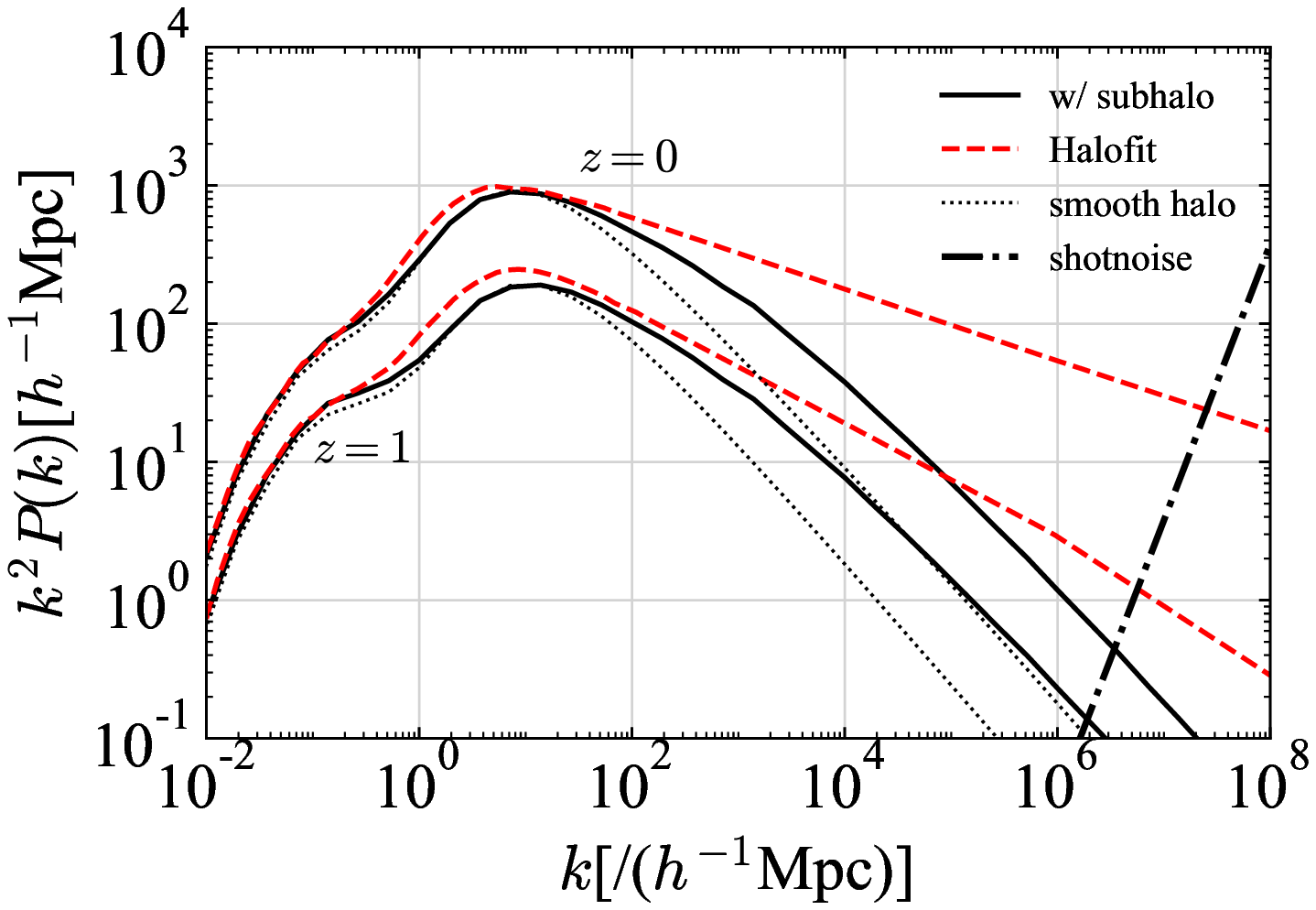} 
\caption{
    The power spectrum for the halo model with subhalos included (solid) is significantly enhanced by the presence of subhalos at small scales compared to the power spectrum of the halo model without subhalos(dotted).
    At large scales, the halo model is fairly accurate with the Halofit (red dashed).
    The Halofit is a fitting formula given in \cite{Takahashi_2012} whose parameters are calibrated on the $N$ body simulation results at $k<30h\mathrm{Mpc}^{-1}$.
    The shot noise effect (compact object with mass $0.5M_{\odot}$ and mass fraction $f_{p}=0.01$) on the power spectrum is illustrated by the dash-dotted line emerging at $k\sim 10^{6}h\mathrm{Mpc}^{-1}$. 
The hump seen at $k=10^{-2}\sim10^{-1}h\mathrm{Mpc}^{-1}$ represents the peak of the linear power spectrum, while the peak at $k\sim10^{1}h\mathrm{Mpc}^{-1}$ represents the scale of the biggest halos considered in the model.
} 
\label{fig:powerspectrum}
\end{figure}

Fig.~\ref{fig:powerspectrum} shows the matter power spectrum used in this paper evaluated at $z=0, 1$(solid line).
The dotted lines are the power spectrum without the subhalo contributions, illustrating the enhancement of the small-scale power spectrum due to the subhalos. 
We also show with the red dashed lines the power spectrum computed by the Halofit model \cite{Takahashi_2012}
Halofit is a fitting formula whose functional form is motivated by the halo model and is calibrated against the $N$ body simulation at $k\lesssim 30h\mathrm{Mpc}^{-1}$.
The halo model is indeed consistent with the fitting formula within the calibration range.
Note that in this figure, the vertical axis is scaled $k^{2}P_{\delta}(k)$ as this is the contribution to gravitational lensing per $\log{k}$.
At $k\sim10^{-2}h\mathrm{Mpc}^{-1}$, a slight hump can be observed. 
This scale represents the peak of the linear power spectrum, while the peak observed at $k\sim10h\mathrm{Mpc}^{-1}$ indicates the scale of the largest halos.
As the scale moves towards the lower side (high $k$), the power spectrum decreases as the contribution from the larger halos (halos whose radius is greater than the scale of interest) becomes less and less significant.
The dash-dotted straight line emerging at $k=10^{6}h\mathrm{Mpc}^{-1}$ represents the shot noise effect due to the point mass.
The shot noise is evaluated using a simple model, where all point masses are the same type of object and are randomly distributed throughout the Universe. 
This model does not take into account the time evolution of the point mass either.
The power spectrum based on this model is simply given by a constant $P_{\rm shot}=f_{p}^{2}/\overline{n}$, where $f_{p}$ and $\overline{n}$ are the mass fraction of the point mass to the total matter and the number density of the point mass, respectively.
In order to compute $\overline{n}$, we use the relation $\overline{\rho}\Omega_{m}f_{p}=m_{p}\overline{n}$,
and $H_{0}^{2}=\frac{8\pi G}{3}\overline{\rho}$ with mass $m=0.5M_{\odot}$ and mass fraction $f_{p}=0.01$.
The fiducial value of $f_{p}$ is consistent with the measured 
abundance of stars \cite{Fukugita:1997bi}.

Next, we would like to provide an approximation formula for the power spectrum:
\be
\label{Pdelta-largek}
P_{\delta}(k,\chi)= P_{\delta}(k_{0},\chi)\left(\frac{k}{k_{0}}\right)^{\frac{k{0}}{P_{\delta}(k_{0},\chi)}\frac{dP_{\delta}(k_{0},\chi)}{dk_{0}}}=B(\chi)k^{-b}.
\ee
This is equivalent to the Taylor expansion of $\log{P_{\delta}}$ with respect to $\log{k}$ and is useful to approximately evaluate the asymptotic behavior of the
average and the variance of $S$ and $K$ (particularly $\delta_{S^{2},\rm dc}$, the correction term to the variance of the phase modulation).
$k_{0}$ is an arbitrary scale around which $P(k,\chi)$ is expanded, thus we can take $k_{0}=\sqrt{H_{0}\omega}\sim \text{1/(Fresnel scale)}$ so that the variation of the power spectrum around the Fresnel scale is properly evaluated.

In addition to the power spectrum, we also need to evaluate the bispectrum for calculating the lensing dispersion.
In the halo model formulation, the total bispectrum is the sum of so-called 1-halo, 2-halo, and 3-halo terms:
\begin{align}
    \label{eq: mBi}
    B_{\delta}=&B_{\rm 1H}+B_{\rm 2H}+B_{\rm 3H},
\end{align}
where each term is calculated by following the formalism presented in \cite{Cooray:2002dia}.
Since we are interested in the small-scale bispectrum, we need to incorporate the subhalo contribution to this expression. 
To do so, we follow the formulation in \cite{Dolney:2004cn}, where the 1-halo term is separated into seven terms as
\begin{align}
    \label{[eq: mBi1H}
    B_{\rm 1H}=&B_{\rm sss}+B_{\rm ssc}+B_{\rm s1c}+B_{\rm s2c}+B_{\rm 1c}+B_{\rm 2c}+B_{\rm 3c}.
\end{align}
The notation used here is the same as the one in \cite{Dolney:2004cn} (s and c mean smooth and clump, respectively).
We evaluate each term using the same mass function and density profile of the subhalos in \cite{Oguri:2020ldf}.

Note that we have ignored the contributions from subhalos to the 2-halo term and 3-halo term.

The rationale for this assumption is as follows: The 2-halo term refers to the three-point correlation involving two points from the same halo and the third point from a different halo.
In the case of an equilateral or flattened configuration of a triangle, 
this term becomes subdominant compared to the 1-halo term.
For instance, when $k_{1},k_{2},$ and $k_{3}$ are equal (equilateral), and their corresponding scales are smaller than the size of a main halo, the 2-halo term is significantly suppressed due to a very small correlation between these two halos.
However,  the 1-halo term remains relevant because of the matter density fluctuation and the presence of subhalos in a main halo.
The only triangle configuration where the 2-halo term becomes relevant is when the scales corresponding to $k_{1}$ and $k_{2}$ are of the order of the halo size, while the scale corresponding to $k_{3}$ is much larger than the size of a main halo.
In such cases, the contribution of subhalos can be ignored because, at scales much larger than the size of a main halo, 
even the main halos can be treated as point mass objects.
Similar considerations can be made for the 3-halo term. 
The 3-halo term only becomes relevant when the scale corresponding to $k_{1},k_{2},$ and $k_{3}$ are significantly larger than the size of a main halo.
In such cases, the structure of individual main halos becomes irrelevant.

\begin{figure}[tb!]
  \begin{minipage}[b]{0.5\linewidth}
    \centering
    \includegraphics[keepaspectratio, scale=0.55]{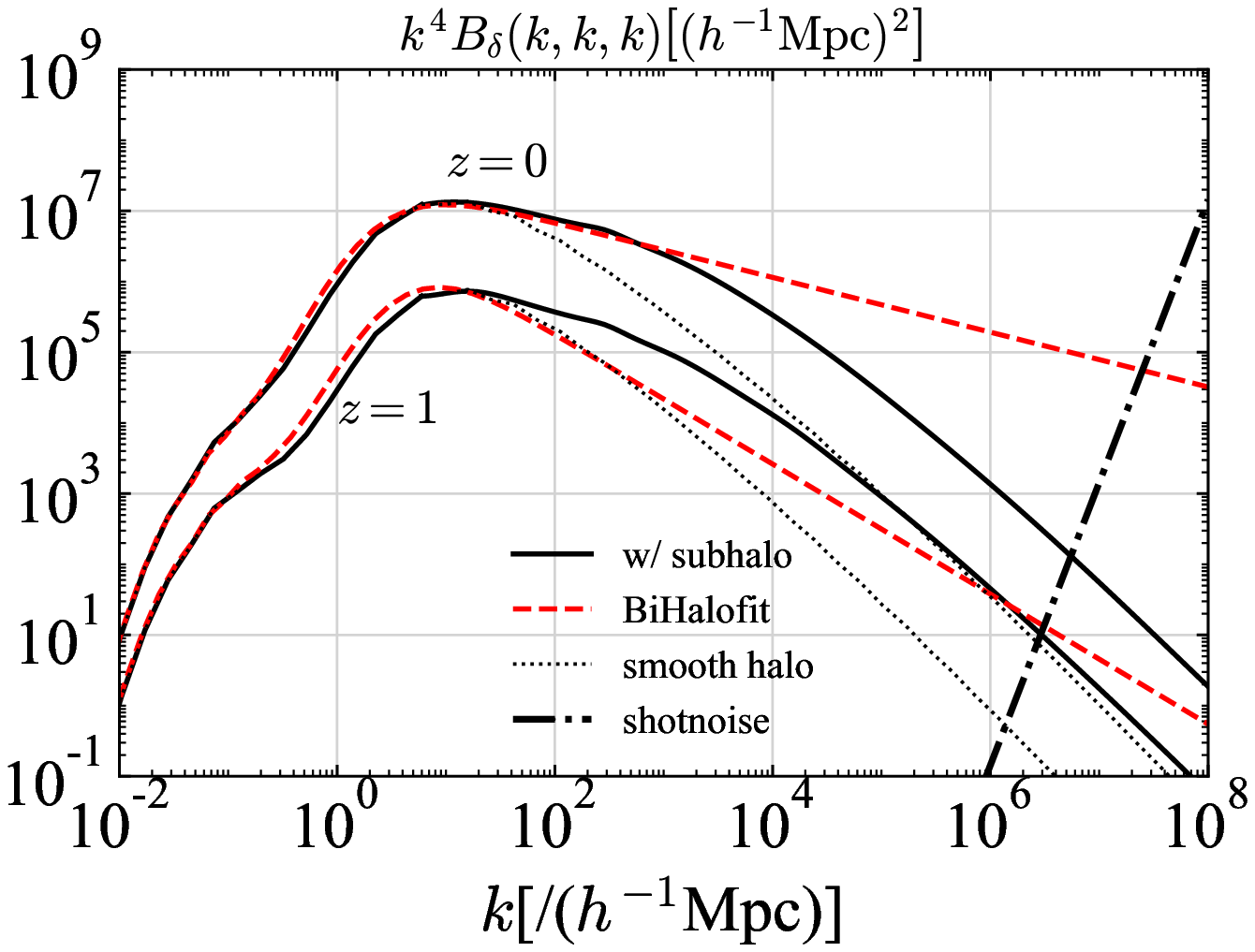}
    \subcaption{Equilateral}
    \label{fig:Bieq}
  \end{minipage}
  \begin{minipage}[b]{0.5\linewidth}
    \centering
    \includegraphics[keepaspectratio, scale=0.55]{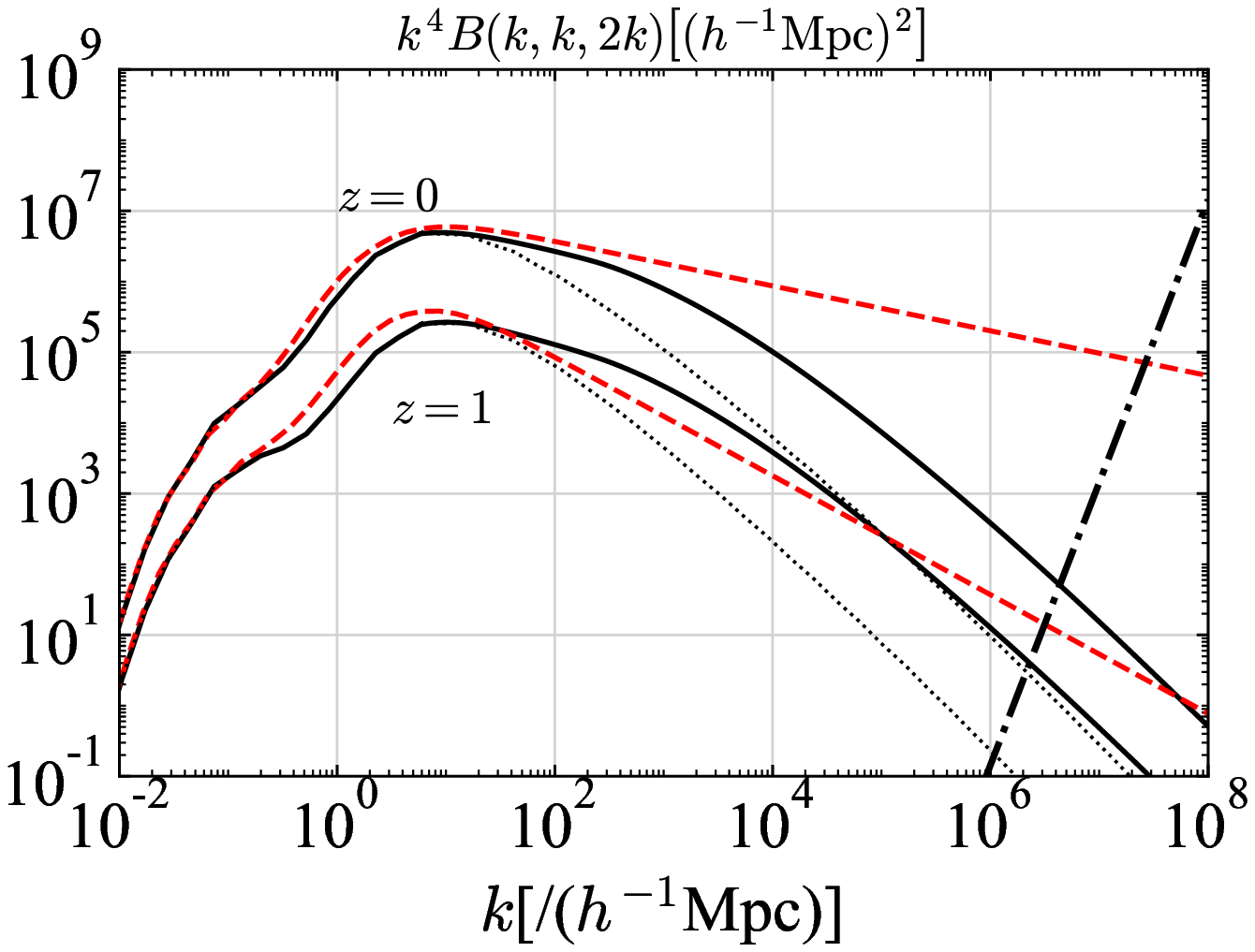}
    \subcaption{Flattened}
    \label{fig:Bifl}
  \end{minipage}
    \bigskip
  \begin{minipage}[b]{0.5\linewidth}
    \centering
    \includegraphics[keepaspectratio, scale=0.55]{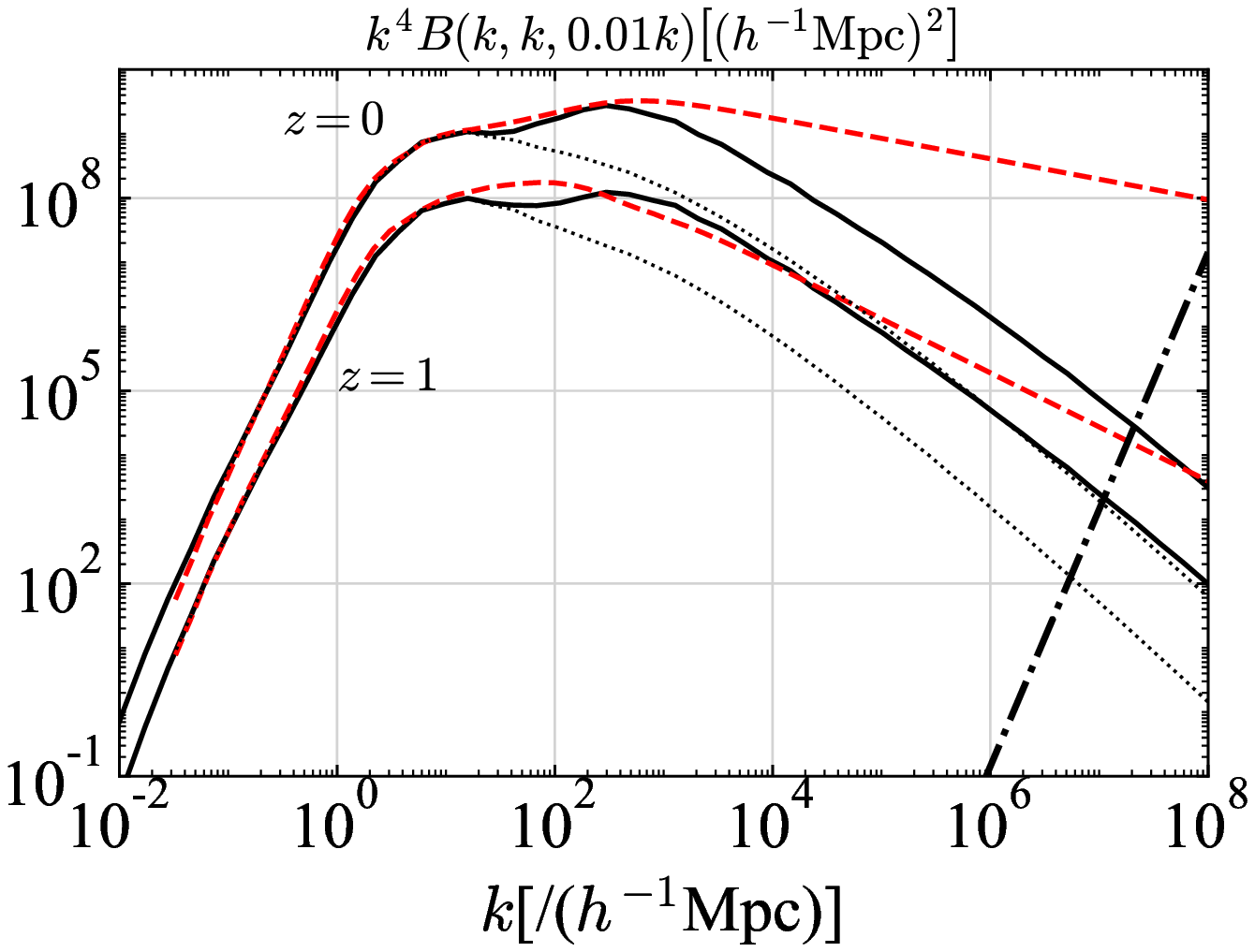}
    \subcaption{Squeezed}
    \label{fig:Bisq}
  \end{minipage}
  \caption{
  In all configurations of a triangle (equilateral, flattened, squeezed), the halo model with subhalos (solid) is significantly enhanced at small scales compared to the smooth (without subhalos) model (dotted).
  At large scales, the halo model is fairly consistent with the BiHalofit model \cite{TakahashiNishimich2019} (red dashed) in all configurations at different redshifts ($z=0,1$).
  The shot noise (dash-dotted) becomes relevant above $k\sim10^{7}h\mathrm{Mpc}^{-1}$.
  In this squeezed bispectrum($k_{1}=k_{2}=k,k_{3}=0.01k$), there is a peak at $k\sim10^{3}h\mathrm{Mpc}^{-1}$ due to the subhalo enhancement.
  }
  \label{fig:Bispectrum}
\end{figure}

In Fig.~\ref{fig:Bispectrum}, we show the bispectrum evaluated at three different configurations (equilateral, flattened, squeezed) and two different redshifts ($z=0,1$).
The squeezed configuration is set to $k_{1}=k_{2}=k, k_{3}=0.01k$.
The solid (dotted) lines show the bispectrum with (without) subhalos. 
The red dashed lines are the BiHalofit model given in \cite{TakahashiNishimich2019} (a fitting formula for the matter bispectrum calibrated on the simulations at $k<30h\mathrm{Mpc}^{-1}$).
The trend observed in the bispectrum is the same as the one in the power spectrum: The inclusion of subhalos enhances the small-scale bispectrum, while the large-scale behavior (the small hump representing the peak linear bispectrum and the peak representing the largest scale of halos) is compatible with the fitting formula.
However, there is a notable difference from the power spectrum which is intrinsic to the bispectrum. 
In one squeezed configuration($k1=k2=k, k_{3}=0.01k$), it is observed that the subhalo enhancement to the bispectrum has the peak at $k_{1}=k_{2}\sim10^{3}h\mathrm{Mpc}^{-1}$ and $k_{3}\sim10h\mathrm{Mpc}^{-1}$.
This represents the correlation within the main halo where the larger scale corresponds to the scale of the largest halos, while the smaller scale involves the subhalo scale.
The straight dash-dotted lines are the shot noise effect, which is calculated by assuming that it is given by a constant $B_{\rm shot}=f_{p}^{3}/\overline{n}^{2}$ with $m=0.5M_{\odot}$ and $f_{p}=0.01$.
Precisely speaking, the shot noise effect on the bispectrum contains not just the constant term, instead, it is given by \cite{Oddo:2019run}:
\begin{align}
    \label{eq:BishotAc}
    B_{\rm shot} = \frac{f_{p}}{\overline{n}}
    \left(
    P_{\delta}(k_{1})+P_{\delta}(k_{2})+P_{\delta}(k_{3})
    \right)
    +\frac{f_{p}^{3}}{\overline{n}^{2}},
\end{align}
where $P{\delta}(k)$ is the power spectrum without the shot noise effect.
In fact, the first three terms in Eq.~(\ref{eq:BishotAc}) come from the coupling of the non-shot noise effect and the shot noise effect.
However, as mentioned earlier, we treat the shot noise separately, and the isolation of the shot noise from the rest of the terms results in ignoring the coupling terms.

Up to this point, we have assumed that the halo model can be a reasonable estimation for such a small scale as $k\sim 10^{6}h\mathrm{Mpc}^{-1}$.
However, it is not clear whether the uncertainty of the spectra based on the halo model is reasonably suppressed even at scales much smaller than the ones covered by the simulations. 
Since our aim in this paper is to provide formalism to compute the post-Born corrections given that the matter power spectrum and bispectrum are properly evaluated, we simply take it for granted that  our matter power spectrum and bispectrum models are valid at all scales.

Thus, the numerical values of the post-Born corrections that will be given later should not be understood as the robust quantitative prediction of the post-Born effects 
\footnote{In addition to the uncertainty of the matter power spectrum caused by the use of the halo model, there are other types of uncertainties coming from our ignorance about the microscopic properties of dark matter and primordial power spectrum on small scales both of which affect the shape of the matter power spectrum at small scales.} 
. 
Once the correct matter power spectrum and bispectrum are obtained, it is immediate to evaluate the post-Born corrections by using the formulation given in this paper. 
We emphasize that our general conclusion that the post-Born corrections are subdominant compared to the Born approximation remains unaffected by the refinement of the matter power spectrum in the absence of the shot noise.

The cosmological parameters we use to compute the halo model, as well as the lensing dispersion, are $h=0.6739, \Omega_{m}=0.3147, \Omega_{\Lambda}=0.6888,n_{s}=0.9665, \sigma_{8}=0.8102$.

\section{Post-Born effect on $S$ and $K$}
\subsection{Numerical computation}
In order to evaluate the lensing signal, it is essential to compute the multivariable integral over a considerably wide range for highly oscillatory functions.
To achieve this, we use the Monte Carlo algorithm to compute the lensing dispersion. 
We ensure that the error of our calculation is smaller than 10 \%  by progressively increasing the number of sample points until the fluctuation in the results is less than 10 \%.
In the case of $\braket{S^{(3)}}$ and $\braket{S_{\rm Born}S^{(2)}}$, which exhibit a slow convergence rate, the computation error is ensured to be less than 30\%.
Additionally, we assess the impact of the integration range on the final result by varying its width and  we confirm that the results remain unchanged (the default integration range is set to $10^{-4}$~$h$Mpc$^{-1}<k <10^{12}$~$h$Mpc$^{-1}$).
By undertaking these checks, we can guarantee that the results are indeed converged. 
We have to keep in mind that the error in our calculation, which we estimate to be 10\%, arises from the slow convergent rate of these integrals.
Thus, this number cannot be interpreted as the error of the actual observed $S$ and $K$.

\subsection{Born approximation}
\label{3-Ba}
For completeness, we first evaluate the variance of $S$ and $K$ under the Born approximation.
Formally, $\braket{K_{\rm Born}^{2}}$ is given by Eq.~(\ref{eq: magn vari Born}) and 
$\braket{S_{\rm Born}^{2}}$ is given by the same equation with $\sin$ function being 
replaced with $1-\cos$.
In Fig.~\ref{fig: S variance},
we show $\braket{S_{\rm Born}^{2}}$ as a function of $f$ with the corresponding Fresnel scale $1/\overline{r}_{\rm F}=\sqrt{H_{0}\omega}$ represented by the second horizontal axis at the top.
Irrespective of the source redshift, $\braket{S_{\rm Born}^{2}}$ exhibits a general behavior pertaining to the power spectrum. 
As the frequency increases from $f=10^{-18}$~Hz, it initially rises and forms a hump around $f=10^{-15}$~Hz ($k\sim 10^{-2}h\mathrm{Mpc}^{-1}$). It then reaches a peak at $f=10^{-11}$~Hz ($k\sim 10^{1}h\mathrm{Mpc}^{-1}$) and gradually decreases as the frequency moves towards higher values. 
As seen in Fig.~\ref{fig:powerspectrum}, these scales correspond to the peak of the linear power spectrum and the largest scale of the halos, respectively.

In order to understand this feature, let us write $\braket{S_{\rm Born}^{2}}$ in terms 
of the matter power spectrum as
\begin{align}
\label{3:S-Born}
\braket{S_{\rm Born}^{2}}=4\omega^{2} \left(
    \frac{3H_{0}^{2}\Omega_{m}}{2}
    \right)^{2}
    \int_{0}^{\chi_{s}} \frac{d\chi}{a^2(\chi)}
    \int\frac{dk}{2\pi k^3}
    {\left( 1-\cos {\left[\frac{(\chi_{s}-\chi)\chi}{2\chi_{s}\omega}k^{2}\right]} \right)}^2
    P_{\delta}(k,\chi).
\end{align}
The integration over $k$ is dominated by the integrand around the scale where
the argument of the cosine function becomes ${\cal O}(1)$.
Thus, approximating $P_\delta$ at that scale as a single-power law $P_\delta \propto k^{-b}$,
we find the scaling $\braket{S_{\rm Born}^{2}} \propto \omega^{1-\frac{b}{2}}$,
which explains the behavior of the purple curve in Fig.~(\ref{fig: S average}) and the reason why it reflects the matter power spectrum at the corresponding Fresnel scale.
Notice that the Taylor-expansion of the cosine in $1/\omega$ in the high-frequency regime,
which naively gives the scaling $\braket{S_{\rm Born}^{2}} \propto \omega^{-2}$, does not make sense
due to the divergence of $k$ integration stemming from the coefficient of $\omega^{-2}$.

The fact that the scaling of $\braket{S_{\rm Born}^{2}}$ depends on $b$ is one of the advantages of the weak lensing of GWs as the measurement of
the frequency dependence of $\braket{S_{\rm Born}^{2}}$ at a detectable frequency range of GWs provides a direct probe of the slope of the matter spectrum at small scales (as small as or smaller than $k\sim10^{5}\sim10^{6}h\mathrm{Mpc}^{-1}$).

The orange in Fig.~\ref{fig: K variance} shows $\braket{K_{\rm Born}^{2}}$ as a function of $f$ with the corresponding Fresnel scale at the top.
Contrary to the phase modulation, the variance of $K$ approaches a constant value in the high-frequency limit, which is nothing but the variance of the convergence $\kappa$ in geometrical optics.
Also, we can observe a hump at $f\sim10^{-15}$~Hz in the same way as the phase modulation. 
At frequencies below $f=10^{-11}~{\rm Hz}$, $\braket{K_{\rm Born}^{2}}$ decreases as the
frequency is lowered.
In order to understand this feature, let us write $\braket{K_{\rm Born}^{2}}$ in terms 
of the matter power spectrum as
\begin{align}
\braket{K_{\rm Born}^{2}}=4\omega^{2} \left(
    \frac{3H_{0}^{2}\Omega_{m}}{2}
    \right)^{2}
    \int_{0}^{\chi_{s}} \frac{d\chi}{a^2(\chi)}
    \int\frac{dk}{2\pi k^3}
    \sin^2 {\left[\frac{(\chi_{s}-\chi)\chi}{2\chi_{s}\omega}k^{2}\right]}
    P_{\delta}(k,\chi).
\end{align}
The difference between the magnification and the phase arises from the difference in the filter function.
As mentioned above, $\braket{S_{\rm Born}^{2}}$ is predominantly determined by the power spectrum at the Fresnel scale due to the filter function being relevant only when $k$ is comparable to the Fresnel scale.
On the other hand, the filter function for $\braket{K_{\rm Born}^{2}}$ indicates that it provides relevant contributions whenever the argument of $\sin(\cdots)$ is comparable to or smaller than unity. 
Thus,$\braket{K_{\rm Born}^{2}}$ computes the weighted sum of the power spectrum at all scales above the Fresnel scale.
For this reason, the magnification in wave optics can be viewed as the convergence in geometric optics with the smoothing radius being replaced by the Fresnel scale.
Since the scale above the Fresnel scale contributes to the magnification, the strong lensing due to galaxies might cause relevant effects, even if the Fresnel scale for typical GWs ($k\sim10^{6}h\mathrm{Mpc}^{2}$) is much smaller than the  size of a galaxy ($k\sim10^{3}h\mathrm{Mpc}^{-3}$).
However, as partially mentioned in \cite{Oguri:2020ldf}, the removal of strong lensing by galaxies allows us to extract the underlying lensing signal primarily due to the dark matter halos and the point mass.

\subsection{Average}
\begin{figure}[tb!]
  \centering
  \includegraphics[width=16cm]{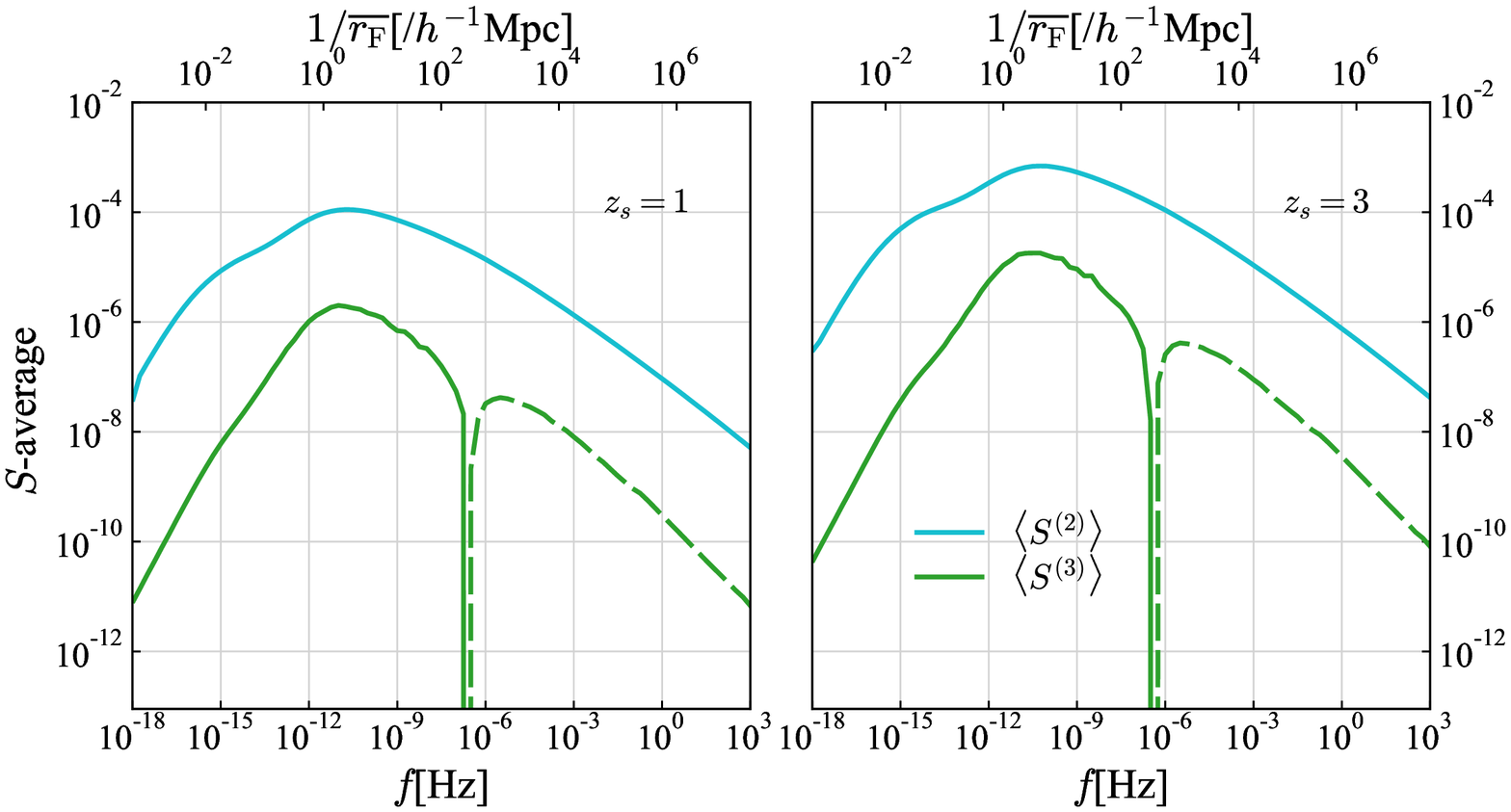} 
  \caption{
    It is clear that $\braket{S^{(3)}}$ (green) is a few orders of magnitude smaller than $\braket{S^{(2)}}$ (cyan), confirming the validity of using $\braket{S^{(2)}}$ as an approximation of $\braket{S}$. 
  In both figures, $\braket{S^{(2)}}$ experiences a hump at $f\sim10^{-15}$~Hz ($k\sim10^{-2}h\mathrm{Mpc}^{-1}$) and reaches the peak at $f\sim10^{-11}$~Hz ($k\sim10^{1}h\mathrm{Mpc}^{-1}$), tracing the behavior of the power spectrum.
  $\braket{S^{(3)}}$ also exhibits a similar behavior (the hump at $f\sim10^{-15}$~Hz and the peak at $f\sim10^{-11}$~Hz ). 
  However, the change of signature at $f\sim10^{-7}$~Hz seems to be a reflection of enhancement on the squeezed bispectrum caused by subhalos.
  The solid(dashed) line indicates the $+(-)$ value.
  }
  \label{fig: S average}
\end{figure}

Having understood the behavior of the magnification and phase modulations in the Born approximation,
let us proceed to the post-Born corrections.
The cyan curve in Fig.~\ref{fig: S average} shows the non-trivial leading order term of $\braket{S}(=\braket{S^{(2)}})$ and the green line shows their non-Gaussian correction $\braket{S^{(3)}}$.
They are both shown as a function of GW frequency $f$ and are numerically computed based on
Eq.~(\ref{eq: phase average}) and Eq.~(\ref{eq: S3 ave}) combined with Eq.~(\ref{eq: Ppotential Pmatter}) and Eq.~(\ref{eq: Bpotential Bmatter}).
The source redshift is taken to be $z_{s}=1$ for the left and $z=3$ for the right.
As mentioned in the previous section,
$\braket{S^{(2)}}$ should be always positive and our numerical result also confirms this property.

Fig.\ref{fig: S average} implies that $\braket{S}$ shows close similarities to $\braket{S_{\rm Born}^{2}}$: it has a hump around $f=10^{-15}$~Hz and a peak at $f=10^{-11}$~Hz, representing the hump and peak of the power spectrum as $\braket{S_{\rm Born}^{2}}$ does.
Based on the same analysis, we found that $\braket{S^{(2)}}$ scales as $\omega^{(2-b)/4}$ thus, the scaling behavior is the same as $\braket{S_{\rm Born}^{2}}$ as well as $\braket{S^{(2)}}$ and $\braket{S_{\rm Born}}$ having the same order of magnitude $\braket{S}\sim\braket{S_{\rm Born}^{2}}$.
We also evaluate $\braket{S^{(3)}}$ (the correction term to $\braket{S}$) that originates due to the bispectrum contribution.
$\braket{S^{(3)}}$ is shown to be about two orders of magnitude smaller than $\braket{S^{(2)}}$, which guarantees that the correction to $\braket{S}$ due to the non-Gaussina effect is negligible and the estimation of $\braket{S}$ solely by the power spectrum term is still reliable.

Based on the result of $\braket{S_{\rm Born}^{2}}$ and $\braket{S^{(2)}}$, it is expected that the frequency dependence of $\braket{S^{(3)}}$ also reflects the behavior of the bipsetrum.
Indeed, $\braket{S^{(3)}}$ traces the general behavior of the bispectrum: it initially increases with a small hump at $f=10^{-15}$~Hz and reaches a peak at $f=10^{-11}$~Hz, after which it decreases.
However, there is a frequency scale at $f\sim10^{-7}$~Hz (corresponding Fresnel scale is $10^{2}\sim 10^{3}h\mathrm{Mpc}^{-1}$) where the signature of $\braket{S^{(3)}}$ flips.
Our numerical computation reveals that this change of signature $\braket{S^{(3)}}$ is attributed to the squeezed bispectrum.
Nevertheless, the precise influence of different squeezed bispectrum configurations, such as the ratio of the larger and smaller scales, on the change of signature has not been elucidated yet.
However, it is worth noting that this scale significantly exceeds the scale of our interest ($f\sim0.01\sim1000$Hz), and would also be predominantly influenced by galactic structures.
For this reason, the specific behavior of $\braket{S^{(3)}}$ around this frequency is not particularly relevant in our analysis compared to the scale around $f\sim1$~Hz
Also, as we will see later, the hump in $\braket{S^{(3)}}$ at $f\sim10^{-15}$~Hz is relatively smooth compared to the hump seen in $\delta_{S^{2},\rm c}$, though its specific physical meaning is unclear to us yet.

\begin{figure}[tb!]
  \centering
  \includegraphics[width=16cm]{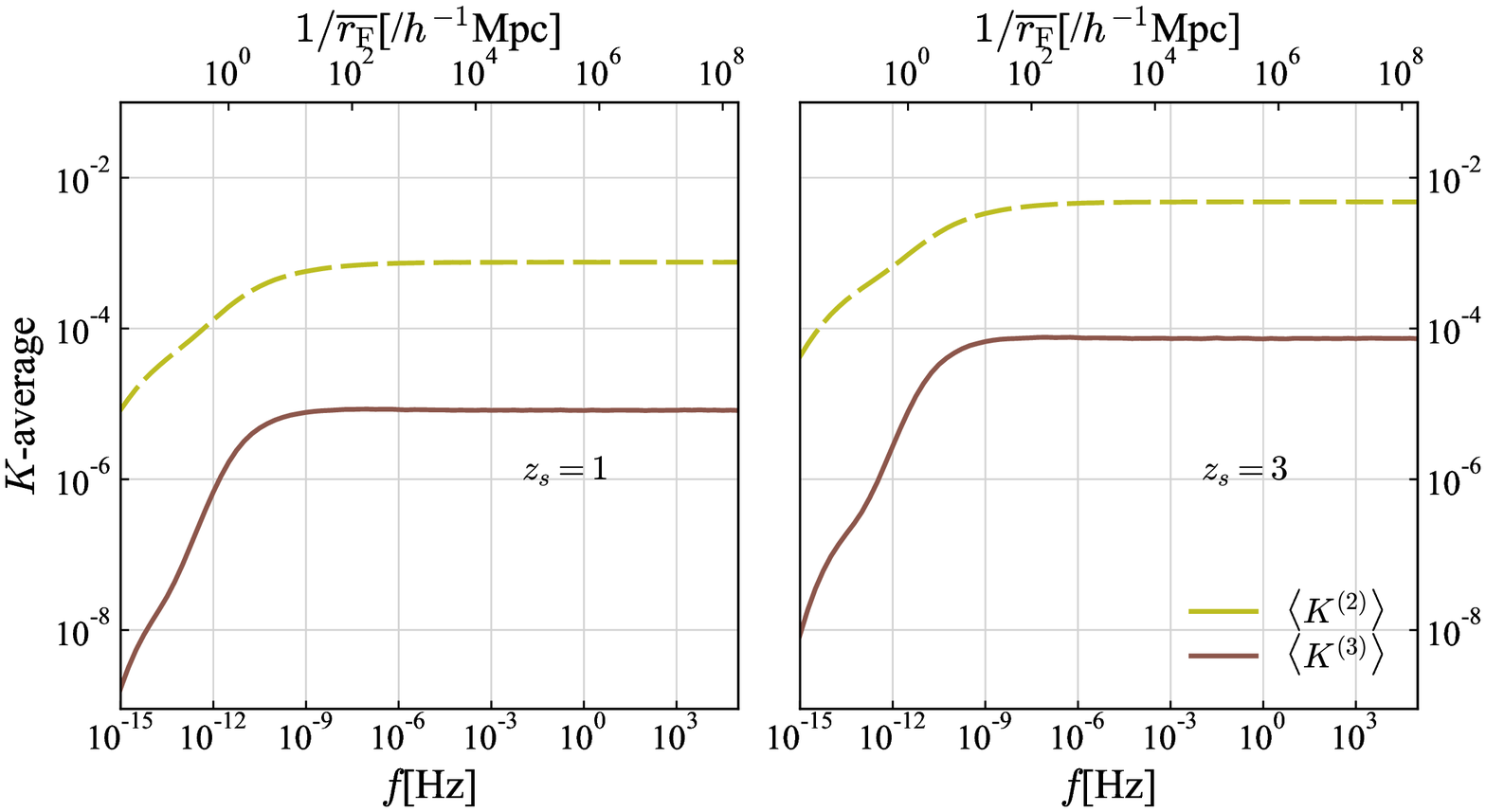} 
  \caption{
  As expected, both $\braket{K^{(2)}}$ (olive) and $\braket{K^{(3)}}$ (brown) approaches a constant value as the frequency increases.
  Since $\braket{K^{(3)}}$ is more than an order of magnitude smaller than $\braket{K^{(2)}}$, 
  it is reasonable to use $\braket{K^{(2)}}$ as an approximation of $\braket{K}$.
  At the low end of these figures, $K^{(3)}$ decreases more rapidly than $\braket{K^{(2)}}$, indicating the small non-Gaussian effect at large scales.
  We can see that the hump also exist ($f\sim10^{-15}$~Hz) for the average of $K$
  The solid(dashed) line indicates the $+(-)$ value.
  }
  \label{fig: K average}
\end{figure}
The olive curve in Fig.~\ref{fig: K average} shows the non-trivial leading order correction to $\braket{K}$, while the brown dashed line is the next leading order contribution.
As observed in $\braket{K_{\rm Born}^{2}}$, $\braket{K}$initially increases and approaches a constant.

We find that $\braket{K}$ is always negative within the frequency range of our calculation, which is not obvious from the formulation of $\braket{K}$. 
Although the exact reason for this negative sign is still unclear, a similar result appears in geometric optics.
In geometric optics, the mean convergence $\braket{\kappa}$ is shown to follow the simple relation $\braket{\kappa}=-2\braket{\kappa^2}<0$ \footnote{By performing integration by parts for the integration over $\chi_1$ in Eq.~(\ref{ave-K}),
we can verify that $\braket{K}$ and $\braket{K_{\rm Born}^{2}}$ derived in this paper
reproduce the relation $\braket{K}=-\braket{K_{\rm Born}^{2}}$ in the high-frequency limit.}, hence negative, as demonstrated analytically \cite{Kaiser:2016gyr} and in $N$-body simulation \cite{Takahashi:2011fgq}.
Since $K$ is reduced to the convergence $\kappa$ in weak lensing when taking the geometric optics limit ($\omega\to\infty$),
$\braket{K}$ should be negative at least when $\omega \to \infty$ is taken.

While $\braket{K^{(3)}}$ (the correction to $\braket{K}$) also shows similar behavior to $\braket{K^{(2)}}$, it is smaller than $\braket{K^{(2)}}$ by about two orders of magnitude.
Similar to the case of $\braket{S}$, the correction to $\braket{K}$ from the non-Gaussian effect is small, and therefore the estimation of $\braket{K}$ solely by the power spectrum term is reliable.


\subsection{Variance}
Before discussing the significance of the post-Born corrections to the variance solely due to the  presence of the dark matter halos, it is important to clarify that we have successfully calculated the post-Born corrections numerically for almost all cases using the expressions for $\delta_{S^{2},\rm dc},\delta_{S^{2},\rm c},\delta_{K^{2},\rm dc},\delta_{K^{2},\rm c}$ presented in Appendix~\ref{app:postBornvariance}.
However, there is one case that poses computational challenges when computing the Gaussian correction to the phase $\delta_{S^{2}\rm dc}$, particularly above frequency $f=10^{-7}$~Hz.
The reason for this difficulty arises from the cancellation of significant digits.
Specifically, $\braket{(S^{(2)})^{2}}_{\rm dc}$ and $2\braket{S^{(1)}S^{(3)}}_{\rm dc}$  in $\delta_{S^{2}\rm dc}$ have very similar values but differ in their signature at high frequency, causing the cancellation of almost all contributions from each term and leaving very small differences.
It is known that, in geometric optics, 
the translation invariance of the correlation functions plays a key role 
in this type of cancellation for $K$ \cite{Shapiro:2006em, Hilbert:2008kb, Krause:2009yr, Pratten:2016dsm, Petri:2016qya}. 
To overcome this issue, we employ an alternative approximation method to obtain $\delta_{S^{2}\rm dc}$ at frequencies above $f\sim10^{-7}$~Hz.
In appendix \ref{app: high freq S}, we develop a method to approximately obtain $\delta_{S^{2},\rm dc}$ given the power spectrum following a singe-power law at high wavenumber.
On the other hand, the power spectrum can be approximated by a single-power law around an arbitrary reference scale $k_{0}$ using Eq.~(\ref{Pdelta-largek}).
Due to the expectation that the power spectrum at the Fresnel scale dominates the contribution to $\delta_{S^{2},\rm dc}$, we chose the reference scale to be the approximated value of the corresponding Fresnel scale, namely $k_{0}=\sqrt{H_{0}\omega}$. 
By utilizing the approximated power spectrum, which takes the form of a power law around the Fresnel scale and our developed computation method, we can effectively compute $\delta_{S^{2},\rm dc}$ with reasonable accuracy using the following expression:

\begin{align}
    \label{eq: postSvar Asymp main}
    &\delta_{S^{2},\rm dc}
    =
    \frac{3}{4}\left(\frac{3H_{0}^{2}\Omega_{m}}{2}\right)^{4}
    \int_{0}^{\chi_{s}}d\chi_{1}
    \int_{0}^{\chi_{1}}d\chi_{2}
    \frac{
    W^{4}(\chi_{1},\chi_{s})
    \chi_{1}^{4}\chi_{2}^{4}}{a^{2}(\chi_{1})a^{2}(\chi_{2})}
    \frac{1}{(2\pi)^{2}}
    \notag
    \\
    &
    \times
    \left\{ 
    \frac{B(\chi_{2})}{\omega^{\frac{b_{2}}{2}-1}} 
     \int_{0}^{\infty}dk_{1}
    k_{1}P_{\delta}(k_{1},\chi_{1})
    \left[\frac{\chi_{2}^{2}W(\chi_{2},\chi_{s})}{2}\right]^{\frac{b_{2}}{2}-1}
    \left(1-\frac{2^{\frac{b_{2}}{2}}}{2}\right)
    \Gamma\left(1-\frac{b_{2}}{2}\right)
    \sin{\left[\frac{b_{2}}{2}\right]}
    +(1\leftrightarrow2)
    \right\}.
\end{align}
where $b_{2}=b(\chi_{2})$ and $B(\chi_{2})$ are calculated using Eq.~(\ref{Pdelta-largek}).
Note that in analyzing the frequency dependency of $\delta_{S^{2},\rm dc}$, it can be regarded as a constant with respect to the variations in $\chi$ due to the relatively small variation of $b(\chi)$ with changing $\chi$.

\begin{figure}[tb!]
  \centering
  \includegraphics[width=16cm]{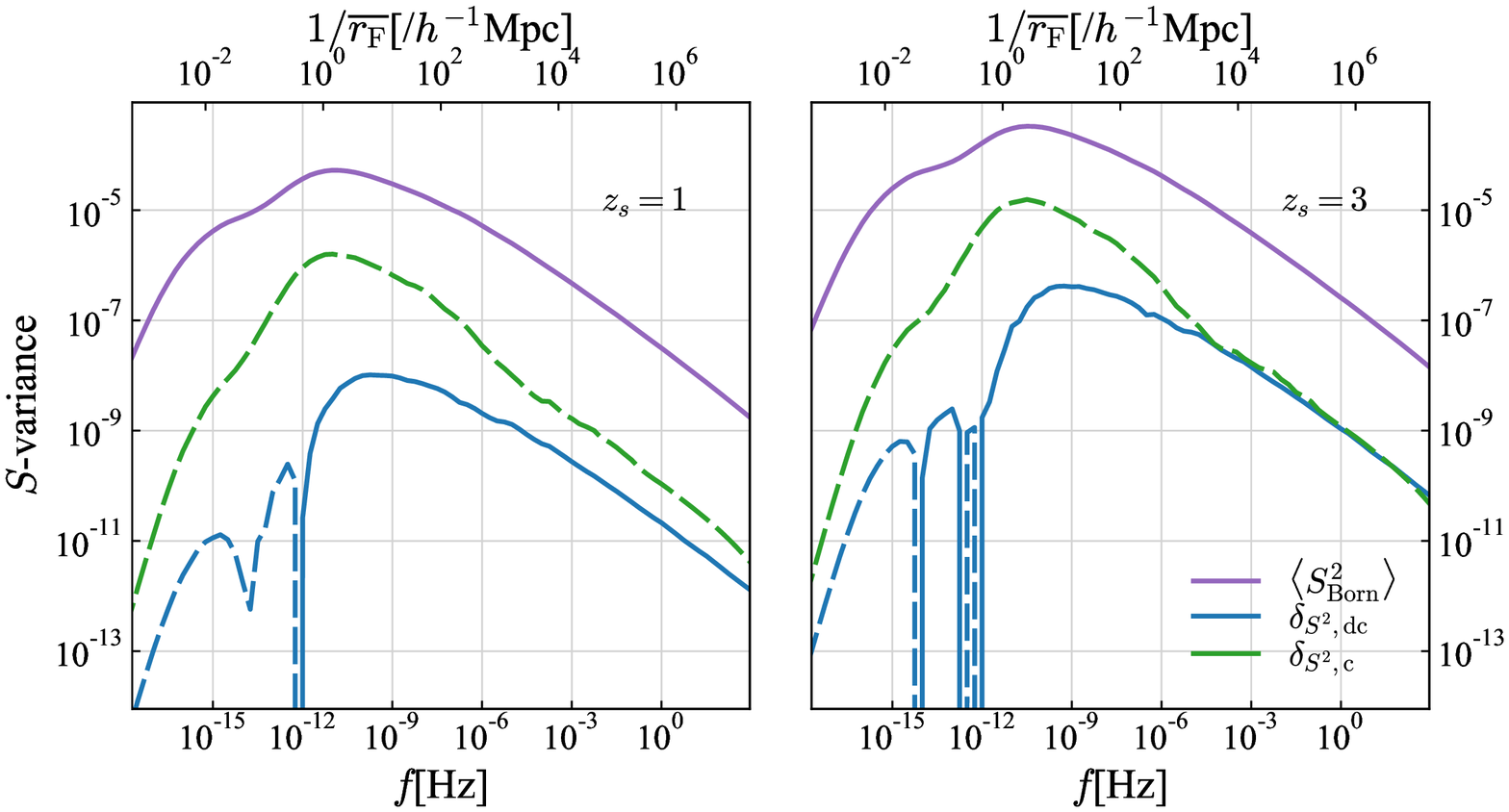} 
  \caption{
    In the frequency range of our interest $f>0.01$~Hz ($k\sim10^{5}h\mathrm{Mpc}^{-1}$),  $\braket{S_{\rm Born}^{2}}$ (purple) is more than two orders of magnitude larger than both $\delta_{S^{2},\rm dc}$ (blue) and $\delta_{S^{2},\rm c}$ (green), showing that the Born approximation remains valid under the halo model. 
    In this frequency range, the post-Born corrections $\delta_{S^{2},\rm dc}$, and $\delta_{S^{2},\rm c}$ mainly enhance the amplitude of $\braket{S_{\rm Born}^{2}}$, which is interpreted as the effect of the halos being distributed unevenly.
    While $\braket{S_{\rm Born}^{2}}$ and $\delta_{S^{2},\rm c}$ show the hump and the peak at $f\sim10^{-15}$~Hz ($k\sim10^{5}h\mathrm{Mpc}^{-1}$, the scale corresponding to the peak of linear spectra) and at $f\sim10^{11}$~Hz ($k\sim10^{5}h\mathrm{Mpc}^{-1}$, the scale corresponding to the size of the largest halos), $\delta_{S^{2},\rm dc}$ changes its signature multiple times in this range.
    Also, at $f\sim10^{-5}$~Hz, there is a sudden change of the slope of $\delta_{S^{2},c}$, respecting the effect of the enhanced squeezed bispectrum by subhalos.
    The solid(dashed) line indicates the $+(-)$ value.
    }
  \label{fig: S variance}
\end{figure}

Fig.~\ref{fig: S variance} shows the variance of $S$ including the Born approximation $\braket{S_{\rm Born}^{2}}$ and two types of post-Born corrections (Gaussian $\delta_{S^{2}, \rm dc}$ and non-Gausian $\delta_{S^{2},\rm c}$ with the non-Gaussian correction only containing the bispectrum contribution).
The source redshift is taken to $z_{s}=1,3$ for the left and right graphs, respectively.
This result suggests that the post-Born corrections to the variance of $S$ are subdominant at all frequency ranges considered in the paper.
Also, it implies that the smallness of the post-Born corrections remains valid regardless of whether the matter density is Gaussian or non-Gaussian, as long as the contribution to the variance is primarily attributed to the dark matter halos.
Note that the computation of $\delta_{S^{2},\rm dc}$ using Eq.~(\ref{eq: postSvar Asymp main})  is reliable as it nicely coincides with the result of the numerical computation of $\delta_{S^{2}.\rm dc}$ at $f\sim 10^{-7}$~Hz.

The behavior of $\delta_{S^{2}\,dc}$ is drastically different from $\braket{S_{\rm Born}^{2}}$ below $f=10^{-9}$~Hz.
While $\braket{S_{\rm Born}^{2}}$ exhibits a hump and peak at $f=10^{-15}$~Hz and $f=10^{-11}$~Hz, the Gaussian correction $\delta_{S^{2},\rm dc}$ changes its signature multiple times in this region.
On the other hand, the non-Gaussian correction $\delta_{S^{2},\rm c}$ shows generally the same behavior as $\braket{S_{\rm Born}^{2}}$: it increases with a small hump at $f\sim10^{-15}$~Hz and reaches the peak at $f=10^{-11}$~Hz, after which  it gradually decreases.
However, there is a frequency scale at $f=10^{-5}$~Hz corresponding to $k\sim10^{4}h\mathrm{Mpc}^{-1}$ where the behavior of  $\delta_{S^{2},\rm c}$  changes.
Similar to $\braket{S^{(3)}}$, it appears that this change of behavior represents the effect of subhalos through the dependence of $\braket{S^{(3)}}$ on the squeezed bispectrum. 
Moreover, by comparing Fig.~\ref{fig: S average} and Fig.~\ref{fig: S variance} at $f\sim10^{-15}$~Hz, it is clear that the hump observed in Fig.~\ref{fig: S average} is relatively less prominent than the one observed in Fig.~\ref{fig: S variance}.

Since the frequencies $f\sim10^{-15}$~Hz and $f\sim10^{-11}$~Hz correspond to the peak that appears in the linear spectrum and the scale of the largest halos in the halo model ($b\sim0$ and $b\sim2$ for each scale), it is suggested that $\delta_{S^{2},\rm dc}$ encodes information about this scale.
In fact, Eq.~(\ref{eq: postSvar Asymp main}) partially captures this behavior.
For example, the sine function in Eq.(\ref{eq: postSvar Asymp main}) enforces this term to be suppressed when $b\sim0$.
Note that this estimation should not be taken too seriously as Eq.(\ref{eq: postSvar Asymp main}) cannot be used when $b<2$.
In addition, identifying the exact scale is also challenging since the error of the numerical calculation inevitably increases at this specific frequency.
Thus, further investigation is needed for understanding the more precise property of $\delta_{S^{2},\rm dc}$ at around this frequency.

On the other hand, the frequency above $f=10^{-9}$~Hz provides important physical insight.
By examining the frequency dependence of $\delta_{S^{2},\rm dc}$ in Eq.~(\ref{eq: postSvar Asymp main}), it is clear $\delta_{S^{2},\rm dc}$ scales as $\omega^{1-b/2}$, the same scaling property as $\braket{S_{\rm Born}^{2}}$.
The underlying reason for this scaling can be understood as follows.
In the case of $\braket{S_{\rm Born}^{2}}$, the main contribution arises solely from the power spectrum (thus two-point correlation function) evaluated at the Fresnel scale. 
However, as Eq.~(\ref{eq: postSvar Asymp main}) implies, the main contribution to $\delta_{S^{2},\rm dc}$ comes from the product of the power spectrum evaluated at two scales: the Fresnel scale and the scale that primarily contributes to $\int dk kP(k)$.
This leads to the power spectrum evaluated at the Fresnel scale producing $\omega^{1-b/2}$, while the other one contributes mainly to the amplitude of $\delta_{S^{2},\rm dc}$. 

The significance of this cross-term contribution is that information about large-scale density fluctuation is encoded in the variance of $S$ through higher-order terms.
The above discussion about the behavior of the Gaussian correction $\delta_{S^{2},\rm dc}$ immediately implies that the correlations between density fluctuations at the Fresnel scale and much larger scales are the main contributing factor to $\delta_{S^{2},\rm dc}$.
Physically, this correlation arises from the fact that small-scale density fluctuations are more likely to grow in regions with a high matter density (i.e. large-scale density fluctuations are significant). 
In fact, the main contribution to the non-Gaussian correction $\delta_{S^{2},\rm c}$ at high frequencies also comes from this correlation, since $\delta_{S^{2},\rm c}$ is particularly affected by the squeezed bispectrum which is by definition the correlation between large and small-scale density fluctuation.

In order to understand this effect more intuitively, let us consider a universe in which lensing only occurs by dark matter halos.
In this case,  $\braket{S_{\rm Born}^{2}}$ fails to capture the uneven distribution of the halos.
The rationale for this is that $\braket{S_{\rm Born}^{2}}$ is solely determined by the abundance of halos at the Fresnel scale, and information about unevenness (such as bispectrum) is absent.
However, it is expected that the true $\braket{S^{2}}$ would deviate from $\braket{S_{\rm Born}^{2}}$ simply because the lensing effect is enhanced(suppressed) in regions with a high(low) halo number density compared to regions with average number density.
Our analysis indicates that the post-Born approximation can capture the uneven distribution of lens objects by accounting for the correlation between the density fluctuation at the Fresnel scale and the density fluctuations at much larger scales.

In terms of the frequency range of our interest, which is above $f=0.01$~Hz and is corresponding to the scale dominated by the dark low-mass halos, our result suggests that the post-Born corrections to the variance are estimated to be $\delta_{S^{2}}/\braket{S_{\rm Born}^{2}}\lesssim\mathcal{O}(10^{-2})$.
Also, it is important to mention that the Gaussian correction term $\delta_{S^{2},\rm dc}$, and the non-Gaussian correction term $\delta_{S^{2},\rm c}$ are relatively the same order of magnitude in this frequency range.

\begin{figure}[tb!]
  \centering
  \includegraphics[width=16cm]{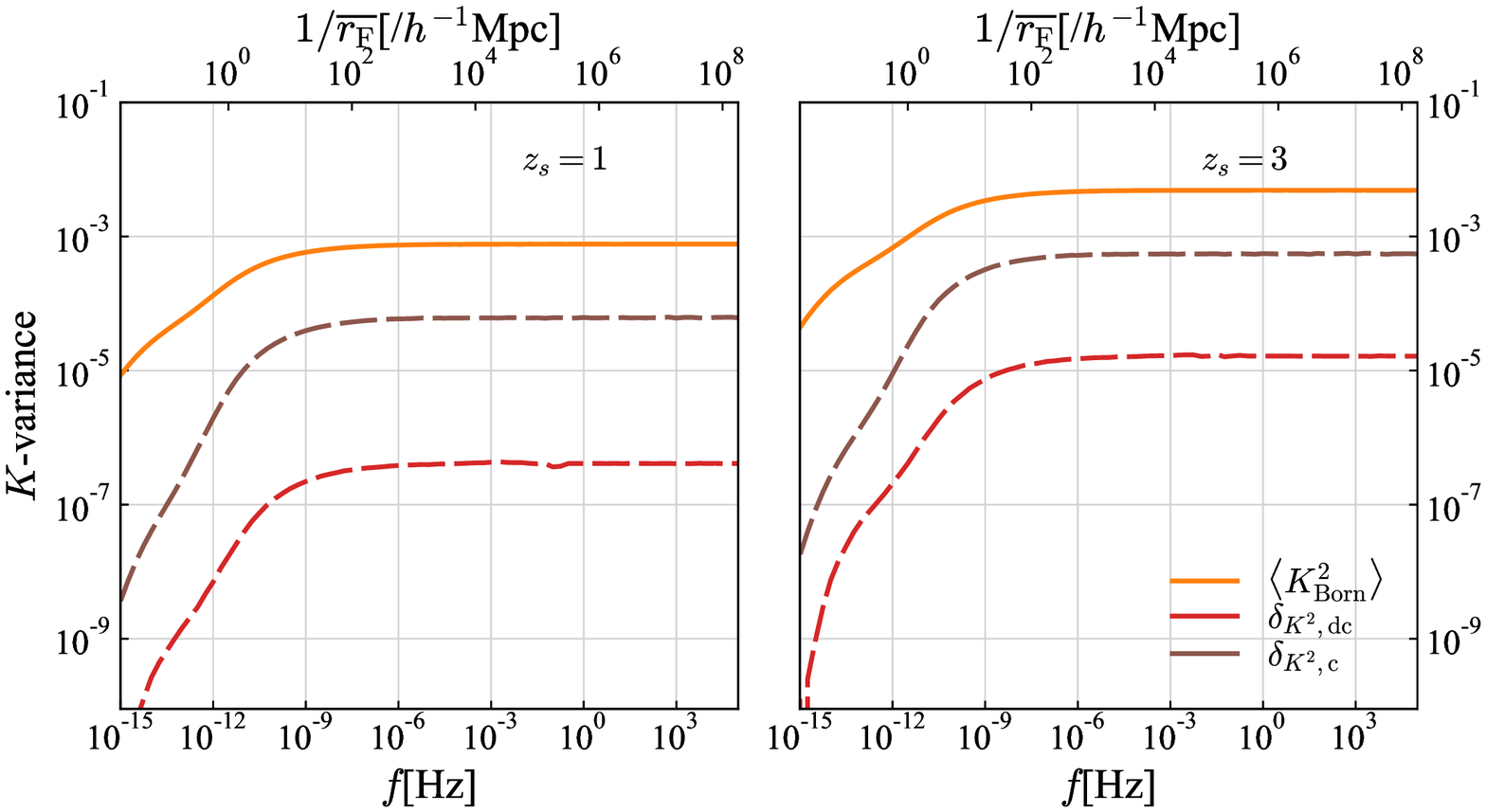}
  \caption{
  The Born approximation $\braket{K^{2}_{\rm Born}}$ (orange) is the dominant effect on the $K$, with $\delta_{K^{2},\rm c}$ (brown) being the second.
  Gaussian contribution $\delta_{K^{2},\rm c}$ (red) is more than an order of magnitude smaller than the non-Gaussian term.
  The solid(dashed) line indicates the $+(-)$ value.
  }
  \label{fig: K variance}
\end{figure}

Let us next investigate the variance of $K$.
Fig.~\ref{fig: K variance} shows the correction to $\braket{K^{2}_{\rm Born}}$ in the case where the source redshift is $z_{s}=1$ and $z_{s}=3$, respectively.
Similar to $\braket{K^{2}_{\rm Born}}$ and $\braket{K}$, the correction to the variance
is suppressed at low frequency and approaches a constant value in the high-frequency limit.
One difference between the post-Born correction to the magnification and the phase modulation is the relative magnitude of the non-Gaussianity term to the Gaussianity term.
In the case of the magnification, the non-Gaussian term $\delta_{K^{2}\rm c}$ is almost two orders of magnitude larger than the Gaussian term $\delta_{K^{2},\rm dc}$ in the frequency range $f=0.01\sim 1000$~Hz, while those terms are relatively the same order of magnitude for the phase modulation.
Due to this effect, the post-Born correction to the variance of $K$ in the high-frequency range is mainly caused by the non-Gaussian term $\delta_{K^{2}}\approx\delta_{K^{2},\rm c}$.
The relative magnitude of the post-Born correction is given by $\delta_{K^{2}}/\braket{K_{\rm Born}^{2}}\lesssim \mathcal{O}(10^{-1})$ in the frequency range where the magnification can be treated as a constant.
On the other hand, when the frequency is smaller than $f=10^{-9}$~Hz, the non-Gaussian term exhibits a faster decrease compared to the Gaussian correction term, which can be seen by the reduction of the difference between $\delta_{K^{2}\rm c}$ and $\delta_{K^{2}\rm dc}$ in Fig.~(\ref{fig: K variance}).
This behavior reflects the fact that the non-Gaussianity is less significant at larger scales.

\subsection{Shot noise contribution}
\label{shot-noise}
Up to this point, we have not considered the effect of the shot noise 
coming from the Poisson distributed stars (or any other dark compact objects)
because i) the low-frequency GWs are not strongly affected by the point masses and 
ii) numerical integrations face technical challenges associated with the highly oscillatory behavior of the integrand.
However, the shot noise may become important at high frequencies, as suggested in \cite{Oguri:2020ldf}, and the effect of the post-Born correction needs to be considered.
Thus we discuss the contribution of the shot noise  in this subsection.
At the level of the post-Born approximation, the lensing signals have a dependency on not only the terms purely representing the shot noise but also the cross term between the shot noise terms and the smooth halos terms.
This is due to the fact that the higher-order corrections to $S$ and $K$ contain the bispectrum and the product of the power spectrum.
The evaluation of the cross terms is beyond the scope of this paper and we will only focus on the pure contribution of the shot noise.

Formally, the shot noise is given by adding a constant to the matter power spectrum and the bispectrum, namely,
\begin{align}
    \label{P-shotnoise}
    P_{\delta}=\frac{f_{p}^{2}}{\overline{n}},
    \\
    \label{B-shotnoise}
    B_ {\delta} =\frac{f_{p}^{3}}{\overline{n}^{2}},
\end{align}
where $\overline{n}$ is the average number density of an individual star 
and $f_{p}$ is the mass fraction of the point mass to total matter density.
For simplicity, we ignore the time variation of $\overline{n}$ due to stellar evolution.
As mentioned earlier, we chose $f_{p}=0.01$ and $m=0.5M_{\odot}$ as a fiducial value, which is consistent with observations\cite{Fukugita:1997bi}.
The shot noise effect on the lensing signal is caused by the point mass, which in reality, possesses a finite physical size.
To account for this finite size, it is reasonable to introduce a cutoff scale $k_{c}$, which characterizes the regime where the point mass approximation holds. 
Since the cutoff scale represents the size of stars in this case, and we chose $k_{c}=4\times10^{13}$~$h$Mpc$^{-1}$($\sim10^{-6}$~km$^{-1}$).

We first evaluate the shot noise contribution to $\braket{S^{2}_{\rm Born}}$
and $\braket{K^{2}_{\rm Born}}$.
As long as it is smaller than the Fresnel scale, the size of the stars $k_{c}$ does not have a relevant contribution to the results, allowing us to effectively take $k_{c}\to \infty$.
Substituting Eq.~(\ref{P-shotnoise}) for the power spectrum that appears in Eq.~(\ref{3:S-Born}), we can analytically
perform the integral over $k$ by using the formula 
$\int_0^\infty \frac{dx}{x^3} {(1-\cos x^2)}^2=1/(4\pi)$.
Then, we obtain
\begin{align}
     \label{eq: SKborn shot}
    \braket{S^{2}_{\rm Born}}_{\rm shot}
    =&
     \frac{1}{4}\left(\frac{3H_{0}\Omega_{m}}{2}\right)^{2}
    \int_{0}^{\chi_{s}}
    \frac{d\chi}{a^{2}(\chi)}
     \chi^{2} W(\chi,\chi_{s})
    \times\omega\frac{f_{p}^{2}}{\overline{n}}.
\end{align}
Similarly, using the formula $\int_0^\infty \frac{dx}{x^3} \sin^2 x^2=1/(4\pi)$,
we find that $\braket{K^{2}_{\rm Born}}_{\rm shot}=\braket{S^{2}_{\rm Born}}_{\rm shot}$. 
Thus, both the variance of $S$ and $K$ under the Born approximation diverge in the high-frequency limit.
Note that this divergence comes from neglecting the size of stars.
In reality, the point mass approximation breaks down when the Fresnel scale becomes smaller than the size of stars.
By properly incorporating the size of stars, $\braket{S^{2}_{\rm Born}}_{\rm shot}$ and $\braket{K^{2}_{\rm Born}}_{\rm shot}$ can be shown to remain finite in the high-frequency limit.

\subsubsection{Shot noise contribution to average}
Let us next investigate the average $\braket{S}$ and $\braket{K}$.
Plugging Eqs.~(\ref{eq: Ppotential Pmatter}) and (\ref{P-shotnoise}) into the
expression $\braket{S}$ given by Eq.~(\ref{ave-S}),
we find that the integration over $k$ diverges logarithmically at large $k$.
Hence, we need the cutoff wavenumber $k_c$ which physically represents 
the inverse of the size of the stars.
With this cutoff, we can perform the integration over $k$ and the result is given by
\begin{align}
    \braket{S}_{\rm shot}=&2\left(\frac{3H_{0}\Omega_{m}}{2}\right)^{2}
    \int_{0}^{\chi_{s}}d\chi
    \int_{0}^{\chi}d\chi'
    \frac{\chi'^{2}}{\chi^{2}}
    \frac{1}{a^{2}(\chi')}
    \frac{1}{4\pi}  \mathrm{Cin}\left(k_{c}^{2}\frac{\chi'^{2}W(\chi',\chi)}{\omega}\right)
    \times\omega\frac{f_{p}^{2}}{\overline{n}}
\end{align}
where $\mathrm{Cin}(x)$ is the cosine integral defined as $\mathrm{Cin}(x)=\int_{0}^{x}dt\frac{1-\cos{t}}{t}$.
When $x$ is sufficiently large, the cosine integral is approximated as $\mathrm{Cin}(x)\approx \log{(e^{\gamma}x)}$, where $\gamma $ is Euler's constant.
We usually consider the case where $\chi_{s}$ takes the cosmological distance ($\chi_{s}\approx1/H_{0}$).
In this specific case, we can further approximate this expression by 
\begin{align}
\label{eq: SAve shot}
    \braket{S}_{\rm shot}
    \sim&
     \frac{1}{2\pi}\left(\frac{3H_{0}\Omega_{m}}{2}\right)^{2}
    \int_{0}^{\chi_{s}}d\chi
    \int_{0}^{\chi}d\chi'
    \frac{\chi'^{2}}{\chi^{2}}
    \frac{1}{a^{2}(\chi')}
    \times
    \omega
    \frac{f_{p}^{2}}{\overline{n}}
    \log{\left(\frac{k_{c}^{2}}{H_{0}\omega}\right)}.
\end{align}
Since the cutoff scale $k_{c}$ only appears in $\log(\cdots)$, it can be eliminated by the weighted subtraction of $S$ evaluated at different frequencies. 
For example, when $\Braket{S(\omega_{1})/\omega_{1}-S(\omega_{2})/\omega_{2}}$ is computed using the shot noise power spectrum, $k_{c}^{2}/H_{0}\omega$ appeared as an argument of $\log{(\cdots)}$ is replaced by $\omega_{2}/\omega_{1}$. 

As for $\braket{K}$, the integral over $k$ does not diverge and we can practically take $k_c \to \infty$.
Then, the result is given by
\begin{align}
    \label{eq: KAve shot}
    \braket{K}_{\rm shot}=&
   -\frac{1}{4}\left(\frac{3H_{0}\Omega_{m}}{2}\right)^{2}
    \int_{0}^{\chi_{s}}d\chi
    \int_{0}^{\chi}d\chi'
    \frac{\chi'^{2}}{\chi^{2}}
    \frac{1}{a^{2}(\chi')}
    \times
    \omega
    \frac{f_{p}^{2}}{\overline{n}}.
\end{align}
The third-order terms  $\braket{S^{(3)}}$ and $\braket{K^{(3)}}$ can be calculated in a similar way.
Using  Eq.~(\ref{eq: Bpotential Bmatter}) and Eq.~(\ref{B-shotnoise}), we obtain the following expressions:
\begin{align}
    \label{eq: S3 shot}
    &\braket{S^{(3)}}_{\rm shot}
    =0
    \\
    \label{eq: K3 shot}
    &\braket{K^{(3)}}_{\rm shot}
    =
    8\omega^{2}\left(\frac{3H_{0}^{2}\Omega_{m}}{2}\right)^{3}
    \int_{0}^{\chi_{s}}\frac{d\chi }{a(\chi)^{3}}\chi^{2}W(\chi,\chi_{s})J\times\frac{f_{p}^{3}}{\overline{n}^{2}}
\end{align}
where $J$ is just a number given by the following integral:
\begin{align}
         &J=\frac{1}{12}\int_{0}^{\infty}\frac{d\xi_{1}}{2\pi}
    \int_{0}^{\infty}\frac{d\xi_{2}}{2\pi}
    \int_{0}^{2\pi}\frac{d\phi}{2\pi}\notag
    \\
    &\times\frac{\sin{(\xi_{1}^{2})}+\sin{(\xi_{2}^{2})}+\sin{(\xi_{1}^{2}+\xi_{2}^{2}+2\xi_{1}\xi_{2}\cos{\phi})}-2\sin{(\xi_{1}^{2}+\xi_{2}^{2}+\xi_{1}\xi_{2}\cos{\phi})}}{\xi_{1}\xi_{2}(\xi_{1}^{2}+\xi_{2}^{2}+2\xi_{1}\xi_{2}\cos{\phi})}
\end{align}
which is found to be $J\simeq0.0021$.
Note that $\braket{S^{(3)}}_{\rm shot}$ becomes exactly zero.
In fact, $\braket{S^{(3)}}_{\rm shot}$ is formally written in the same way as $\braket{K^{(3)}}_{\rm shot}$ is given:
    \begin{align}
        \braket{S^{(3)}}_{\rm shot}=&  8\omega^{2}\left(\frac{3H_{0}^{2}\Omega_{m}}{2}\right)^{3}
    \int_{0}^{\chi_{s}}\frac{d\chi }{a(\chi)^{3}}\chi^{2}W(\chi,\chi_{s})L\times\frac{f_{p}^{3}}{\overline{n}^{2}}
    \end{align}
    where $L$ is given by an integral
    \begin{align}
        &L=\frac{1}{6}\int_{0}^{\infty}\frac{d \xi_{1}}{2\pi}
    \int_{0}^{\infty}\frac{d \xi_{2}}{2\pi}
    \int_{0}^{2\pi}\frac{d \phi}{2\pi}\notag
    \\
    &\times
    \frac{
    2\sin^{2}\left(\xi_{1}^{2}+\xi_{2}^{2}+\xi_{1}\xi_{2}\cos{\phi}\right)
    -\sin^{2}(\xi_{1}^{2})
    -\sin^{2}(\xi_{2}^{2})
    -\sin^{2}(\xi_{1}^{2}+\xi_{2}^{2}+2\xi_{1}\xi_{2}\cos{\phi})
    }{\xi_{1}\xi_{2}(\xi_{1}^{2}+\xi_{2}^{2}+2\xi_{1}\xi_{2}\cos{\phi})}
\end{align}
However,  this integral is found to be exactly zero.
This implies that the corrections to $\braket{S}$ come from much higher order terms but it is not clear the rationale for $\braket{S^{(3)}}$ being zero.

\begin{figure}[tb!]
  \centering
  \includegraphics[width=16cm]{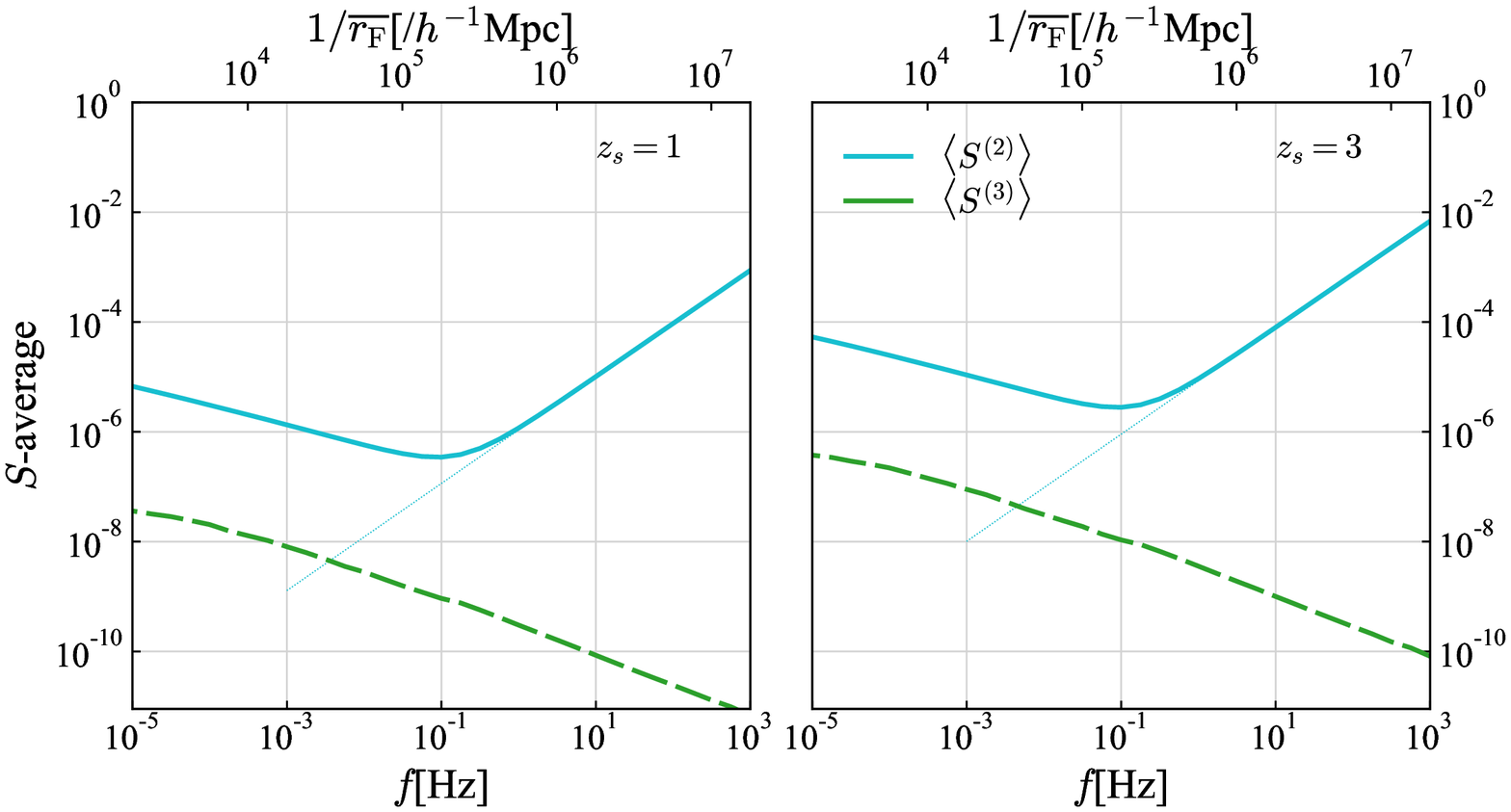} 
  \caption{
  The cyan line shows the value of $\braket{S}$ including the halo and the shot noise terms.
  The shot noise is represented by the thin cyan straight line ascending in the upper right direction.
  The shot noise term overcomes the halo term at $f\sim10^{-1}$~Hz.
  Compared to the shot noise effect on $\braket{S_{\rm Born}^{2}}$ in Fig.~(\ref{fig: S variance shot}), the onset of the shot noise effect for $\braket{S}$ occurs at a lower frequency than $\braket{S_{\rm Born}^{2}}$ due to $\braket{S^{(2)}}$ depending on the size of the point mass $k_{c}$.
  The shot noise effect on $\braket{S^{(3)}}$ (green) vanishes, indicating that the higher-order terms are necessary to provide the corrections to $\braket{S^{(2)}}$.
  The solid(dashed) line indicates the $+(-)$ value.
  }
  \label{fig: S average shot}
\end{figure}

Fig.\ref{fig: S average shot} shows the shot noise contribution to the average of $S$. 
For both cases (the source redshift $z_{s}=1,3$), the shot noise effect becomes dominant at slightly above $f=0.1$~Hz.
This frequency that the shot noise takes over is lower than the one for the variance.
This is due to the enhancement of the shot noise effect on $\Braket{S}$, which arises from the logarithmic factor in Eq.(\ref{eq: SAve shot}).

\begin{figure}[tb!]
  \centering
  \includegraphics[width=16cm]{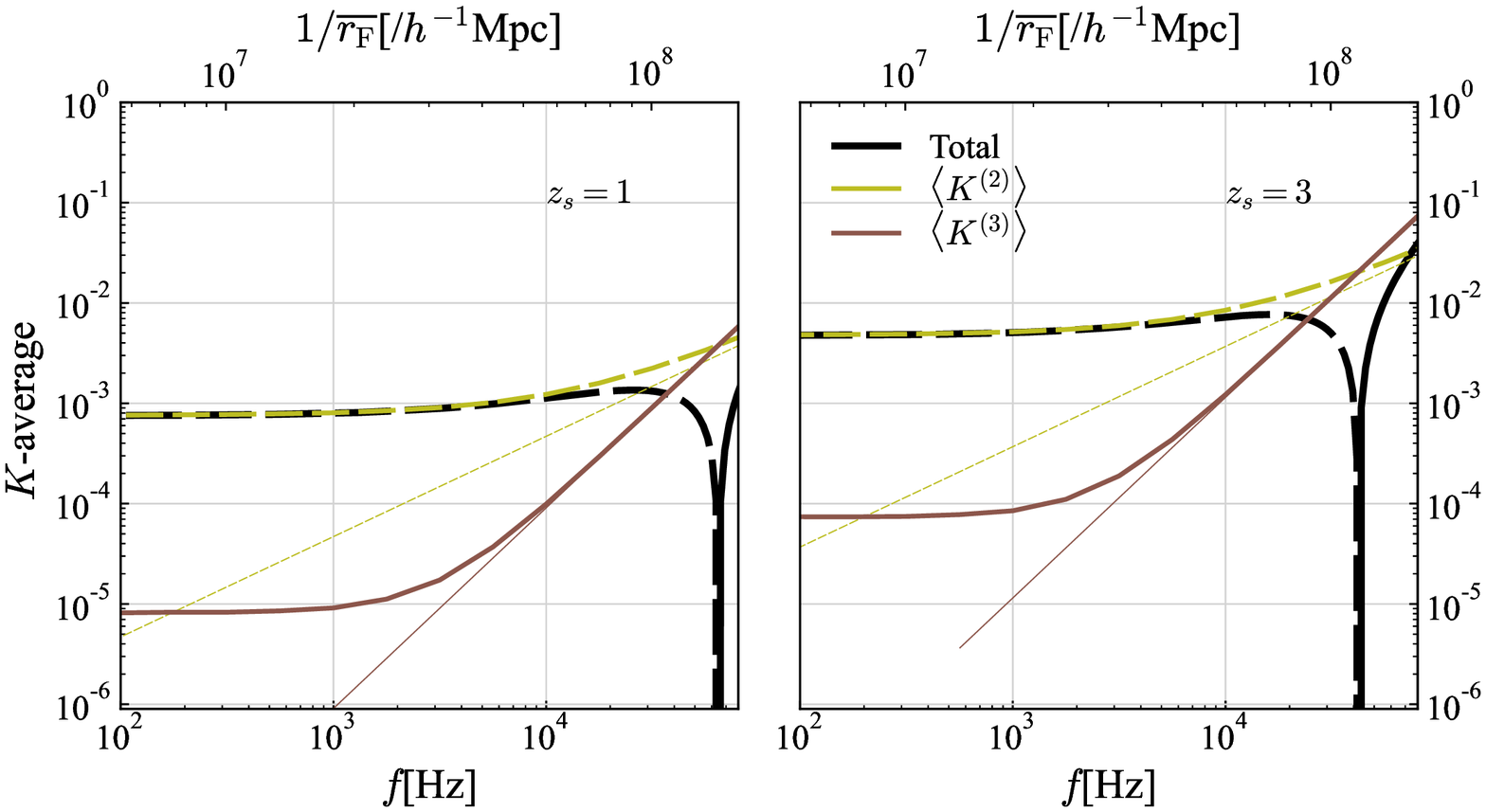} 
  \caption{
  The back lines show the total $\braket{K}$.
   $\braket{K^{(2)}}$ (olive) is dominated by the halo term until $f\sim10^{4}$~Hz. 
  However, $\braket{K^{(3)}}$ (brown) exceeds $\braket{K^{(2)}}$ around $f\sim5\times 10^{4}$~Hz as well, indicating that the perturbative approach fails around this frequency.
  Eq.~(\ref{eq:Kavecond}) provides a condition under which $\braket{K^{(2)}}$ is a reliable approximation of $\braket{K}$, which is equivalent to the weak lensing condition presented in \cite{Oguri:2020ldf} up to a constant prefactor.
  Note that the precise behavior of $\braket{K}$ around frequencies where the halo and the shot noise are of similar magnitude (transition frequency) is not captured in these figures due to the exclusion of cross terms.
  The solid(dashed) line indicates the $+(-)$ value.
  }
  \label{fig: K average shot}
\end{figure}

Fig.\ref{fig: K average shot} shows the shot noise contribution to the average of $K$.
The shot noise effect is hidden until the frequency becomes $f=10^{4}$~Hz.
Thus, the frequency lower than this is not affected by the shot noise effect.
When the frequency is above $f\sim10^{4}$, the shot noise effect becomes dominant.
However, soon after the shot noise effect dominates $\braket{K}$, the higher order contribution $\braket{K^{(3)}}$ overcomes the lower order term $\braket{K^{(2)}}$.
When the higher-order terms dominate the lower-order terms, it is an indication that the perturbative approach fails.
Taking the ratio of Eq.~(\ref{eq: KAve shot}) to Eq.~(\ref{eq: K3 shot}) and assuming the redshift is not too large (truncating the second or higher order terms in $z_{s}$), we find the following condition for the validity of the perturbative approach for the computation of $\braket{K}$:
\begin{align}
    \label{eq:Kavecond}
    \left|\frac{\braket{K^{(3)}}_{\rm shot}}{\braket{K^{(2)}}_{\rm shot}}\right|=1.7Gm_{p}\omega \left(1+\frac{z_{s}}{2}\right)<1
\end{align}
As we will show later, this condition is analogous to the condition under which the Born approximation for the magnification can be reliably applied. 

\subsubsection{Shot noise contribution to variance}
Finally, we evaluate the post-Born correction to the variance.  
First, we consider the corrections to the phase modulation.
For the same reason as $\braket{S}_{\rm shot}$, the integral for
$\delta_{S^{2},\rm dc,shot}$ and $\delta_{S^{2},\rm c,shot}$ require the cutoff scale $k_c$ to avoid divergence.

We can compute the approximation of the Gaussian correction $\delta_{S^{2},\rm dc}$, using the property of Eq.~(\ref{eq: FS12}).
When $k_{1}$ and $k_{2}$ are large enough,
the main contribution to $\delta_{S^{2},\rm dc,shot}$ comes from the factor $\frac{1}{k_{1}k_{2}}$ due to the cancellation by oscillations.
In other words, it is possible to make the following approximation:
\begin{align}
    &\left[
    \frac{1}{k_{1}k_{2}}\mathcal{F}_{S,12}
    -\frac{1}{k_{1}^{2}}\mathcal{F}_{S,1}
    -\frac{1}{k_{2}^{2}}\mathcal{F}_{S,2}
    \right]\notag
    \\
    &\sim
    -\frac{1}{2k_{1}k_{2}}
    \left(1-\cos{\left(\frac{\chi_{1}^{2}W(\chi_{1},\chi_{s})}{2\omega}k_{1}^{2}\right)}\right)
    \left(1-\cos{\left(\frac{\chi_{2}^{2}W(\chi_{2},\chi_{s})}{2\omega}k_{2}^{2}\right)}\right)
\end{align}
where $k_{1}$ and $k_{2}$ are taken to be sufficiently large due to the dominant contribution to $\delta_{S^{2},\rm dc, shot}$ coming from such a region.
Using this approximation as well as other simplifications used in the computation of
$\braket{S}_{\rm shot}$, we obtain
\begin{align}
\label{eq: SVar shot}
    \delta_{S^{2},\rm dc,\rm shot}
    \sim&
    -\frac{1}{2\pi^{2}}\left(\frac{3H_{0}^{2}\Omega_{m}}{2}\right)^{4}
    \int_{0}^{\chi_{s}}\frac{d\chi}{\chi^{2}}
    \int_{0}^{\chi}\frac{d\chi'}{\chi'^{2}}
    \int_{0}^{\chi'}d\chi_{1}
    \int_{0}^{\chi'}d\chi_{2}
    \notag
    \\
    &\times
    \frac{1}{a^{2}(\chi_{1})}
    \frac{1}{a^{2}(\chi_{2})}
    \chi_{1}^{2}\chi_{2}^{2}\omega^{2}
    {\left( \frac{f_{p}^{2}}{\overline{n}} \right)}^2
   \left( \log\left[\frac{k_{c}^{2}}{H_{0}\omega}\right]\right)^{2}.
\end{align}

The correction from the bispectrum term can be calculated using Eq.~(\ref{eq: S1S2shot}), and the expression for the bispectrum Eq.(\ref{B-shotnoise}).
Since we consider the bispectrum term as the main contribution to the non-Gaussian correction $\delta_{S^{2},\rm c, shot}=2\braket{S_{\rm Born}S^{(2)}}$ holds.
Adopting the same approximation we used to calculate $\braket{S}_{\rm shot}$ and $\delta_{S^{2}\rm dc, shot}$, we have the following result:
\begin{align}
    \label{eq: S1S2shot}
        \delta_{S^{2},\rm c, shot}
        \sim&
        \frac{1}{4\pi^{2}}\left(\frac{3H_{0}^{2}\Omega_{m}}{2}\right)^{3}
    \int_{0}^{\chi_{s}}\frac{d\chi}{a^{3}}\chi^{2}W(\chi,\chi_{s})
    \times\omega^{2}\frac{f_{p}^{3}}{\overline{n}^{2}}
    \left( \log\left[\frac{k_{c}^{2}}{H_{0}\omega}\right]\right)^{2}.
\end{align}
Similarly, the logarithmic factor arises due to the presence of the cutoff.

As for the Gaussian correction $\delta_{K^{2},\rm dc}$, it does not diverge even if we take $k_c \to \infty$.
Taking this limit, the formal expression is given by
\begin{align}
    \label{eq: KVar shot}
    \delta_{K^{2},\rm dc,\rm shot}
    =&
    16\left(\frac{3H_{0}^{2}\Omega_{m}}{2}\right)^{4}
    \int_{0}^{\chi_{s}}\frac{d\chi}{\chi^{2}}
    \int_{0}^{\chi}\frac{d\chi'}{\chi'^{2}}
    \int_{0}^{\chi'}d\chi_{1}
    \int_{0}^{\chi'}d\chi_{2}
    \frac{1}{a^{2}(\chi_{1})}
    \frac{1}{a^{2}(\chi_{2})}
    \frac{1}{(2\pi)^{2}}\notag
    \\
    &\times 
    \int_{0}^{\infty}dk_{1}
    \int_{0}^{\infty}dk_{2}
    \left[
    \frac{1}{k_{1}k_{2}}\mathcal{F}_{K,12}
    -\frac{1}{k_{1}^{2}}\mathcal{F}_{K,1}
    -\frac{1}{k_{2}^{2}}\mathcal{F}_{K,2}
    \right]
    \times
     \omega^{2}\frac{f_{p}^{2}}{\overline{n}}.
\end{align}
The definition of $\mathcal{F}_{K,12},\mathcal{F}_{K,1},\mathcal{F}_{K,2}$
is given in Appendix \ref{app:postBornvariance}.
We have not been able to find an analytic method to approximately compute 
the integration over $k_1, k_2$ because it requires careful analytical treatment.
Although it is, in principle, possible to compute the integral numerically, 
it turned out to be quite complicated to achieve it.
This arises from the fact that the expression is highly oscillatory at large $k$.
However, it is expected that the non-Gaussian correction $\delta_{K^{2},\rm dc}$ is sufficiently smaller than the Gaussian correction $\delta_{K^{2}\rm c, shot}$ due to the reason we discuss below

On the other hand, $\delta_{K^{2},\rm c, shot}=2\braket{K_{\rm Born}K^{(2)}}$ is computed using Eq.~(\ref{eq; magnification 12 bispectrum}) and Eq.~(\ref{B-shotnoise}) as,
\begin{align}
    \label{eq: K1K2shot}
    \delta_{K^{2},\rm c, shot}
    =&16\left(\frac{3H_{0}^{2}\Omega_{m}}{2}\right)^{3}
    \int_{0}^{\chi_{s}}\frac{d\chi}{a^{3}}\chi^{2}W(\chi,\chi_{s}) I\times \omega^{2}\frac{f_{p}^{3}}{\overline{n}^{2}},
\end{align}
where $I$ is just a number given by the following integral:
  \begin{align}
        I=&\frac{1}{2}\int_{0}^{\infty}\frac{d\xi_{1}}{2\pi}
    \int_{0}^{\infty}\frac{d\xi_{2}}{2\pi}
    \int_{0}^{2\pi}\frac{d\phi}{2\pi}\notag
   \\ &\times\frac{\sin{(\xi_{1}^{2}+\xi_{2}^{2}+2\xi_{1}\xi_{2}\cos{\phi})}\sin{(\xi_{1}^{2}+\xi_{2}^{2}+\xi_{1}\xi_{2}\cos{\phi})\sin{(\xi_{1}\xi_{2}\cos{\phi})}}}{\xi_{1}\xi_{2}(\xi_{1}^{2}+\xi_{2}^{2}+2\xi_{1}\xi_{2}\cos{\phi})}
    \end{align}
which is found to be $I\sim -0.0038$.

\begin{figure}[tb!]
    \centering
  \includegraphics[width=16cm]{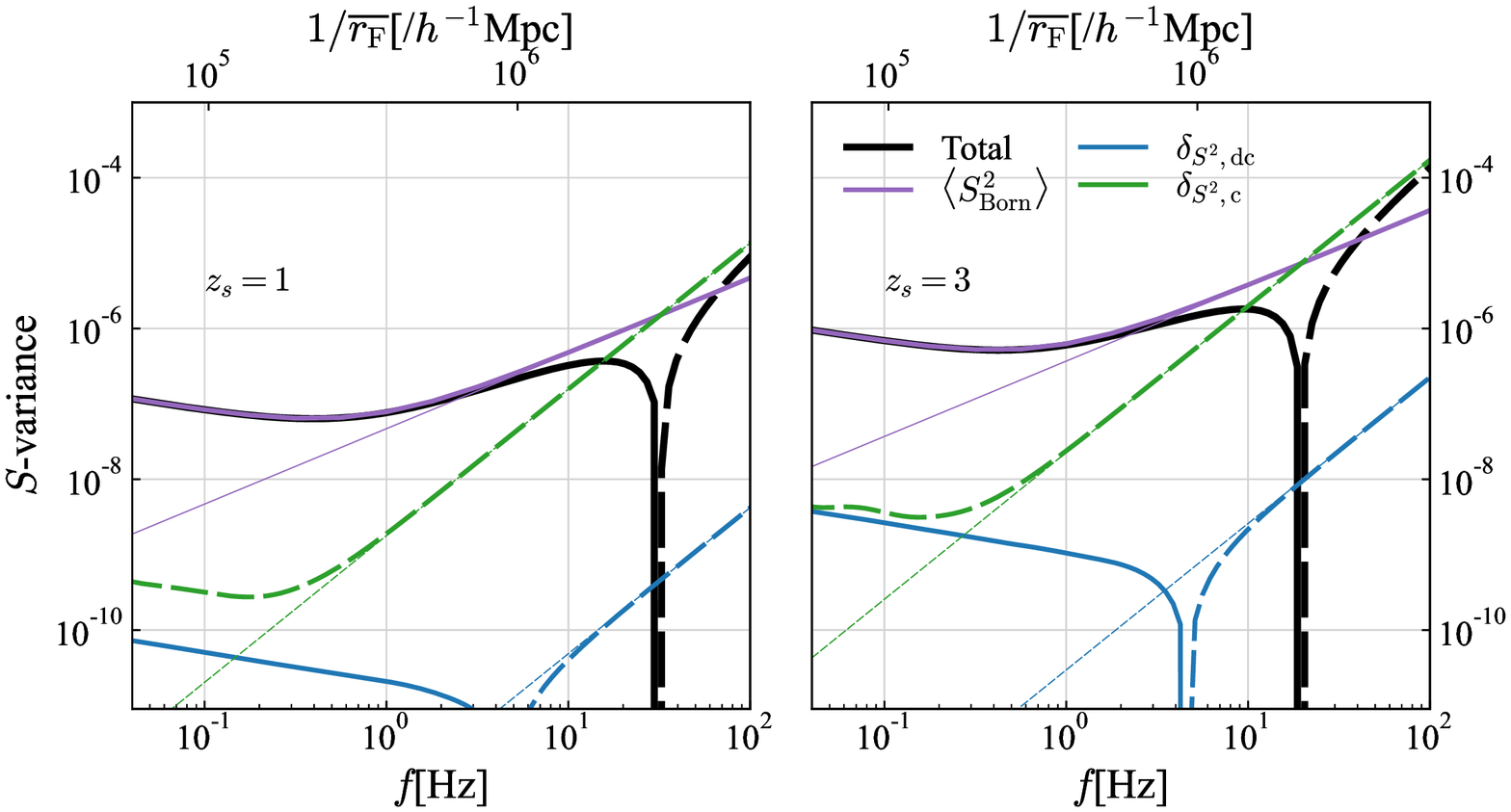} 
  \caption{
  The black lines show the total $\braket{S^{2}}$.
  The shot noise effect dominates $\braket{S_{\rm Born}^{2}}$ (purple) above $f\sim1$~Hz, while for $\delta_{S^{2},c}$ (green) the shot noise takes over the halo term at $f\sim0.3$~Hz.
  The shot noise from the Gaussian correction $\delta_{S^{2},\rm dc}$ (blue) is subdominant compared to  $\delta_{S^{2},\rm c}$ due to $B_{\rm shot}\gg P_{\rm shot}^{2}$. 
  At $f\sim20$~Hz, $\delta_{S^{2}}(=\delta_{S^{2},\rm c}+\delta_{S^{2},\rm dc})$ exceeds $\braket{S_{\rm Born}^{2}}$, indicating the breakdown of the Born approximation.
  Since $\delta_{S^{2}}$ is enhanced by the  $\log{(\cdots)}$ factor which reflects the physical size of the point mass, the breakdown frequency for $\braket{S^{2}}$ is lower than that for $\braket{K^{2}}$ under the same point mass scenario.
  Note that the precise behavior of $\delta_{S^{2},\rm dc}$ and  $\delta_{S^{2},\rm c}$ evaluated at the transition frequency are imprecise due to the exclusion of the cross term.
  The solid(dashed) line indicates the $+(-)$ value.
  }
  \label{fig: S variance shot}
\end{figure}

Fig.~\ref{fig: S variance shot} shows the shot noise effect on the variance of the phase modulation, indicating that the shot noise is subdominant until $f$ becomes greater than $f\sim1$~Hz.
As for the post-Born corrections, it can be seen that the non-Gaussian contribution $\delta_{S^{2},\rm c}$ is the dominant contribution compared to the Gaussian contribution $\delta_{S^{2},\rm dc}$.
This can be understood by considering that $\delta_{S^{2}\rm c, shot}$ is proportional to $B_{\rm shot}$ while $\delta_{S^{2},\rm  dc, shot}$ is proportional to $P_{\rm shot}^{2}$.
Since $P_{\rm shot}^{2}$ is smaller than $B_{\rm shot}$ by a factor of $f_{p}(=0.01)$, as we can see in Eq.~(\ref{P-shotnoise}) and Eq.~(\ref{B-shotnoise}), the effect from the non-Gaussian term $\delta_{S^{2},\rm c, shot}$ is dominant compared to the Gaussian term $\delta_{S^{2},\rm dc, shot}$.
Thus, it can be concluded that the post-Born correction is primarily determined by the non-Gaussian term ($\delta_{S^{2},\rm shot}\approx \delta_{S^{2},\rm c, shot}$)

An important observation is that, in this point mass scenario ($m_{p}=0.5M_{\odot}, f_{p}=0.01, k_{c}=4\times10^{13}h\mathrm{Mpc}^{-1}$), the post-Born term $\delta_{S^{2},\rm shot}$ surpasses the Born approximation $\braket{S_{\rm Born}^{2}}_{\rm shot}$ at around $f\sim20$~Hz.
This indicates the breakdown of the Born approximation around this frequency.

\begin{figure}[tb!]
    \centering
  \includegraphics[width=16cm]{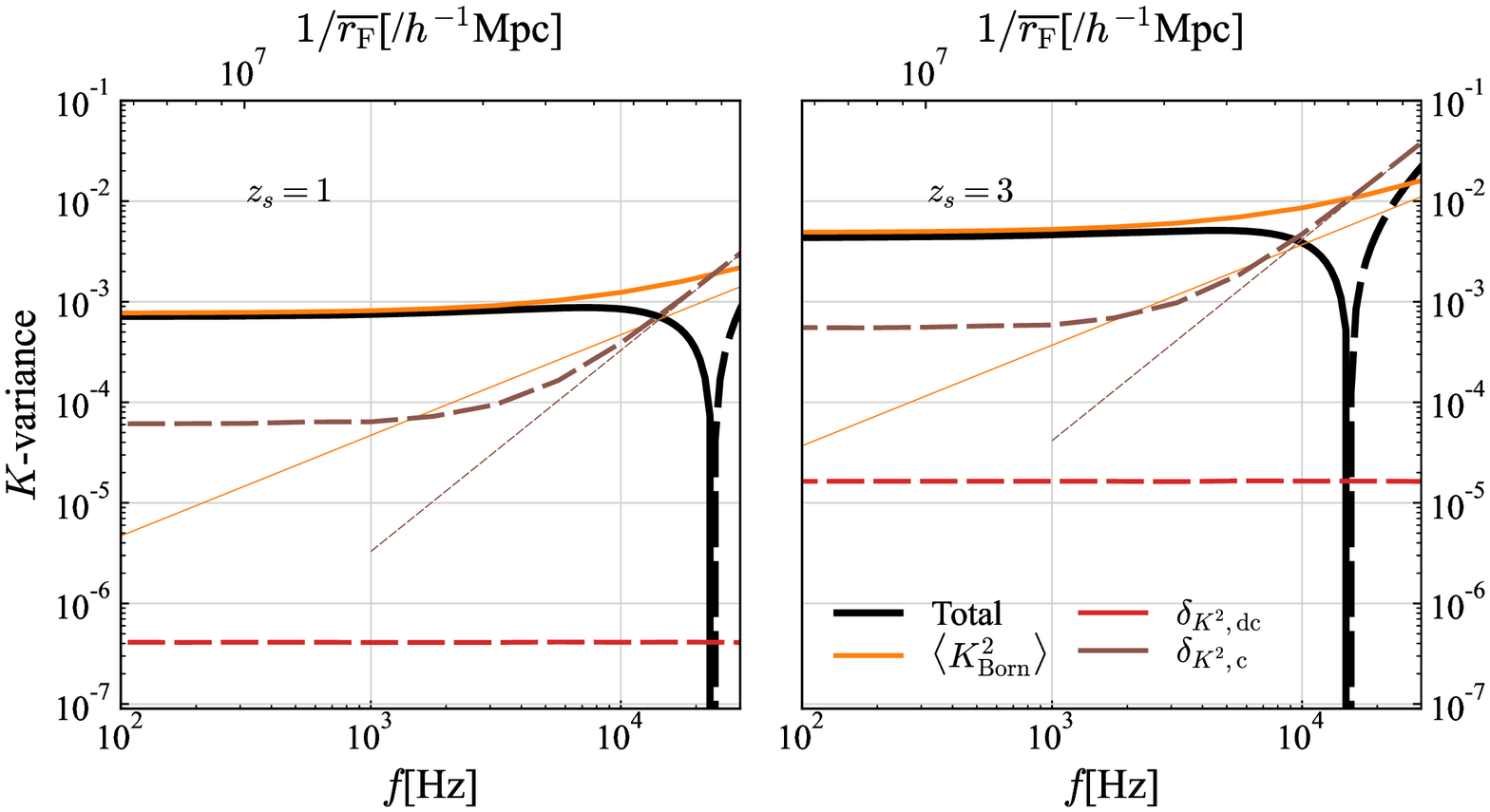} 
  \caption{
  The black lines show the total $\braket{K^{2}}$.
  The shot noise effect on $\braket{K_{\rm Born}^{2}}$ (orange) is subdominant until $f\sim10^{4}$~Hz, while the non-Gaussian correction $\delta_{K^{2},\rm c}$ (brown) takes over $\braket{K_{\rm Born}^{2}}$ at a similar frequency, implying the breakdown of the Born approximation.
  We do not show the shot noise effect on the Gaussiaon correction $\delta_{K^{2},\rm dc}$ due to computational challenges but it is expected to be smaller than the non-Gaussian correction $\delta_{K^{2},\rm c}$ because $B_{\rm shot}\gg P_{\rm shot}^{2}$.
  For the magnification,  the condition under which the Born approximation holds is given by Eq.~(\ref{eq: Weaklenscond}), which ensures that the point mass does not cause strong lensing.
  Thus, the breakdown of the Born approximation can be attributed to the variance of $K$ being dominated by rare strong lensing events to which the weak lens approximation cannot be applied.
  Note that the precise behavior of $\delta_{K^{2},\rm dc}$ and $\delta_{K^{2},\rm c}$ evaluated at the transition frequency are not accurate due to the exclusion of the cross terms.
  The solid(dashed) line indicates the $+(-)$ value.
  }
  \label{fig: K variance shot}
\end{figure}

A similar trend can be observed for the variance of $K$ in Fig.\ref{fig: K variance shot}.
In this case, the post-Born term $\delta_{K^{2},\rm c, shot}$ exceeds the Born result $\braket{K_{\rm Born}^{2}}$ at around $f=10000$~Hz.
It is expected that $\delta_{K^{2},\rm dc, shot}$ is smaller than $\delta_{K^{2},\rm c, shot}$ for similar reasons as $\delta_{S^{2},\rm dc, shot}$ being smaller than $\delta_{S^{2},\rm c, shot}$ (bispectrum is much larger than the square of the power spectrum). 
Therefore, we can consider the correction to the variance of $K$ to be dominated by the non-Gaussian term $\delta_{K^{2},\rm shot} \approx \delta_{K^{2},\rm c, shot}$.

Based on these considerations, it is possible to provide the general condition under which the Born approximation holds for the shot noise.
This condition can be derived by computing the relative magnitude of the post-Born corrections $\delta_{S^{2},\rm shot},\delta_{K^{2},\rm shot}$ to the Born approximation $\braket{S_{\rm Born}^{2}}_{\rm shot}, \braket{K^{2}_{\rm Born}}_{\rm shot}$.
By assuming that the source redshift is not exceedingly large and truncating the second or higher-order terms in $z_{s}$ is justified, we obtain the following conditions:
\begin{align}
    \label{eq: WeaklenscondPhase}
     \left|\frac{\delta_{S^{2},\rm shot}}{\braket{S_{\rm Born}^{2}}_{\rm shot}}\right|=&c_{S}Gm_{p}\omega\left(1+\frac{z_{s}}{2}\right)\left(\log{\left[\frac{k_{c}^{2}}{H_{0}\omega}\right]}\right)^{2}< 1
     \\
    \label{eq: Weaklenscond}
    \left|\frac{\delta_{K^{2},\rm shot}}{\braket{K_{\rm Born}^{2}}_{\rm shot}}\right|=&c_{K}Gm_{p}\omega\left(1+\frac{z_{s}}{2}\right)< 1
\end{align}
where a factor of order unity $c_{S}$ and $c_{K}$ are found to be approximately $c_{S}=4/\pi$ and $c_{K}=3.1$ in this study.
Note that the presence of the Hubble parameter $H_{0}$ in Eq.~(\ref{eq: WeaklenscondPhase}) arises from the assumption that the source and the lens redshifts are of cosmological order $\chi(z_{s})\sim1/H_{0}$. 
The general trend observed in Eq.~(\ref{eq: WeaklenscondPhase}) and Eq.~(\ref{eq: Weaklenscond}) is that the Born approximation provides a reliable estimation when $Gm_{p}\omega$ is small.
Physically,  there are two ways to interpret the factor  $Gm_{p}\omega$. 
One is to consider this as a ratio of the Schwarzschild radius of the point mass to the wavelength of GWs. while the other views it as a square of the ratio of the Einstein radius of the point mass to the Fresnes scale of GWs.
In the second interpretation, the distances to the source and lens from the observer are assumed to be the same order of magnitude.
Also, Eq.~(\ref{eq: Weaklenscond}) is the same as the one derived in \cite{Oguri:2020ldf} up to constant, which is based on the requirement that strong lensing by the point mass does not occur.

\section{Discussion}
In this section, we summarize our main findings and discuss the possibility of the applications and the detectability of the post-Born effect.

\subsection{Validity of the Born approximation}
The weak lensing of gravitational waves offers the advantage of probing the scale corresponding to the Fresnel scale. 
In the case of typical GWs seen by the ground-based detectors ($f=10\sim1000$Hz), the Fresnel scale can reach values as small as a few parsecs.
At such a small scale, a high degree of non-Gaussianity is expected.
However, strong non-Gaussianity does not automatically indicates that the post-Born corrections are large.
To gain a better understanding of this, let us begin by examining the general case of statics before delving into our specific cases.

Suppose $X[\delta]$ is a physical quantity such as $K$ and $S$ evaluated by a random variable $\delta$ (in this case, it is the matter density fluctuation $\delta$).
Here, the Born approximation $X_{\rm Born}[\delta]$ is usually interpreted as an approximation of $X[\delta]$ by the leading order terms of its Taylor series, 
thus $X[\delta]=X_{\rm Born}[\delta]+\delta X[\delta]$, where $\delta X[\delta]$ is the post-Born correction.

For non-Gaussianity, it can be characterized by comparing the skewness of $X$ with its variance while ensuring that they have the same dimension. 
In other words, the dimensionless parameter $\braket{X^{3}}^{2/3}/\braket{X^{2}}$ can be used to quantify the degree of non-Gaussianity in $X$. 
If this quantity is sufficiently small compared to 1, it suggests that $X$ is almost Gaussian.
When this is comparable to 1, it indicates a strong deviation from Gaussianity.

On the other hand, the post-Born corrections to $\braket{X_{\rm Born}^{2}}$ are given by the higher-order terms, which in this case are $2\braket{X_{\rm Born}\delta X}+\braket{\delta X^{2}}$.
Since non-Gaussianity and the post-Born corrections are intrinsically different, even if $X_{\rm Born}$ exhibits strong non-Gaussianity ($\braket{X_{\rm Born}^{3}}^{2/3}\sim\braket{X_{\rm Born}^{2}}$), 
the post-Born corrections can be still small ($\braket{X_{\rm Born}^{2}}\gg 2\braket{X_{\rm Born}\delta X}+\braket{\delta X^{2}}$).

Now,  let us turn to more specific cases, particularly those involving the presence of dark matter halos and the shot noise.

\subsubsection{Without the shot noise}
When considering only the dark matter halos, we found that the corrections to the Born approximation are significantly small, especially for the phase modulation.
For frequencies higher than $f=0.01$Hz, the ratio of the correction terms to the leading order term is $\delta_{S^{2}}/\braket{S_{\rm Born}^{2}}\lesssim\mathcal{O}(10^{-2})$.
Similarly, the corrections to the magnification are small, although not excessively so: $\delta_{K^{2}}/\braket{K_{\rm Born}^{2}}\lesssim\mathcal{O}(10^{-1})$.
Note that the frequencies below $f=0.01$Hz correspond to scales larger than $k\sim10^{5}h\mathrm{Mpc}^{-1}$ and may be strongly influenced by baryonic matter, specifically in the form of galaxies \cite{Oguri:2020ldf}. 

Our findings indicate that in the absence of the shot noise, the Born approximation remains still valid across the frequency ranges where the primary contributions to $\braket{S^{2}}$ and $\braket{K^{2}}$ are attributed to dark matter halos.
Since the post-Born corrections account for the effect of the halos being unevenly distributed, the suppression of the post-Born corrections implies that halos can be treated as though they are uniformly distributed when computing the lensing signal.

\begin{figure}[tb!]
    \centering
  \includegraphics[width=16cm]{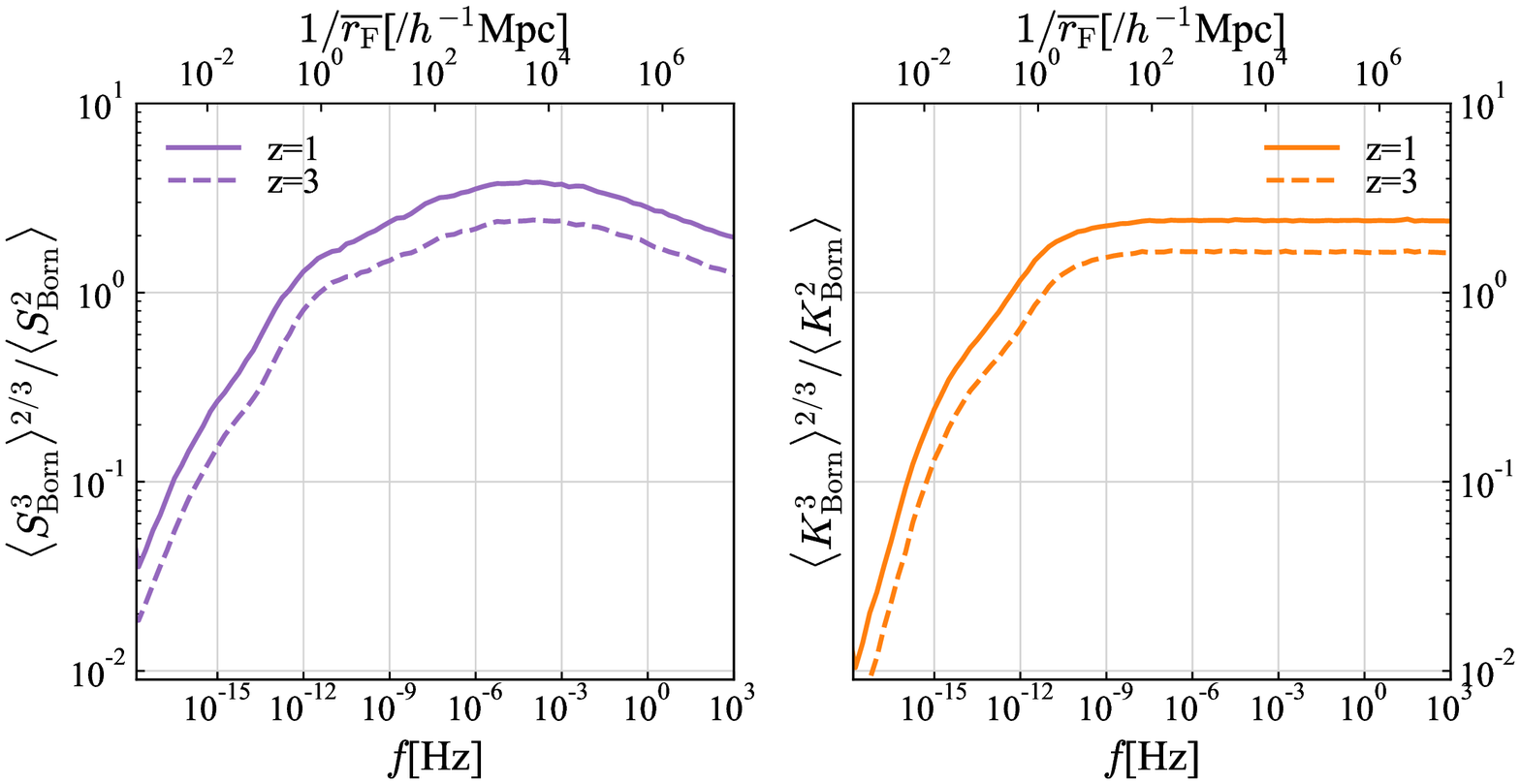} 
  \caption{
  For the phase modulation (left), the degree of non-Gaussainity exhibits a rapid increase, followed by a deceleration at $f\sim10^{-12}$~Hz, and reaching a peak at $f\sim10^{-4}$~Hz.
  After that, it gradually decreases.
  A similar trend can be observed for the non-Gaussianity of $K$ (right). 
  However, $\braket{K_{\rm Born}^{3}}^{3/2}/\braket{K_{\rm Born}^{2}}$ reaches a constant value once the frequency exceeds $f\sim10^{-11}$~Hz. 
  In both cases, the degree of non-Gaussianity is large at $f>10^{-10}$~Hz ($k\sim10^{1}h\mathrm{Mpc}^{-1}$), which is intuitively true.
  Also, it is observed that the non-Gaussianity in $S$ and $K$ decreases as the source redshift becomes larger.
  This is because $S$ and $K$ become more Gaussian when they are able to pass many dark matter halos on average due to the central limit theorem.
  This behavior is also consistent with the discussion about the non-Gaussianity of $S$ and $K$ found in \cite{Oguri:2020ldf}.
  }
  \label{fig: Born skewness}
\end{figure}

This result does not contradict the expectation that the matter distribution is highly non-Gaussian at small scales.
In fact, our analysis revealed that $S$ and $K$ show significant non-Gaussianity behavior.
Fig.~\ref{fig: Born skewness} shows the degree of non-Gaussianity in $S$ and $K$.
As these figures indicate, $\braket{S_{\rm Born}^{3}}^{3/2}/\braket{S_{\rm Born}^{2}}$  and $\braket{K_{\rm Born}^{3}}^{3/2}/\braket{K_{\rm Born}^{2}}$ exceed unity at high frequencies, exhibiting a strong deviation from Gaussian behavior.
Note that, in Fig.~\ref{fig: Born skewness}, it can be seen that the degree of non-Gaussianity decreases as the source of GWs moves further away (at higher redshift)
This behavior is consistent with \cite{Oguri:2020ldf}.
Essentially,  when the source of GWs is distant, they are more likely to traverse multiple halos. 
When they pass through many halos, the overall lensing effect is described by the sum of each individual lensing event that occurs during their propagation.
Due to the central limit theorem, this summation process leads to the reduction of non-Gaussianity.

\subsubsection{With the shot noise}
In the presence of the shot noise, we found that the applicability of the Born approximation depends on two factors: the satisfaction of the weak lensing condition (Eq.~(\ref{eq: WeaklenscondPhase}) for $S$ and Eq.~(\ref{eq: Weaklenscond}) for $K$) and the dominance of the shot noise contribution to the variance.
For simplicity, let us focus on the validity of the Born approximation for the magnification. 
The same argument can be applied to the phase modulation as well.

For the magnification, the Born approximation remains valid if the shot noise effect is subdominant or Eq.(\ref{eq: Weaklenscond}) is satisfied. 
However, if  Eq.(\ref{eq: Weaklenscond}) is not satisfied and the shot noise is the dominant contribution to the variance, the Born approximation breaks down.
In this context, the breakdown of the Born approximation specifically refers to the situation where the variance of $S$ and $K$ computed using the leading order terms of $S$ and $K$ in $\Phi$ no longer provide a reliable estimate of the true variance.
This also implies that the perturbative approach fails since adding up a finite number of higher-order terms does not necessarily improve the accuracy of variance estimation. 
At this stage, a full-order analysis or simulation is required to obtain the true distribution of $S$ and $K$.
The simulation approach has been taken in geometric optics \cite{Takahashi:2011fgq}, but further 
investigation is needed within the framework of wave optics.

A similar condition to Eq.~(\ref{eq: Weaklenscond}) is also derived in \cite{Oguri:2020ldf} (the only difference is a unity order prefactor) under the requirement for the absence of strong lensing.
Thus, the violation of Eq.~(\ref{eq: Weaklenscond}) implies the existence of a specific configuration of lenses and a GW source that can lead to strong lensing such as the point mass being very close to the line of sight.
It is important to note that the violation of Eq.~(\ref{eq: Weaklenscond})  does not by itself imply the breakdown of the use of the Born approximation for the variance.
As mentioned above, the breakdown requires not only the violation of Eq.~(\ref{eq: Weaklenscond}) but also the dominance of the shot noise effects on the variance over other effects.
Therefore, even if there are objects that can potentially cause strong lensing, 
the Born approximation is still valid as long as the contribution of these lenses to the variance is subdominant compared to the contribution of other lenses that do not cause strong lensing (such as  dark low-mass halos).

Due to this property, the shot noise signal can help constrain the nature of point masses.
For example, if Eq.(\ref{eq: Weaklenscond}) is satisfied and the shot noise effect dominates the variance (this is true for the phase modulation when the parameters for the shot noise is $f_{p}=0.01, m=0.5M_{\odot}$  and the frequency of GWs is around $f=1$Hz), it is possible to estimate the parameters of the point masses such as $m$ and $f_{p}$ \cite{Oguri:2020ldf}.
In addition to this, our analysis provides a method to include the correction to the Born approximation.
By using the post-Born terms calculated in this study, the shot noise contribution to the variance is modified as 
\begin{align}
    \label{eq:SBornshotmodi}
    \braket{S^{2}}_{\rm shot}=&\braket{S_{\rm Born}^{2}}_{\rm shot}\left\{1-\frac{4}{\pi}Gm_{p}\omega\left(1+\frac{z_{s}}{2}\right)\left(\log{\left[\frac{k_{c}^{2}}{H_{0}\omega}\right]}\right)^{2}\right\},
    \\
    \label{eq:KBornshotmodi}
    \braket{K^{2}}_{\rm shot}=&\braket{K_{\rm Born}^{2}}_{\rm shot}\left\{1-3.1Gm_{p}\omega\left(1+\frac{z_{s}}{2}\right)\right\}.
\end{align}
In the moderately high-frequency region where the perturbative approach is still useful but the accuracy of the Born approximation is uncertain, this modification will enable us to more accurately estimate the variance of $S$ and $K$ produced by the point masses with specific parameters $m_{p}$, $k_{c}$, and $f_{p}$.

On the other hand, the scarcity of strongly lensed signals can place constraints on the abundance of lens objects capable of causing strong lensing.
For example, if we consider a scenario where $m=50M_{\odot}$ instead of $m=0.5M_{\odot}$ while maintaining $f_{p}=0.01$, Eq.~(\ref{eq: Weaklenscond}) indicates that the frequency at which the breakdown of the Born approximation for the magnification shifts from $f\sim10000$Hz to $f\sim100$Hz. 
This scenario ($m=50M_{\odot}, f_{p}=0.01$) corresponds to the universe in which the fifty solar mass black holes as part of dark matter are as prevalent as the stellar components.

As this frequency range falls within the sensitivity of current ground-based detectors, there is a possibility of detecting the strong lensing signal caused by such black holes. 
If the number of strong lensing events is small enough so that their impact on the variance is subdominant, the abundance of such black holes can be constrained by this information.

Such a scenario ($m=50M_{\odot}, f_{p}=0.01$) has been attracting great interest recently after the observations of such massive black holes by the GW experiments.
It is under active investigation whether the abundance of primordial black holes 
comparable to $f_p \simeq 0.01$ is consistent with the existing observations \cite{Villanueva-Domingo:2021spv}. 
GL of GWs studied in this paper provides an alternative path to test this possibility 
(see also \cite{Oguri:2020ldf}).

\subsection{Average as an additional probe}
We found that the ensemble average of $K$ and $S$ is no longer zero at the level of the post-Born approximation. 
This provides the possibility to detect the average of $S$ and $K$ and use them as an additional probe for matter abundance.
However, it is crucial to assess the validity of the approximation of the average by only considering the power spectrum term.
Therefore, we will now examine the reliability of this approximation.

In the absence of the shot noise, the main contribution to the average of $S$ and $K$ comes from the power spectrum and the contribution from the higher-order terms containing the bispectrum is subdominant.
This is consistent with the behavior observed in the variance, where the correction terms to the Born approximation are found to be subdominant.
In this case, the average of $S$ and $K$ is roughly of the same order as their variance ($\braket{S}\sim \braket{S^{2}},\braket{K}\sim -\braket{K^{2}}$).

In the presence of the shot noise, we found that there are cases where the computation of the average of $K$ by accounting only for the matter power spectrum breaks down, which is the same condition as the breakdown of the Born approximation for $\braket{K^{2}}$ up to a constant prefactor.
Therefore, if the Born approximation for the variance of $K$ is valid, then the computation of the average of $K$ by accounting only for the power spectrum contribution remains valid.

Regarding the phase modulation, the contribution from the bispectrum term due to the shot noise becomes exactly zero $\braket{S^{(3)}}=0$.
This suggests that including higher-order terms such as trispectrum would be necessary to capture the corrections to $\braket{S}$ in the presence of the shot noise.
However, as long as the Born approximation for $\braket{S^{2}}$ holds, it is expected that the approximation of $\braket{S}$ using the power spectrum contribution alone is valid.
This presumption is reasonable because if the Born approximation for the variance holds, it implies that the lensing signal is weak and the first term in the perturbative approach offers a reliable approximation.

Based on these considerations, the average calculation is valid as long as the Born approximation for the variance also holds.
Now,  let us shift our focus to the average of $S$, as it can play a significant role in probing the properties of the point masses.
Specifically, by combining   $\braket{S}$ and $\braket{S^{2}}$, we can probe the size of the shot noise constituent, as well as its mass and abundance, since the shot noise has different effects on  $\braket{S}$ and $\braket{S^{2}}$.
This analysis cannot be performed by considering the variance alone because the size dependency does not appear in the variance.
The obtained properties of the shot noise can be compared with the properties of stars inferred by other astronomical observations.
This provides the test of whether the sources causing the shot noise
in the gravitational lensing of GWs are stars or other types of compact objects that have not been detected by non-GW observations.

It is also important to mention that, according to our formulation,  $\braket{S}$ is always positive.
If the negative value of $\braket{S}$ is detected, it means an indication of the presence of something outside the lensing effect. It could mean the presence of new matter that interacts with gravity in an unusual way or the violation of GR, which could lead to new physics.

\subsection{Detectability of the post-Born effect}
In \cite{Lidbolm:2008abc}, it is suggested that the amplitude and phase fluctuation of GWs can be measured with an accuracy of $1/\rm SNR$, where SNR is the signal-to-noise ratio. 
According to \cite{Oguri:2020ldf}, the accuracy of measurement is improved by combining many gravitational events.
In \cite{Oguri:2020ldf}, it is argued that the required accuracy is written as $\sim (2/N_{\rm event})^{1/4}(1/\rm SNR)$, which yields $N_{\rm event}\sim3\times10^{5}$
as the number of GW events (with $\mathrm{SNR}=50$) required for detecting the lensing signal.
We would like to perform a similar estimation of $N_{\rm event}$ required to detect the post-Born effect.

First, let us consider the number of GW events required for detecting the average of $S$ and $K$.
Since our purpose is to estimate $N_{\rm event}$ by the back-of-the-envelope calculations,
we will consider the following toy model which simplifies the situation without 
losing the essential point.
Suppose we have succeeded in inferring the source parameters and hence the unlensed waveform from the GW measurement.
Then, the residual signal, which we denote by $s$ and remains after subtracting the unlensed waveform from the measured waveform consists of the uncertainties $n$ of the unlensed waveform and the lensing signal $X$, namely 
\begin{align}
    \label{eq:s=n+X}
    s_{X,i}=n_{i}+X_{i},
\end{align}
where $i$ labels the GW events, while $X$ takes either $S$ or $K$.
For simplicity, we assume that both $n_{i}$ and $X_{i}$ are Gaussian random variables
and each GW event is independent of the others.
In this case, the ensemble average of the quantities computed from $n_{i}$ and $X_{i}$ is given by
\begin{align}
    \label{eq:n Var}
    \braket{n_{i}n_{j}}=&\left(\frac{1}{\rm SNR}\right)^{2}\delta_{ij},
    \\
    \label{eq:X Ave}
    \braket{X_{i}}=&\mu_{X},
    \\
    \label{eq:X Var}
    \braket{X_{i}X_{j}}=&\sigma_{X}^{2}\delta_{ij}+\mu_{X}^{2},
    \\
    \label{eq:nX}
    \braket{n_{i}X_{j}}=&0.
\end{align}
Here, $\mu_{X}$ and $\sigma_{X}$ are the values of both the average and the standard deviation of the phase modulation and the magnification.
All the other quantities can be computed from the combination of these relations.
The first relation $\braket{n_{i}n_{j}}=\delta_{ij}/\rm SNR^{2}$ is about the accuracy of detecting the phase and magnification fluctuation mainly discussed in \cite{Lidbolm:2008abc}.

In reality, we are only able to detect a finite number of GW events. 
Thus, it is convenient to introduce the estimator of the average $\mu_X$ defined as 
\begin{align}
    \label{eq:Estimator Ave}
    \mathcal{E}_{\mu}=&\frac{1}{N_{\rm event}}\sum_{i=1}^{N_{\rm event}}s_{X,i}.
\end{align}
This quantity is an approximated version of the ensemble average, thus 
taking $N_{\rm event} \to\infty$ reproduces $\mu_{X}$.
Indeed, computing the average and the variance of $\mathcal{E}_{\mu}$, 
we obtain
\begin{align}
    \label{eq:EAve Ave}
    \braket{\mathcal{E}}_{\mu}
    =&\mu_{X},
    \\
    \label{eq:EAve Var}
    \braket{\mathcal{E}_{\mu}^{2}}-\braket{\mathcal{E}_{\mu}}^{2}
    \sim&
    \left(\frac{1}{\rm SNR}\right)^{2}\frac{1}{N_{\rm event}}.
\end{align}
It is important to mention that we have used the assumption $1/\mathrm{SNR}\gg \mu_{X}, \sigma_{X}$ 
to derive the second equation.
This result shows that $\mathcal{E}_{\mu}$ fluctuates around $\mu_{X}$ with a fluctuation width of about $\left(\frac{1}{\rm SNR}\right)\frac{1}{\sqrt{N_{\rm event}}}$.
In order to confidently conclude that the average is nonzero,  $\mu_{X}>\left(\frac{1}{\rm SNR}\right)\frac{1}{\sqrt{N_{\rm event}}}$ needs to be satisfied.
Using this restriction, we can estimate that $N_{\mathrm{event},\mu_{X}}\sim \left(\frac{1}{\rm SNR}\right)^{2}\frac{1}{\mu_{X}^{2}}$ is at least necessary to detect the average of $K$ and $S$.

Next, we consider the number of events for detecting the variance. 
In this case, we need at least two independent measurements of the residual for the same GW event if it is difficult to distinguish the lensing signal from the uncertainty associated with the unlensed waveform by using one measurement alone. In the following, we assume measurements by two detectors.
For this purpose, we denote the signals from two different measurements (1 and 2) to be $s_{X,1,i}=n_{1,i}+X_{i}$, $s_{X,2,i}=n_{2,i}+X_{i}$
and assume that one measurement noise is independent of the other's $\braket{n_{1,i}n_{2,j}}=0$.
The detectability is calculated in the same way above by introducing the estimator of the variance
\begin{align}
    \label{eq:Estimator Var}
    \mathcal{E}_{\sigma_{X}^{2}}
    =&
    \frac{1}{N_{\rm event}}
    \sum_{i=1}^{N_{\rm event}}s_{X,1,i}s_{X,2,i}
    -\frac{1}{N_{\rm 
    event}^{2}}
    \sum_{i=1}^{N_{\rm event}}s_{X,1,i}
    \sum_{j=1}^{N_{\rm event}}s_{X,2,j}.
\end{align}
From this expression, we obtain the expressions of the ensemble average of $\mathcal{E}_{\sigma_{X}^{2}}$:
\begin{align}
    \label{eq:EVar Ave}
    \braket{\mathcal{E}_{\sigma_{X}^{2}}}
    =&
    \sigma^{2}_{X},
    \\
    \label{eq:EVar Var}
    \braket{\mathcal{E}_{\sigma_{X}^{2}}^{2}}-\braket{\mathcal{E}_{\sigma_{X}^{2}}}^{2}
    \sim&
    \left(\frac{1}{\rm SNR}\right)^{4}\frac{1}{N_{\rm event}}.
\end{align}
The interpretation of this result is exactly the same as $\mathcal{E}_{\mu_{X}}$ that $\mathcal{E}_{\sigma_{X}^{2}}$ fluctuates around $\sigma_{X}^{2}$ with a width of about $\left(\frac{1}{\rm SNR}\right)^{2}\frac{1}{\sqrt{N_{\rm event}}}$.
Therefore, the number of gravitational wave events required to detect the variance is given by $N_{\mathrm{event},\sigma_{X}^{2}}\sim\left(\frac{1}{SNR}\right)^{4}\frac{1}{\sigma_{X}^{4}}$.

Now, let us examine the detectability of the phase modulation and the magnification.
Table.\ref{tab: postBornError} presents the order of magnitude for the average and variance of $S$ and $K$, along with the post-Born corrections to the variance.
We consider the scenario where the source redshift is $z_{s}=3$, and the signal-to-noise ratio is $\mathrm{SNR}=50$.
The shot noise effect we include corresponds to lensing by point masses with $m=0.5M_{\odot}, f_{p}=0.01, k_{c}=4\times10^{13}h\mathrm{Mpc}^{-1}$.

For the phase modulation, we focus on the frequency range of $f=0.01\sim10$~Hz, which falls within the range where the Born approximation is valid.
Within this frequency range, the lensing signal is dominated by both the dark low-mass halos and the point masses, with specific dominance depending on the frequency.
At the lower end of this range ($f=0.01$~Hz), the signal is primarily attributed to the dark matter halos.
However,  as the frequency of GWs increases, the shot noise effect becomes more significant, taking over the halo contribution at around $f=1$~Hz for the Born variance and $f=0.3$Hz for the average.

In this frequency range, the variance remains relatively stable, while the average increases moderately.
The typical order of the average is around $\mathcal{O}(10^{-5})$, but at the higher end of this range $f=10$~Hz, it can be enhanced by up to $\mathcal{O}(10^{-4})$.
On the other hand, the order of the Born variance remains $\mathcal{O}(10^{-6})$, even at the higher end of the range.
This difference arises from the dependency of the average on the size of the point mass.
Using the formalism we developed above, the number of GW events required to detect $\braket{S}$ is estimated to be $\mathcal{O}(10^{6})$ in the middle of this frequency range.
However, at the higher-frequency end, the required number reduces to $\mathcal{O}(10^{4})$.
On the other hand, the number of events required to detect $\braket{S^{2}}$ is $\mathcal{O}(10^{5})$ in the middle-frequency range and $\mathcal{O}(10^{4})$ at the high end.
 As a result, the detection cost for the average is comparable to the detection cost for the variance at high frequencies in which the shot noise dominates. 
 In a slightly different scenario, the detection of the average might be easier than the detection of the variance.
 For instance, if the point masses we considered here are not ordinary stars but black holes with the same mass and mass fraction ( thus, $k_{c}$ becomes much bigger), the required number for detecting the average decreases while the number for the variance remains the same.
 
Note that, if the signal-to-noise ratio is much larger than $\mathrm{SNR}=50$, the number of GW events required to detect the variance becomes significantly smaller compared to the number required for the detection of the average. 
 This is because the number of required events for the variance scales as $1/\rm SNR^{4}$, while the number for the average scales as $1/\rm SNR^{2}$. 
 Hence, the situation where the average might be easier to detect is when $\rm SNR$ is not excessively high.

 In the case of the post-Born corrections to the variance, their relative magnitude compared to the Born variance prior to the onset of the shot noise is $\mathcal{O}(10^{-3})$.
 However, once the shot noise effect becomes dominant, their relative magnitude is described by $f/20$~Hz. 
 In this case, the ratio of the number of GW events required to resolve this correction  to the number required to detect the Born variance scales as $(20\mathrm{Hz}/f)^{2}$.
 This means that even if the corrections to the Born variance are 10\%, resolving it would require 100 times more GW events than those needed to detect the Born variance.
If $\rm SNR$ is 100, which is expected to be achieved in the future\cite{Abbott_2017}, the number of events to resolve the Born variance reduces to $\mathcal{O}(10^{-3})$.
Assuming a total of $10^{5}$ GW events are observed, it would be possible to resolve the post-Born corrections that exceed 1\% of the Born variance. 
This corresponds to frequencies around $f\sim2$~Hz, which is already close to the breakdown frequency ($f\sim20$~Hz).

These considerations indicate that the post-Born corrections are challenging to detect except in the vicinity of the breakdown frequency.
In the frequency range where the perturbative approach holds but the accuracy of the Born approximation becomes less reliable, including the post-Born correction (Eq.~(\ref{eq:SBornshotmodi}) and Eq.~(\ref{eq:KBornshotmodi})) can yield a more accurate estimate of the variance caused by the point mass lens.

Next,  let us consider the magnification. 
In table.\ref{tab: postBornError}, we consider the frequency range of $f=0.01\sim 1000$~Hz. 
This frequency range is chosen based on the validity of the Born approximation, which holds until Eq.(\ref{eq: Weaklenscond}) breaks down, which occurs at around $f=10000$~Hz. 
As shown above, the magnification has a broader frequency range within which the Born approximation is valid compared to the phase modulation.

The order of magnitude of the magnification is much larger than that of the phase, making $\rm SNR =50$ sufficient to resolve $\braket{K^{2}}$, while around 100 GW events are required to resolve $\braket{K}$.
Even the post-Born correction to the variance can be resolved with just $\mathcal{O}(10)$ GW events.

However, there is an important consideration to make. 
The magnification approaches a constant value as the frequency increases, representing the geometric optics limit. 
Since the geometric optics limit lacks frequency dependence, it cannot be used to probe the matter abundance at the Fresnel scale.
In order to extract the pure wave effect, which can be used to probe the scale corresponding to the Fresnel scale, the constant term in the magnification needs to be subtracted.
However, as calculated in \cite{Oguri:2020ldf}, this pure frequency-dependent part is of the same order as the phase. 
This can be also understood by considering the consistency relation for the variance of the phase modulation and the magnification, namely $\braket{K^{2}_{\rm Born}(2f)}-\braket{K^{2}_{\rm Born}(f)}=\braket{S^{2}_{\rm Born}(f)}$\cite{Inamori:2021tlx}.
Therefore, the magnification needs to be determined at the same level of accuracy as the phase to extract the wave-dependent part that is superimposed on the constant part.
Consequently, a similar number of GW events is required to make the magnification as useful as the phase in extracting information about the matter abundance at the Fresnel scale.

\begin{table}[tb!]
    \centering
    \begin{tabular}{|c|c|c|}
         \hline
         & $10^{-2}\leq f\leq 10^{1}$~Hz & $N_{\rm events}(\mathrm{SNR}=50)$ \\\hline
         $\braket{S}$   & $\mathcal{O}(10^{-5}\sim 10^{-4})$ & $\mathcal{O}(10^{6}\sim10^{4})$
         \\\hline
         $\braket{S^{2}_{\rm Born}}$ & $\mathcal{O}(10^{-6})$ & $\mathcal{O}(10^{5}\sim10^{4})$
         \\\hline 
         $\delta_{S^{2}}/\braket{S^{2}_{\rm Born}}$ &  $\sim \frac{f} {20\mathrm{Hz}}$ & $\gtrsim \mathcal{O}(10^{5})\times \left(\frac{20\mathrm{Hz}}{f}\right)^{2}$
         \\\hline\hline
         &$10^{-2}\leq f\leq 10^{3}$~Hz & $N_{\rm events}(\mathrm{SNR}=50)$\\\hline
         $\braket{K}$ & $-5\times10^{-3}$ & $\mathcal{O}(10^{2})$
         \\\hline
         $\braket{K^{2}_{\rm Born}}$ & $5\times10^{-3}$ & $\mathcal{O}(1)$
         \\\hline
         $\delta_{K^{2}}/\braket{K^{2}_{\rm Born}}$ & $5\times10^{-2}$ &$\mathcal{O}(10)$
         \\\hline
    \end{tabular}
    \caption{
    In the scenario where the shot noise consists of point masses with $m_{p}=M_{\odot}, f_{p}=0.01, k_{c}=4\times10^{13}h\mathrm{Mpc}^{-1}$, with $z_{s}=3$ and $\mathrm{SNR}=50$.
    In this case, the number of GW events required to detect $\braket{S^{2}}$ and $\braket{S}$ can be comparable $\mathcal{O}(10^{4})$ at $f\sim10$~Hz due to the enhancement of $\braket{S}$ by its dependence on the physical size of the point mass.
    On the other hand, the magnification can be much more easily observed. 
    However, a similar number of GW evens is expected to be required to extract the wave-dependent part from $K$.
    }
    \label{tab: postBornError}
\end{table}

\section{Conclusion}
In this paper, we have investigated the weak lensing of GWs beyond the Born approximation by including the higher-order terms in the gravitational potential $\Phi$.
To do this, we adopted a new formulation for the equation governing the GL of GWs.
Instead of using the amplification factor $F$ defined as the ratio of the lensed to unlensed waveform, we introduced a new variable $J$ defined as $F=e^{i\omega J}$.
This process allows us to partially include the non-linear effect of $\Phi$ and reduces the complexity of calculating the higher-order terms.
We then derived the expression of the phase modulation $S$ and the magnification $K$ up to third order in $\Phi$ and calculated the post-Born corrections to the average and variance.
In computing the post-Born corrections, we considered both Gaussian (product of the bispectrum) and non-Gaussian (bispectrum) terms up to the lowest non-trivial order in $\Phi$.
To evaluate the validity of the Born approximation, 
we numerically computed $\braket{S},\braket{K},\delta_{K^{2}},\delta_{S^{2}}$ 
by using the matter power spectrum and bispectrum obtained by the phenomenological halo model including subhalos.

We found that, at the level of the post-Born approximation, $\braket{S}$ and $\braket{K}$ are no longer zero.
We also confirmed, by computing the contribution to $\braket{S}$ and $\braket{K}$ from the bispectrum terms, that evaluating $\braket{S}$ and $\braket{K}$ by solely using the power spectrum still provides a reliable estimation.
While $\braket{S}$ and $\braket{K}$ typically have the same order as $\braket{S^{2}}$ and $\Braket{K^{2}}$, the presence of the point masses (shot noise) can particularly enhance $\braket{S}$, due to the dependency of $\braket{S}$ on the physical size of the point masses.
We then estimate the number of GW events required to observe $\braket{S}$ and $\braket{K}$ and found that, while detecting the average generally requires a larger number of events than the variance, the number required to observe $\braket{S}$ can be of the order of $\mathcal{O}(10^{4})$ at $f\sim10$~Hz with $\mathrm{SNR}=50$.
This number is comparable to, or potentially even smaller than, the number required to detect $\braket{S^{2}}$, depending on the nature of the point masses.

As for the post-Born corrections to the variance, we found that their primary contribution comes from uneven distributions of the target halos with the corresponding Fresnel scale.
Our findings show that the corrections to $\braket{S^{2}}$ in the absence of the shot noise are two orders of magnitude smaller than the Born approximation at $f>0.01$~Hz and $z_{s}\leq3$.
This also indicates that the halos can be treated as if they are uniformly distributed when computing $\braket{S^{2}}$ and $\braket{K^{2}}$.
In addition, the post-Born corrections do not pose relevant issues in the absence of the shot noise unless $\mathrm{SNR}$ for GWs is excessively high.

In the presence of the shot noise, we determined the conditions under which the Born approximation fails. 
The validity of the Born approximation is guaranteed when the point mass does not dominate $\braket{S^{2}}$ and $\braket{K^{2}}$ or when strong lensing by the point mass does not occur.
However, when these conditions are violated simultaneously, the variance is predominantly determined by rare strong lensing events, and the Born approximation no longer predicts the true variance.
Furthermore, the breakdown frequency for $\braket{S^{2}}$ is lower compared to the one for $\braket{K^{2}}$ due to the enhancing factor pertaining to the physical size of the point masses.
Since the breakdown frequency may fall within the sensitivity range of current detectors in certain scenarios (such as $f\sim 20$~Hz for $\braket{S^{2}}$ with $m_{p}=0.5M_{\odot},f_{p}=0.01, k_{c}=4\times 10^{13}h\mathrm{Mpc}^{-1}$),
careful analysis of the lensing signal is required.
For example, when the frequency of GWs approaches the breakdown frequency from below, and the accuracy of the Born approximation becomes less trustable, the modification to the Born approximation given in Eq.~(\ref{eq:SBornshotmodi}) and Eq.~(\ref{eq:KBornshotmodi}) can be used to provide a more accurate estimation of $\braket{S^{2}}$ and $\braket{K^{2}}$.
When the frequency is above the breakdown frequency, a perturbative approach fails to provide a reliable result. 
Thus, in this case, a separate study involving a full-order analysis is needed to effectively constrain the property of the point masses.

\section*{Acknowledgements}
We would like to thank Ryuichi Takahashi and Adrean Webb for discussions which were quite helpful.
This work is supported by the MEXT KAKENHI Grant Number 17H06359~(TS), JP21H05453~(TS),  
and the JSPS KAKENHI Grant Number JP19K03864~(TS).

\clearpage
\appendix

\section{Geometric optics limit}
\label{app: geometric optics limit}

In \cite{Shapiro:2006em, Hilbert:2008kb, Krause:2009yr, Pratten:2016dsm, Petri:2016qya},
the post-Born approximation is discussed under geometric optics.
Although geometric optics has been widely used in the gravitational lensing, 
fundamentally more accurate description for the GL of GWs is wave optics.
In this sense, wave optics should be able to encompass everything that could be derived in geometric optics.

In geometric optics, we take the large frequency limit ($\omega\to\infty$) from the outset 
and start from the geodesic equation which does not contain $\omega$.
In this appendix, we demonstrate explicitly that the magnification in the high frequency limit
under the post-Born approximation in wave optics coincides with the one derived based on geometric optics.
In order to calculate the magnification under geometric optics, 
we need the convergence $\kappa$ and shear $\gamma$ up to second order and first order in $\Phi$, respectively.
According to \cite{Shapiro:2006em, Hilbert:2008kb, Krause:2009yr, Pratten:2016dsm, Petri:2016qya}, they are given by
\begin{align}
    \kappa^{(1)}(\bm\theta_{0},\chi_{s})=&
    \int_{0}^{\chi_{s}}d\chi
    \chi^{2}W(\chi,\chi_{s})
    \Phi_{ii}(\chi),
    \\
    \kappa^{(2)}(\bm\theta_{0},\chi_{s})
    =&
    -2\int_{0}^{\chi_{s}}d\chi
    \int_{0}^{\chi}d\chi'
    \chi^{2}\chi'^{2}W(\chi,\chi_{s})W(\chi',\chi)
    \Phi_{ij}(\chi)\Phi_{ij}(\chi')
    \notag
    \\
    &\quad-2\int_{0}^{\chi_{s}}d\chi
    \int_{0}^{\chi}d\chi'
    \chi^{3}\chi'W(\chi,\chi_{s})W(\chi',\chi)
    \Phi_{iik}(\chi)\Phi_{k}(\chi'),
    \\
    \gamma_{1}^{(1)}(\bm\theta_{0},\chi_{s})=&
    \int_{0}^{\chi_{s}}d\chi
    \chi^{2}W(\chi,\chi_{s})
    (\Phi_{11}(\chi)-\Phi_{22}(\chi)),
    \\
    \gamma_{2}^{(1)}(\bm\theta_{0},\chi_{s})=&
    2\int_{0}^{\chi_{s}}d\chi
    \chi^{2}W(\chi,\chi_{s})
    \Phi_{12}(\chi).
\end{align}
The gravitational potential is evaluated at the straight line along which the unlensed ray propagates, namely $\Phi(\chi)=\Phi(\bm\theta_{0},\chi)$.
The magnification $\mu_{\rm geo}(\bm\theta_{0},\chi_{s})$ is the inverse of the determinant of the 
Jacobian matrix $\bm A(\bm \theta_{0},\chi)=\begin{pmatrix}
    1-\kappa-\gamma_{1} & -\gamma_{2}-\Omega \\
    -\gamma_{2}+\Omega & 1-\kappa+\gamma_{1} \\
    \end{pmatrix}$ and,
up to second order in $\Phi$, 
$\mu_{\rm geo}$ is given by
\begin{align}
    \label{eq: mag geo}
    \mu_{\rm geo}(\bm\theta_{0},\chi_{s})=&
    1+2\kappa^{(1)}+2\kappa^{(2)}+3(\kappa^{(1)})^{2}+(\gamma_{1}^{(1)})^{2}+(\gamma_{2}^{(1)})^{2}
    \notag
    \\
    =&1+2\kappa^{(1)}(\bm\theta_{0},\chi_{s})+2(\kappa^{(1)}(\bm\theta_{0},\chi_{s}))^{2}
    \notag
    \\
    &\quad
    -4\int_{0}^{\chi_{s}}d\chi
    \int_{0}^{\chi}d\chi'
    \chi^{3}\chi'W(\chi,\chi_{s})W(\chi',\chi)
    \Phi_{iik}(\chi)\Phi_{k}(\chi')
    \notag
    \\
    &\quad+4\int_{0}^{\chi_{s}}d\chi
    \int_{0}^{\chi}d\chi'
    \chi^{2}\chi'^{2}W(\chi,\chi_{s})^{2}
    \Phi_{ij}(\chi)\Phi_{ij}(\chi')
    .
\end{align}
Up to this order, $\Omega$ does not appear in the magnification as $\Omega$ itself is already second order in $\Phi$.

We now show the magnification computed in wave optics based on the formulation given
in this paper reduces to Eq.~(\ref{eq: mag geo}) in the high frequency limit.
Prior to that, we write the approximated solution of the lens equation up to first order in $\Phi$. 
\begin{align}
    \label{eq: lens ingetral}
    \delta \bm \Theta(\bm\theta_{0},\chi)
    =
    -2\int_{0}^{\chi}d\chi'
    W(\chi',\chi)
    \nabla_{\theta}\Phi(\bm\theta_{0},\chi').
\end{align}
In wave optics, the magnification effect is encoded in $K$ as $\mu_{\rm wave}(\bm\theta,\omega)=e^{2K}$, where $\bm \theta$ is the position of the source on the source plane $\bm \theta=\bm \theta_{0}+\delta\bm \Theta(\bm\theta_{0},\chi_{s}) $.
Taking $\omega\to\infty$ of Eqs.~(\ref{eq: mag K 1st order}) and (\ref{eq: mag K 2nd order}) yields
\begin{align}
    K^{(1)}(\bm\theta,\chi_{s},\omega\to\infty)=&
    -2\int_{0}^{\chi_{s}}d\chi
    \chi^{2}W(\chi,\chi_{s})
    \Phi_{ii}(\bm\theta,\chi)
    =\kappa^{(1)}(\bm \theta,\chi_{s})
    \\
    K^{(2)}(\bm\theta,\chi_{s},\omega\to\infty)
    =&
    \int_{0}^{\chi_{s}}\frac{d\chi}{\chi^{2}}
    \int_{0}^{\chi}d\chi_{1}
    \int_{0}^{\chi}d\chi_{2} 
    \notag
    \\
    &\quad\times
    \left[W(\chi,\chi_{s})\nabla^{2}_{\theta 12}
    +W(\chi_{1},\chi)\nabla^{2}_{\theta1}
    +W(\chi_{2},\chi)\nabla^{2}_{\theta2}
    \right]
    \nabla_{\theta1}\Phi_{1}\cdot\nabla_{\theta2}\Phi_{2}
    \notag
    \\
    =&
    -\int_{0}^{\chi_{s}}d\chi_{1}W(\chi_{1},\chi_{s})
    \nabla_{\theta} (\nabla^{2}_{\theta}\Phi(\chi_{1}))\cdot 
    (-2)
    \int_{0}^{\chi_{s}}d\chi_{2}
    W(\chi_{2},\chi_{s})
    \nabla_{\theta}\Phi(\chi_{2})
    \notag
    \\
    &\quad-2\int_{0}^{\chi_{s}}d\chi
    \int_{0}^{\chi}d\chi'
    \chi^{3}\chi'W(\chi,\chi_{s})W(\chi',\chi)
    \Phi_{iik}(\chi)\Phi_{k}(\chi')
    \notag
    \\
    &\quad+2\int_{0}^{\chi_{s}}d\chi
    \int_{0}^{\chi}d\chi'
    \chi^{2}\chi'^{2}W(\chi,\chi_{s})^{2}
    \Phi_{ij}(\chi)\Phi_{ij}(\chi')
    \notag
    \\
    =&-\nabla_{\theta}K^{(1)}\cdot \delta\bm\Theta
    \notag
    \\
    &
    \quad-2\int_{0}^{\chi_{s}}d\chi
    \int_{0}^{\chi}d\chi'
    \chi^{3}\chi'W(\chi,\chi_{s})W(\chi',\chi)
    \Phi_{iik}(\chi)\Phi_{k}(\chi')
    \notag
    \\
    &\quad+2\int_{0}^{\chi_{s}}d\chi
    \int_{0}^{\chi}d\chi'
    \chi^{2}\chi'^{2}W(\chi,\chi_{s})^{2}
    \Phi_{ij}(\chi)\Phi_{ij}(\chi').
\end{align}
The magnification in wave optics up to second order is then given by
\begin{align}
    \label{eq: mag wave}
     \mu_{\rm wave}(\bm\theta,\chi_{s},\omega\to\infty)
     =&
     1+2K^{(1)}+2(K^{(1)})^{2}+2K^{(2)}
     \notag
     \\
     =&1+2\kappa^{(1)}(\bm\theta-\delta\bm \Theta,\chi_{s})+2(\kappa^{(1)}(\bm\theta_{0},\chi_{s}))^{2}
    \notag
    \\
    &\quad
    -4\int_{0}^{\chi_{s}}d\chi
    \int_{0}^{\chi}d\chi'
    \chi^{3}\chi'W(\chi,\chi_{s})W(\chi',\chi)
    \Phi_{iik}(\chi)\Phi_{k}(\chi')
    \notag
    \\
    &\quad+4\int_{0}^{\chi_{s}}d\chi
    \int_{0}^{\chi}d\chi'
    \chi^{2}\chi'^{2}W(\chi,\chi_{s})^{2}
    \Phi_{ij}(\chi)\Phi_{ij}(\chi')
    \notag
    \\
    =&\mu_{\rm geo}(\bm\theta_{0},\chi_{s}).
\end{align}
Therefore, the result of geometric optics is indeed derived by taking the high-frequency limit of wave optics. 
It is important to mention again that the lens plane $\bm\theta_{0}$ is used in geometric optics whereas, in wave optics, the source plane $\bm \theta$ is the fundamental variable.
This difference manifests itself in the argument of both $\mu_{\rm geo}$ and $\mu_{\rm wave}$.
We have shown that, at least up to second order in $\Phi$,
our formulation reduces to the well-known result in geometric optics. 
This consistency strongly supports the validity of the discussion 
about the post-Born approximation of the lensing in wave optics.

\section{Post-Born variance of $S$ and $K$}
\label{app:postBornvariance}
The correction of the variance of $K$ to the Born approximation is described by Eq.~(\ref{eq: corrcKVar}),
and similar relation holds for $S$. 
This equation is rewritten by using the matter power spectrum given in Eq.~(\ref{eq: Ppotential Pmatter}):
\begin{align}
\label{deltaK^2dc}
    \delta_{X^2,\rm dc}
    =&
    16\left(\frac{3H_{0}^{2}\Omega_{m}}{2}\right)^{4}
    \int_{0}^{\chi_{s}}\frac{d\chi}{\chi^{2}}
    \int_{0}^{\chi}\frac{d\chi'}{\chi'^{2}}
    \int_{0}^{\chi'}d\chi_{1}
    \int_{0}^{\chi'}d\chi_{2}
    \frac{1}{a^{2}(\chi_{1})}
    \frac{1}{a^{2}(\chi_{2})}
    \frac{1}{(2\pi)^{2}}
    \notag
    \\
    &\times\omega^{2}
    \int_{0}^{\infty}dk_{1}
    \int_{0}^{\infty}dk_{2}
    P_{\delta}(k_{1},\chi_{1})
    P_{\delta}(k_{2},\chi_{2})
    \left[
    \frac{1}{k_{1}k_{2}}\mathcal{F}_{12}
    -\frac{1}{k_{1}^{2}}\mathcal{F}_{1}
    -\frac{1}{k_{2}^{2}}\mathcal{F}_{2}
    \right],
\end{align}
where
\begin{align}
    \mathcal{F}_{12}
    =&
    \chi_{1}^{2}\chi_{2}^{2}
    \int_{0}^{2\pi}\frac{d\phi}{2\pi}\cos^{2}{\phi}
    \left\{
    F\left(
     \frac{\chi_{1}^{2}W(\chi_{1},\chi_{s})}{2\omega}k_{1}^{2}
    +\frac{\chi_{2}^{2}W(\chi_{2},\chi_{s})}{2\omega}k_{2}^{2}
    +\frac{\chi_{1}\chi_{2}W(\chi,\chi_{s})}{\omega}k_{1}k_{2}\cos{\phi}
    \right)\right.
    \notag
    \\
    &\qquad\qquad\qquad\qquad\times
     F\left(
     \frac{\chi_{1}^{2}W(\chi_{1},\chi_{s})}{2\omega}k_{1}^{2}
    +\frac{\chi_{2}^{2}W(\chi_{2},\chi_{s})}{2\omega}k_{2}^{2}
    +\frac{\chi_{1}\chi_{2}W(\chi',\chi_{s})}{\omega}k_{1}k_{2}\cos{\phi}
    \right)
    \notag
    \\
    &
    -F\left(
     \frac{\chi_{1}^{2}W(\chi_{1},\chi_{s})}{2\omega}k_{1}^{2}
    \right)
    F\left(
    \frac{\chi_{1}^{2}W(\chi_{1},\chi_{s})}{2\omega}k_{1}^{2}
    +\frac{\chi_{2}^{2}W(\chi_{2},\chi)}{\omega}k_{2}^{2}
    +\frac{\chi_{1}\chi_{2}W(\chi',\chi)}{\omega}k_{1}k_{2}\cos{\phi}
    \right)
    \notag
    \\
    &
    \left.
    -F\left(
     \frac{\chi_{2}^{2}W(\chi_{2},\chi_{s})}{2\omega}k_{2}^{2}
    \right)
    F\left(
    \frac{\chi_{2}^{2}W(\chi_{2},\chi_{s})}{2\omega}k_{2}^{2}
    +\frac{\chi_{1}^{2}W(\chi_{1},\chi)}{\omega}k_{1}^{2}
    +\frac{\chi_{1}\chi_{2}W(\chi',\chi)}{\omega}k_{1}k_{2}\cos{\phi}
    \right)
    \right\}, \label{A-F12}
    \\
    \mathcal{F}_{1}
    =&
    \chi_{1}\chi_{2}^{3}
    \int_{0}^{2\pi}\frac{d\phi}{2\pi}\cos{\phi}
    \notag
    \\
    &\times
    F\left(
     \frac{\chi_{1}^{2}W(\chi_{1},\chi_{s})}{2\omega}k_{1}^{2}
    \right)
    F\left(
    \frac{\chi_{1}^{2}W(\chi_{1},\chi_{s})}{2\omega}k_{1}^{2}
    +\frac{\chi_{2}^{2}W(\chi_{2},\chi)}{\omega}k_{2}^{2}
    +\frac{\chi_{1}\chi_{2}W(\chi',\chi)}{\omega}k_{1}k_{2}\cos{\phi}
    \right),
    \\
    \mathcal{F}_{2}
    =&
    \chi_{2}\chi_{1}^{3}
    \int_{0}^{2\pi}\frac{d\phi}{2\pi}\cos{\phi}
    \notag
    \\
    &\times
    F\left(
     \frac{\chi_{2}^{2}W(\chi_{2},\chi_{s})}{2\omega}k_{2}^{2}
    \right)
    F\left(
    \frac{\chi_{2}^{2}W(\chi_{2},\chi_{s})}{2\omega}k_{2}^{2}
    +\frac{\chi_{1}^{2}W(\chi_{1},\chi)}{\omega}k_{1}^{2}
    +\frac{\chi_{1}\chi_{2}W(\chi',\chi)}{\omega}k_{1}k_{2}\cos{\phi}
    \right).
\end{align}
Here, $X$ is either $K$ or $S$.
$F(x)$ is defined as $F(x)=\sin{x}$ for the magnification $K$ 
and $F(x)=1-\cos{x}$ for the phase modulation $S$.
Even though $\chi_{1}$ and $\chi_{2}$ are symmetrical and can be expressed by either one of two terms,
we explicitly write both terms so that the symmetry can be captured easily.
The integral with respect to $\phi$ can be performed analytically 
by using the identities regarding Bessel functions:
\begin{align}
    \int_{0}^{2\pi}\frac{d\phi}{2\pi}\cos{\phi}\sin(x\cos{\phi})
    =&J_{1}(x),
    \\
    \int_{0}^{2\pi}\frac{d\phi}{2\pi}\cos^{2}{\phi}\cos(x\cos{\phi})
    =&\frac{1}{2}\left[J_{0}(x)-J_{2}(x)\right].
\end{align}
In addition to this, $\int_{0}^{2\pi}\frac{d\phi}{2\pi}\cos{\phi}\cos(x\cos{\phi})=\int_{0}^{2\pi}\frac{d\phi}{2\pi}\cos^{2}{\phi}\sin(x\cos{\phi})=0$ holds by virtue of the anti-symmetric nature of the integrand.

For the magnification, $\mathcal{F}_{K,12},\mathcal{F}_{K,1},\mathcal{F}_{K,2}$ 
are given by
\begin{align}
    \label{eq: FK12}
    \mathcal{F}_{K,12}&=
    \frac{\chi_{1}^{2}\chi_{2}^{2}}{4}\left\{
    \left(
    1
    -\cos{\left(\frac{\chi_{1}^{2}W(\chi_{1},\chi)}{\omega}k_{1}^{2}\right)}
    -\cos{\left(\frac{\chi_{2}^{2}W(\chi_{2},\chi)}{\omega}k_{2}^{2}\right)}\right.\right.
    \notag
    \\
    &\left.
    +\cos{\left(\frac{\chi_{1}^{2}W(\chi_{1},\chi_{s})}{\omega}k_{1}^{2}
    +\frac{\chi_{2}^{2}W(\chi_{2},\chi)}{\omega}k_{2}^{2}\right)}
    +\cos{\left(
    \frac{\chi_{2}^{2}W(\chi_{2},\chi_{s})}{\omega}k_{2}^{2}
    +\frac{\chi_{1}^{2}W(\chi_{1},\chi)}{\omega}k_{1}^{2}\right)}
    \right)
    \notag
    \\
    &\times
    \left[J_{0}\left(\frac{\chi_{1}\chi_{2}W(\chi',\chi)}{\omega}k_{1}k_{2}\right)
    -J_{2}\left(\frac{\chi_{1}\chi_{2}W(\chi',\chi)}{\omega}k_{1}k_{2}\right)\right]
    \notag
    \\
    &
    -\cos{\left(\frac{\chi_{1}^{2}W(\chi_{1},\chi_{s})}{\omega}k_{1}^{2}
    +\frac{\chi_{2}^{2}W(\chi_{2},\chi_{s})}{\omega}k_{2}^{2}\right)}\notag
    \\
    &\times
     \left[J_{0}\left(\frac{\chi_{1}\chi_{2}W(\chi,\chi_{s})+\chi_{1}\chi_{2}W(\chi',\chi_{s})}{\omega}k_{1}k_{2}\right)\right.
      \left.\left.
    -J_{2}\left(\frac{\chi_{1}\chi_{2}W(\chi,\chi_{s})+\chi_{1}\chi_{2}W(\chi',\chi_{s})}{\omega}k_{1}k_{2}\right)\right]\right\},
    \\
    \label{eq: FK1}
    \mathcal{F}_{K,1}&=
    \frac{\chi_{1}\chi_{2}^{3}}{2}
    \left\{
    \sin{\left(\frac{\chi_{1}^{2}W(\chi_{1},\chi_{s})}{\omega}k_{1}^{2}+\frac{\chi_{2}^{2}W(\chi_{2},\chi)}{\omega}k^{2}_{2}\right)}
    -\sin{\left(
    \frac{\chi_{2}^{2}W(\chi_{2},\chi)}{\omega}k_{2}^{2}
    \right)}
    \right\}
    \notag
    \\
    &\times
    J_{1}\left(\frac{\chi_{1}\chi_{2}W(\chi',\chi)}{\omega}k_{1}k_{2}\right),
    \\
    \label{eq: FK2}
    \mathcal{F}_{K,2}&=
    \frac{\chi_{2}\chi_{1}^{3}}{2}
    \left\{
    \sin{\left(\frac{\chi_{2}^{2}W(\chi_{2},\chi_{s})}{\omega}k_{2}^{2}+\frac{\chi_{1}^{2}W(\chi_{1},\chi)}{\omega}k^{2}_{1}\right)}
    -\sin{\left(
    \frac{\chi_{1}^{2}W(\chi_{1},\chi)}{\omega}k_{1}^{2}
    \right)}
    \right\}
    \notag
    \\
    &\times
    J_{1}\left(\frac{\chi_{1}\chi_{2}W(\chi',\chi)}{\omega}k_{1}k_{2}\right).
\end{align}
In exactly the same way, the similar equations are derived for the phase modulation:
\begin{align}
    \label{eq: FS12}
    \mathcal{F}_{S,12}
    &=
    \chi_{1}^{2}\chi_{2}^{2}\left\{
    -\frac{1}{2}
    +\frac{1}{2}\cos{\left(\frac{\chi_{1}^{2}W(\chi_{1},\chi_{s})}{2\omega}k_{1}^{2}\right)}
    +\frac{1}{2}\cos{\left(\frac{\chi_{2}^{2}W(\chi_{2},\chi_{s})}{2\omega}k_{2}^{2}\right)}
    \right.\notag
    \\
    &-\frac{1}{2}\cos{\left(\frac{\chi_{1}^{2}W(\chi_{1},\chi_{s})}{2\omega}k_{1}^{2}+\frac{\chi_{2}^{2}W(\chi_{2},\chi_{s})}{2\omega}k_{2}^{2}\right)}
    \notag
    \\
    &\times
    \left[
    J_{0}\left(\frac{\chi_{1}\chi_{2}W(\chi,\chi_{s})}{\omega}k_{1}k_{2}\right)
    +
     J_{0}\left(\frac{\chi_{1}\chi_{2}W(\chi',\chi_{s})}{\omega}k_{1}k_{2}\right)
    \right.\notag
    \\
    &\left.
    - J_{2}\left(\frac{\chi_{1}\chi_{2}W(\chi,\chi_{s})}{\omega}k_{1}k_{2}\right)
    - J_{2}\left(\frac{\chi_{1}\chi_{2}W(\chi',\chi_{s})}{\omega}k_{1}k_{2}\right)\right]
    \notag
    \\
    &+\frac{1}{4}\cos{\left(\frac{\chi_{1}^{2}W(\chi_{1},\chi_{s})}{\omega}k_{1}^{2}+\frac{\chi_{2}^{2}W(\chi_{2},\chi_{s})}{\omega}k_{2}^{2}\right)}
    \notag
    \\
    &\times
     \left[J_{0}\left(\frac{\chi_{1}\chi_{2}W(\chi,\chi_{s})+\chi_{1}\chi_{2}W(\chi',\chi_{s})}{\omega}k_{1}k_{2}\right)\right.
      \left.
    -J_{2}\left(\frac{\chi_{1}\chi_{2}W(\chi,\chi_{s})+\chi_{1}\chi_{2}W(\chi',\chi_{s})}{\omega}k_{1}k_{2}\right)\right]
    \notag
    \\
    &+\left(
    \frac{1}{4}
    +\sin^{2}{\left(\frac{\chi_{1}^{2}W(\chi_{1},\chi_{s})}{4\omega}k_{1}^{2}\right)}
    \cos{\left(
    \frac{\chi_{1}^{2}W(\chi_{1},\chi_{s})}{2\omega}k_{1}^{2}
    +
    \frac{\chi_{2}^{2}W(\chi_{2},\chi)}{\omega}k_{2}^{2}
    \right)}
    \right.
    \notag
    \\
    &
    \left.
    +\sin^{2}{\left(\frac{\chi_{2}^{2}W(\chi_{2},\chi_{s})}{4\omega}k_{2}^{2}\right)}
    \cos{\left(
    \frac{\chi_{2}^{2}W(\chi_{2},\chi_{s})}{2\omega}k_{2}^{2}
    +
    \frac{\chi_{1}^{2}W(\chi_{1},\chi)}{\omega}k_{1}^{2}
    \right)}
    \right)
    \notag
    \\
    &\left.\times
     \left[J_{0}\left(\frac{\chi_{1}\chi_{2}W(\chi',\chi)}{\omega}k_{1}k_{2}\right)
    -J_{2}\left(\frac{\chi_{1}\chi_{2}W(\chi',\chi)}{\omega}k_{1}k_{2}\right)\right]\right\},
    \\
    \label{eq: FS1}
    \mathcal{F}_{S,1}
    &=
    2\chi_{1}\chi_{2}^{3}
    \sin^{2}{\left(
    \frac{\chi_{1}^{2}W(\chi_{1},\chi_{s})}{4\omega}k_{1}^{2}
    \right)
    \sin{\left(
    \frac{\chi_{1}^{2}W(\chi_{1},\chi_{s})}{2\omega}k_{1}^{2}
    +\frac{\chi_{2}^{2}W(\chi_{2},\chi)}{\omega}k_{2}^{2}
    \right)}}
    \notag
    \\
    &\times
    J_{1}\left(\frac{\chi_{1}\chi_{2}W(\chi',\chi)}{\omega}k_{1}k_{2}\right),
    \\
    \label{eq: FS2}
    \mathcal{F}_{S,2}
    &=
    2\chi_{2}\chi_{1}^{3}
    \sin^{2}{\left(
    \frac{\chi_{2}^{2}W(\chi_{2},\chi_{s})}{4\omega}k_{2}^{2}
    \right)
    \sin{\left(
    \frac{\chi_{2}^{2}W(\chi_{2},\chi_{s})}{2\omega}k_{2}^{2}
    +\frac{\chi_{1}^{2}W(\chi_{1},\chi)}{\omega}k_{1}^{2}
    \right)}}
    \notag
    \\
    &\times
    J_{1}\left(\frac{\chi_{1}\chi_{2}W(\chi',\chi)}{\omega}k_{1}k_{2}\right).
\end{align}
For the non-Gaussian correction, We only consider the bispectrum term as the only relevant contribution, thus $\delta_{X^{2}\rm c}=2\braket{X_{\rm Born}X^{(2)}}_{c}$.
Based on this assumption, Eq.(\ref{eq; magnification 12 bispectrum}) can be rewritten as 
\begin{align}
    \label{deltaS^2c}
    \delta_{S^{2}\rm c}
    =&
     8\omega^{3}\left(\frac{3H_{0}^{2}\Omega_{m}}{2}\right)^{3}
   \int_{0}^{\chi_{s}}\frac{d\chi}{a^{3}}
    \int_{0}^{\infty}\frac{dk_{1}}{2\pi}
    \int_{0}^{\infty}\frac{d k_{2}}{2\pi}
    \int_{0}^{2\pi}\frac{d\phi}{2\pi}
    \frac{B_{\delta}(k_{1},k_{2},k_{3},\chi)}{k_{1}k_{2}k_{3}^{2}}\notag
    \\
    &\times
     \left(1-\cos{\left[\frac{r_{\rm F}^{2}}{2}k_{3}^{2}\right]}\right)
    \left\{
    r_{\rm F}^{2}\bm k_{1}\cdot\bm k_{2}
    -2\sin{\left[\frac{r_{\rm F}^{2}}{2}\bm k_{1}\cdot\bm k_{2}\right]}
    \cos{\left[\frac{r_{\rm F}^{2}}{2}
    \frac{k_{1}^{2}+k_{2}^{2}+k_{3}^{2}}{2}
    \right]}
    \right\}
    \\
    \label{deltaK^2c}
    \delta_{K^{2}\rm c}
    =&
    16\omega^{3}\left(\frac{3H_{0}^{2}\Omega_{m}}{2}\right)^{3}
   \int_{0}^{\chi_{s}}\frac{d\chi}{a^{3}}
    \int_{0}^{\infty}\frac{dk_{1}}{2\pi}
    \int_{0}^{\infty}\frac{d k_{2}}{2\pi}
    \int_{0}^{2\pi}\frac{d\phi}{2\pi}
   \frac{B_{\delta}(k_{1},k_{2},k_{3},\chi)}{k_{1}k_{2}k_{3}^{2}}\notag
    \\
    &\times
    \sin{\left[\frac{r_{\rm F}^{2}}{2}k_{3}^{2}\right]}
    \sin{\left[\frac{r_{\rm F}^{2}}{2}\frac{k_{1}^{2}+k_{2}^{2}+k_{3}^{2}}{2}\right]}
    \sin{\left[\frac{r_{\rm F}^{2}}{2}\bm k_{1}\cdot\bm k_{2}\right]}
\end{align}
where $k_{3}^{2}=k_{1}^{2}+k_{2}^{2}+2k_{1}k_{2}\cos{\phi}$ and $\bm k_{1}\cdot\bm k_{2}=k_{1}k_{2}\cos{\phi}$.
The bispectrum contribution to the average of $S$ and $K$ Eq.~(\ref{eq: SaveBi}) and Eq.~(\ref{eq: KaveBi}) can be rewritten in a similar way as
\begin{align}
    \label{eq: S3 ave}
    \braket{S^{(3)}}
    =&4\omega^{3}\left(\frac{3H_{0}^{2}\Omega_{m}}{2}\right)^{3}
    \int_{0}^{\chi_{s}}\frac{d\chi}{a^{3}}
    \int_{0}^{\infty}\frac{dk_{1}}{2\pi}
    \int_{0}^{\infty}\frac{d k_{2}}{2\pi}
    \int_{0}^{2\pi}\frac{d\phi}{2\pi}
    \frac{B_{\delta}(k_{1},k_{2},k_{3},\chi)}{k_{1}k_{2}k_{3}^{2}}
    \notag
    \\
    &\times
    \left\{
    \frac{4k_{3}^{2}\sin^{2}\left(\frac{r_{\rm F}^{2}}{4}(k_{1}^{2}+k_{2}^{2}+k_{3}^{2})\right)}{(k_{1}^{2}+k_{2}^{2}+k_{3}^{2})}
    -
    2\sin^{2}{\frac{r_{\rm F}^{2} k_{3}^{2}}{2}}
    -\frac{r_{\rm F}^{2}}{2}(\bm k_{1}\cdot\bm k_{2})k_{3}^{2}
    \right\},
    \\
    \label{eq: K3 ave}
    \braket{K^{(3)}}_{c}
    =&
   4\omega^{3}\left(\frac{3H_{0}^{2}\Omega_{m}}{2}\right)^{3}
   \int_{0}^{\chi_{s}}\frac{d\chi}{a^{3}}
    \int_{0}^{\infty}\frac{dk_{1}}{2\pi}
    \int_{0}^{\infty}\frac{d k_{2}}{2\pi}
    \int_{0}^{2\pi}\frac{d\phi}{2\pi}
    \frac{B_{\delta}(k_{1},k_{2},k_{3},\chi)}{k_{1}k_{2}k_{3}^{2}}\notag
    \\
    &\times
    \left\{
    \sin{\left[r_{\rm F}^{2}k_{1}^{2}\right]}+
    \sin{\left[r_{\rm F}^{2}k_{2}^{2}\right]}+
    \sin{\left[r_{\rm F}^{2}k_{3}^{2}\right]}-
    2\sin{\left[\frac{r_{\rm F}^{2}}{2}(k_{1}^{2}+k_{2}^{2}+k_{3}^{2})\right]}
    \right\}
\end{align}
Note that, in computing bispectrum contribution numerically, symmetrizing the wavenumber variables $k_{1},k_{2},k_{3}$ reduces the computational cost.

\section{High frequency behavior of $\delta_{S^{2},\rm dc}$}
\label{app: high freq S}
In this appendix, we would like to approximately derive the high-frequency behavior of $\delta_{S^{2},\rm dc}$ in order to avoid the difficulty of numerical computation associated with the cancellation of significant digits.
When the frequency of GWs is high, $\delta_{S^{2},\rm dc}$ is mainly affected by the large $k$ region of the matter power spectrum.
Given that the corresponding Fresnel scale mainly contributes to the lensing, 
we can expand the power spectrum around the approximated Fresnel scale ($1/\sqrt{H_{0}\omega}$) as  
\begin{align}
    \label{eq: matpse power a}
    P_{\delta}(k,\chi)= P_{\delta}(k_{0},\chi)\left(\frac{k}{k_{0}}\right)^{\frac{d\log{P_{\delta}(k_{0},\chi)}}{d\log{k_{0}}}}=B(\chi)k^{-b(\chi)},
\end{align}
where $(k_{0}=\sqrt{H_{0}\omega})$.
In this way, $B(\chi)$ and $b_{\chi}$ are both functions of redshift and the frequency of GWs.
We compute $B(\chi)$ and $b(\chi)$ numerically using our power spectrum at each frequency.
Keeping this in mind, we only consider the case that is relevant to our discussion.

We usually deal with the GW sources whose distance from the earth is roughly given by $1/H_{0}$, 
so the corresponding Fresnel scale is $1/\sqrt{\omega H_{0}}$.
The high frequency behavior in this context is then interpreted as the satisfaction of  the condition $k_{L}\ll\sqrt{H_{0}\omega}$. 
Defining $\idotsint\equiv
16\left(\frac{3H_{0}^{2}\Omega_{m}}{2}\right)^{4}
    \int_{0}^{\chi_{s}}\frac{d\chi}{\chi^{2}}
    \int_{0}^{\chi}\frac{d\chi'}{\chi'^{2}}
    \int_{0}^{\chi'}d\chi_{1}
    \int_{0}^{\chi'}d\chi_{2}
    \frac{1}{a^{2}(\chi_{1})}
    \frac{1}{a^{2}(\chi_{2})}
    \frac{1}{(2\pi)^{2}}$,
the post-Born approximation of the variance of $S$ is given by
\begin{align}
    \delta_{S^{2},\rm dc}
    =&
    \idotsint
    \omega^{2}
    \int_{0}^{\infty}dk_{1}
    \int_{0}^{\infty}dk_{2}
    P_{\delta}(k_{1},\chi_{1})
    P_{\delta}(k_{2},\chi_{2})
    \left[
    \frac{1}{k_{1}k_{2}}\mathcal{F}_{S,12}
    -\frac{1}{k_{1}^{2}}\mathcal{F}_{S,1}
    -\frac{1}{k_{2}^{2}}\mathcal{F}_{S,2}
    \right],
\end{align}
The definition of $\mathcal{F}_{S,12},\mathcal{F}_{S,1},\mathcal{F}_{S,2}$ is the same as the ones in Appendix \ref{app:postBornvariance}.
Change of variable
$k_{1}=\sqrt{\omega}\xi_{1}, k_{2}=\sqrt{\omega}\xi_{2}$
and separating the integral area at $k_{L}$ yield
\begin{align}
    \label{eq: 4th order S variance }
    \delta_{S^{2},\rm dc}
    =&
    \idotsint \omega^{2}
    \int_{0}^{\frac{k_{L}}{\sqrt{\omega}}}d\xi_{1}
    \int_{0}^{\frac{k_{L}}{\sqrt{\omega}}}d\xi_{2}
    P_{\delta}(\sqrt{\omega}\xi_{1},\chi_{1})
    P_{\delta}(\sqrt{\omega}\xi_{2},\chi_{2})
    \left[
    \frac{1}{\xi_{1}\xi_{2}}\mathcal{F}_{12}
    -\frac{1}{\xi_{1}^{2}}\mathcal{F}_{1}
    -\frac{1}{\xi_{2}^{2}}\mathcal{F}_{2}
    \right]
    \notag
    \\
    &+
    \idotsint \omega^{2-\frac{b}{2}}B(\chi_{2})
    \int_{0}^{\frac{k_{L}}{\sqrt{\omega}}}d\xi_{1}
    \int_{\frac{k_{L}}{\sqrt{\omega}}}^{\infty}d\xi_{2}
    P_{\delta}(\sqrt{\omega}\xi_{1},\chi_{1})
    \xi_{2}^{-b}
   \left[
    \frac{1}{\xi_{1}\xi_{2}}\mathcal{F}_{12}
    -\frac{1}{\xi_{1}^{2}}\mathcal{F}_{1}
    -\frac{1}{\xi_{2}^{2}}\mathcal{F}_{2}
    \right]
    \notag
    \\
    &+
    \idotsint \omega^{2-\frac{b}{2}}B(\chi_{1})
    \int_{\frac{k_{L}}{\sqrt{\omega}}}^{\infty}d\xi_{1}
    \int_{0}^{\frac{k_{L}}{\sqrt{\omega}}}d\xi_{2}
   P_{\delta}(\sqrt{\omega}\xi_{2},\chi_{2})
    \xi_{1}^{-b}
    \left[
    \frac{1}{\xi_{1}\xi_{2}}\mathcal{F}_{12}
    -\frac{1}{\xi_{1}^{2}}\mathcal{F}_{1}
    -\frac{1}{\xi_{2}^{2}}\mathcal{F}_{2}
    \right]
    \notag
    \\
    &+
    \idotsint \omega^{2-b}B(\chi_{1})B(\chi_{2})
    \int_{\frac{k_{L}}{\sqrt{\omega}}}^{\infty}d\xi_{1}
    \int_{\frac{k_{L}}{\sqrt{\omega}}}^{\infty}d\xi_{2}
    \xi_{2}^{-b}\xi_{1}^{-b}
    \left[
    \frac{1}{\xi_{1}\xi_{2}}\mathcal{F}_{12}
    -\frac{1}{\xi_{1}^{2}}\mathcal{F}_{1}
    -\frac{1}{\xi_{2}^{2}}\mathcal{F}_{2}
    \right].
\end{align}
Since these four terms contribute to $\delta_{S^{2},\rm dc}$ in a different way, 
we will compute the contribution from each term separately.
To begin with, we consider the first term.
In the high frequency limit, the integral range$ \int_{0}^{\frac{k_{L}}{\sqrt{\omega}}}$
is restricted in a very small area so the contribution from the first term in Eq.~(\ref{eq: 4th order S variance }) comes from the region where $\xi_{1},\xi_{2}\ll1$ holds.
Since $\xi_{1}$ and $\xi_{2}$ are both order $1/\sqrt{\omega}$ in this integral range,
the expansion of $\mathcal{F}$ in $1/\sqrt{\omega}$ up to leading order yields
$\mathcal{F}=\mathcal{O}(1/\omega^{4})$.
Considering that $\omega^{2}$ is multiplied in the expression, 
we can conclude that the first term is proportional to $\omega^{-2}$.

The second and third terms in Eq.~(\ref{eq: 4th order S variance }) are symmetrical with respect to the subscript 1,2, 
so they have the same contribution.
In the second term, $\xi_{1}$ is still restricted in the area where $\xi_{1}\ll1$ 
whereas $\xi_{2}$ is no longer small.
In this case, we can expand $\mathcal{F}$ only in terms of $\xi_{1}$ and keep $\xi_{2}$ term untouched then we have
\begin{align}
    \label{eq: Fexpand_only_xi1}
     \mathcal{F}=&
    \xi_{1}\left[
    \xi_{2}\left(
        \cos{\left(\frac{\chi_{2}^{2}W(\chi_{2},\chi_{s})}{2}\xi_{2}^{2}\right)}
        -\cos^{2}{\left(\frac{\chi_{2}^{2}W(\chi_{2},\chi_{s})}{2}\xi_{2}^{2}\right)}
        \right)C\right.
        \notag
        \\
       & +\left(
            \frac{ \sin{\left(\frac{\chi_{2}^{2}W(\chi_{2},\chi_{s})}{2}\xi_{2}^{2}\right)}}{2\xi_{2}}
            -\frac{ \sin{\left(\chi_{2}^{2}W(\chi_{2},\chi_{s})\xi_{2}^{2}\right)}}{4\xi_{2}}
            \right)D_{1}
            +\left.C\xi_{2}\sin^{2}{\left(\frac{\chi_{2}^{2}W(\chi_{2},\chi_{s})}{2}\xi_{2}^{2}\right)}\right]
        +\mathcal{O}(\xi_{1}^{3}).
\end{align}
where $C,D_{1}$ are
\begin{align}
    C=&\frac{3}{8}\chi_{1}^{4}\chi_{2}^{4}
    W(\chi,\chi_{s})W(\chi',\chi_{s}),
    \\
    D_{1}=&\chi_{1}^{6}\chi_{2}^{2}\left\{
    -W(\chi',\chi)+W(\chi,\chi_{s})
    \right\}.
\end{align}
Combining these notations, the second term is calculated as
\begin{align}
   (2\rm nd)=& \idotsint \omega^{2-\frac{b}{2}}B(\chi_{2})
    \int_{0}^{\frac{k_{L}}{\sqrt{\omega}}}d\xi_{1}
    P_{\delta}(\sqrt{\omega}\xi_{1},\chi_{1})
    \int_{\frac{k_{L}}{\sqrt{\omega}}}^{\infty}d\xi_{2}
   \xi_{2}^{-b}
   \left[
    \frac{1}{\xi_{1}\xi_{2}}\mathcal{F}_{12}
    -\frac{1}{\xi_{1}^{2}}\mathcal{F}_{1}
    -\frac{1}{\xi_{2}^{2}}\mathcal{F}_{2}
    \right]
    \notag
    \\
    =&
    \idotsint  \frac{1}{\omega^{\frac{b}{2}-1}}B(\chi_{2}) 
     \int_{0}^{\infty}dk_{1}
    k_{1}P_{\delta}(k_{1},\chi_{1})
     \int_{0}^{\infty}d\xi_{2}
     \frac{1}{\xi_{2}^{b}}
     \left[
    (\cdots)C+(\cdots)D_{1}+(\cdots)C
    \right]
    \notag
    \\
    =&\idotsint 
    \frac{1}{\omega^{\frac{b}{2}-1}}B(\chi_{2}) \int_{0}^{\infty}dk_{1}
    k_{1}P_{\delta}(k_{1},\chi_{1})
    \left(\frac{\chi_{2}^{2}W(\chi_{2},\chi_{s})}{2}\right)^{\frac{b-2}{2}}
    \left[
     C I_{c}
    +\frac{2}{\chi_{2}^{2}W(\chi_{2},\chi_{s})}D_{1}I_{d}
    +C I_{e}
    \right].
\end{align}
From the first line to the second, we used Eq.~(\ref{eq: Fexpand_only_xi1}), and took the integral range from zero to infinity. 
We can safely perform this approximation due to the fact that the integral converges.
Here, $I_{c}, I_{d},I_{e}$ are just numbers defined as
\begin{align}
    I_{c}=&
    \int_{0}^{\infty}dx
    \frac{\cos{x^{2}}-\cos^{2}{x^{2}}}{x^{b-1}},
    \\
    I_{d}=&
    \int_{0}^{\infty}dx
    \frac{2\sin{x^{2}}-\sin{2x^{2}}}{4x^{b+1}},
    \\
    I_{e}=&
    \int_{0}^{\infty}dx
    \frac{\sin^{2}{x^{2}}}{x^{b-1}}.
\end{align}
The forth term in Eq.~(\ref{eq: 4th order S variance }) is calculated in a similar way,
\begin{align}
      (4\rm th)=&
   \idotsint 
    \frac{1}{\omega^{\frac{b}{2}-1}}B(\chi_{1})B(\chi_{2}) 
    \frac{k_{L}^{-b+2}}{b-2}
    \notag
    \\
    &\times
    \left\{
    \left(\frac{\chi_{2}^{2}W(\chi_{2},\chi_{s})}{2}\right)^{\frac{b-2}{2}}
    \left[
     C I_{c}
    +\frac{2}{\chi_{2}^{2}W(\chi_{2},\chi_{s})}D_{1}I_{d}
    +C I_{e}
    \right]
    +(1\Longleftrightarrow2)
    \right\}.
\end{align}
Note that the fourth term is essentially determined by $k_{L}$, which is the arbitrary scale.
However, the second term and the third term are determined by the scale at which $P_{\delta}$ changes from an increasing function to a decreasing function
due to the dependence on
$\int_{0}^{\infty}dk_{1}k_{1}P_{\delta}(k_{1})$.
Since we can take $k_{L}$ to be sufficiently larger than this scale,
it is justified to ignore the contribution from the fourth term, and we have the following expression for $ \delta_{S^{2},\rm dc}$:
\begin{align}
    \delta_{S^{2},\rm dc}
    =&
    2\times
    \times 16\left(\frac{3H_{0}^{2}\Omega_{m}}{2}\right)^{4}
    \int_{0}^{\chi_{s}}\frac{d\chi}{\chi^{2}}
    \int_{0}^{\chi}\frac{d\chi'}{\chi'^{2}}
    \int_{0}^{\chi'}d\chi_{1}
    \int_{0}^{\chi'}d\chi_{2}
    \frac{1}{a^{2}(\chi_{1})}
    \frac{1}{a^{2}(\chi_{2})}
    \frac{1}{(2\pi)^{2}}\notag
    \\
    &
    \times 
    \frac{B(\chi_{2}) }{\omega^{\frac{b}{2}-1}} \int_{0}^{\infty}dk_{1}
    k_{1}P_{\delta}(k_{1},\chi_{1})
    \left[
     C I_{c}
    +\frac{2}{\chi_{2}^{2}W(\chi_{2},\chi_{s})}D_{1}I_{d}
    +C I_{e}
    \right]
     \left(\frac{\chi_{2}^{2}W(\chi_{2},\chi_{s})}{2}\right)^{\frac{b_{2}-2}{2}}
\end{align}
This expression can be further simplified by changing the order of integral and using some formula for the gamma function, and finally, we have
\begin{align}
    \label{eq: postSvar Asymp}
   &\delta_{S^{2},\rm dc}
    =
    \frac{3}{4}\left(\frac{3H_{0}^{2}\Omega_{m}}{2}\right)^{4}
    \int_{0}^{\chi_{s}}d\chi_{1}
    \int_{0}^{\chi_{1}}d\chi_{2}
    \frac{
    W^{4}(\chi_{1},\chi_{s})
    \chi_{1}^{4}\chi_{2}^{4}}{a^{2}(\chi_{1})a^{2}(\chi_{2})}
    \frac{1}{(2\pi)^{2}}
    \notag
    \\
    &
    \times
    \left\{
    \frac{B(\chi_{2}) }{\omega^{\frac{b}{2}-1}} \int_{0}^{\infty}dk_{1}
    k_{1}P_{\delta}(k_{1},\chi_{1})
    \left[\frac{\chi_{2}^{2}W(\chi_{2},\chi_{s})}{2}\right]^{\frac{b_{2}}{2}-1}
    \left(1-\frac{2^{\frac{b_{2}}{2}}}{2}\right)
    \Gamma\left(1-\frac{b_{2}}{2}\right)
    \sin{\left[\frac{b_{2}}{2}\right]}
    +(1\leftrightarrow2)
    \right\}.
\end{align}
It is clear from this expression that this term depends not only on the scale corresponding to the Fresnel scale but also on the factor $\int_{0}^{\infty}dk_{1} k_{1}P_{\delta}(k_{1},\chi_{1})$ that is mainly contributed by the large scale matter fluctuation.
This means that the information pertaining to the larger scale fluctuation is encoded within the small scale through the higher-order terms. 
In the physics context, the correlation between the large and the small-scale matter fluctuations arises from the fact that the regions where the large-scale matter fluctuation is significant have higher matter density than areas with small fluctuation, and in this region, the small-scale matter fluctuation is more likely to grow and be amplified simply due to the abundance of matter available.

\bibliography{ref}
\end{document}